\documentclass[%
 aip,
 amsmath,amssymb,
 reprint,%
 floatfix,%
]{revtex4-1}

\usepackage{graphicx}
\usepackage{dcolumn}
\usepackage{bm}
\usepackage{wrapfig}
\usepackage{float}
\usepackage[utf8]{inputenc}
\usepackage[T1]{fontenc}
\usepackage{mathptmx}
\usepackage{etoolbox}
\usepackage{soul}
\usepackage[table]{xcolor}
\usepackage{gensymb}
\usepackage{csquotes}
\usepackage{array}
\usepackage{booktabs}
\usepackage{multirow}
\usepackage{makecell}
\usepackage{rotating}
\usepackage{csquotes}
\providecommand{\contentsname}{Contents}
\usepackage{lineno}

\makeatletter
\renewcommand{\tableofcontents}{%
  \section*{\contentsname}%
  \@starttoc{toc}%
}
\makeatother

\renewcommand{\thesection}{\arabic{section}}
\renewcommand{\thesubsection}{\arabic{section}.\arabic{subsection}}
\renewcommand{\thesubsubsection}{\arabic{section}.\arabic{subsection}.\arabic{subsubsection}}
\makeatletter
\renewcommand{\p@subsection}{}
\renewcommand{\p@subsubsection}{}
\def\@email#1#2{%
 \endgroup
 \patchcmd{\titleblock@produce}
  {\frontmatter@RRAPformat}
  {\frontmatter@RRAPformat{\produce@RRAP{*#1\href{mailto:#2}{#2}}}\frontmatter@RRAPformat}
  {}{}
}%
\makeatother
\begin{document}

\preprint{AIP/123-QED}

\title{Active Learning for Interatomic Potentials: A Review of Uncertainty-Driven Data Efficiency}

\author{Ilya V. Chepkasov}
\affiliation{Skolkovo Institute of Science and Technology, Bolshoy Boulevard 30, bld. 1, Moscow 121205, Russian Federation}
\affiliation{Bauman Moscow State Technical University, 2nd Baumanskaya Str., 5, Moscow 105005, Russian Federation}
\email{I.Chepkasov@skoltech.ru}

\author{Alexander V. Shapeev}%
\affiliation{Skolkovo Institute of Science and Technology, Bolshoy Boulevard 30, bld. 1, Moscow 121205, Russian Federation}
\affiliation{Digital Materials LLC, Kutuzovskaya, str. 4A, Odintsovo 143001, Russia}
\email{a.shapeev@skoltech.ru}

\author{Alexander G. Kvashnin}%
\affiliation{Skolkovo Institute of Science and Technology, Bolshoy Boulevard 30, bld. 1, Moscow 121205, Russian Federation}
\email{A.Kvashnin@skoltech.ru}

\date{\today}

\begin{abstract}
Developing a machine learning interatomic potential (MLIP) that is accurate for the specific conditions under which it will be applied, rather than simply for the predefined configurations chosen by the model developer, is a significant challenge in the area of atomistic simulation.
Active learning (AL) addresses this challenge by closing the loop between candidate generation, data selection, reference calculation, and model updating.
In the data-selection step, an acquisition criterion based on uncertainty, extrapolation, or structural coverage selects configurations for density-functional theory (DFT) labeling, directing expensive quantum-mechanical calculations toward the needs of the target task.
Here, we survey more than seventy studies, with an emphasis on work from 2020 to 2026, to understand how active learning has evolved from a niche, on-the-fly training technique for query-by-committee and Gaussian process force fields into a diverse family of data-efficient strategies.
We organize these workflows by four design choices: feedback-loop organization, acquisition criterion, exploration strategy, and model training and updating regime.
The reviewed examples include on-the-fly molecular dynamics, active data selection for fine-tuning pretrained models, kernel- and representation-based acquisition using frozen pretrained models, and reaction-aware sampling for chemical systems far from equilibrium.
We explore the use of active learning in elemental and alloy mechanics, hydrogen and ion diffusion, ferroelectrics and effective Hamiltonians of functional materials, reactive and photochemical dynamics, high-entropy and compositionally disordered materials, as well as generative and autonomous material discovery pipelines.
These applications demonstrate substantial and reproducible savings in the reference calculations needed to reach a target prediction accuracy, often by an order of magnitude for system-specific potentials, alongside gains in materials-discovery efficiency.
Drawing on a growing body of critical, methodologically self-aware literature, we also show that this achievement carries systematic caveats. 
Namely, actively selected training data are generally not independent and identically distributed samples of the target application distribution, which can introduce active-learning bias.
Uncertainty estimates are not interchangeable across model architectures and are frequently the least reliable exactly where they matter most.
In the unlabeled candidate pool, emphasizing informativeness without considering representativeness or relevance to the target task can degrade downstream accuracy.
We close by discussing the outlook for standardized, uncertainty-calibrated, and observable-targeted active-learning workflows in the foundation-model era.
\end{abstract}


\maketitle
\newpage
\tableofcontents
\newpage

\section{Introduction}

The accuracy of a machine-learned interatomic potential is often limited more by the range of data it was trained on than by the complexity of its architecture.
A neural network with millions of parameters can nevertheless extrapolate, sometimes silently and catastrophically, when a molecular dynamics (MD) simulation, a defect relaxation, or a reaction pathway encounters an atomic environment that is unlike anything it has experienced.
Building a training set that anticipates every such environment by brute-force \emph{ab initio} molecular dynamics (AIMD) sampling is not merely expensive but combinatorially unreasonable.
For example, capturing the four low-lying conformers of a molecule as simple as glycine required 1400 reference points in the Accurate Neural Network Engine for Molecular Energies (ANI) training protocol, and capturing the eight low-lying crystal structures of elemental aluminum required 6300 DFT calculations on cells of more than 100 atoms each \cite{Kulichenko2024}.
Multiply this by every element, every defect type, every reaction channel, and every thermodynamic condition a simulation might explore, and naive, undirected sampling becomes the true bottleneck of atomistic machine learning, not model architecture.

Active learning addresses this issue by using a simple loop.
Candidate configurations, generated during simulation or assembled into a pool, are assessed using an acquisition criterion based on model uncertainty, extrapolation, or structural coverage.
Selected configurations are sent to a reference method (often DFT) for labeling, and the resulting data are used to update the model.
The loop repeats until a task-specific stopping criterion is met, such as uncertainty below a chosen threshold along the trajectories of interest.
The idea of this approach is not new in the machine learning literature, but its application to interatomic potentials draws on several foundational approaches.
An early example is the iterative training strategy of Artrith and Behler \cite{Artrith2012}, which uses disagreement between neural-network potentials to identify configurations requiring additional reference calculations.
Other foundational approaches include the D-optimality / maximum-volume (MaxVol) extrapolation-grade criterion developed for moment tensor potentials (MTPs) \cite{Shapeev2016} by Podryabinkin and Shapeev \cite{Podryabinkin2017}, the Bayesian linear regression on-the-fly approach implemented in VASP by Jinnouchi and Kresse \cite{Jinnouchi2020}, and the Gaussian process (GP) epistemic plus noise uncertainty of the FLARE framework presented by Vandermause \emph{et al.}\ in 2020 \cite{Vandermause2020}.
These frameworks provide the foundation for the review discussed here.

Over the following six years, and with particular acceleration since 2023, this active learning loop has been adapted, extended, and in several cases fundamentally reconsidered across essentially every subfield of atomistic simulation.
Applications and complementary strategies include on-the-fly training of system-specific potentials for defects, dislocations, and rare events \cite{Hodapp2020,Kulichenko2023,Zaverkin2024,Kang2025,Felis2025,bidoggia2026aiida,Unglert2026}, targeted fine-tuning, with or without uncertainty-guided selection, of pretrained foundation MLIPs that already encode broad chemical knowledge \cite{Hwang2026,Radova2025,HanserothB2026,Liu2026,Lee2026}, kernel- and representation-based acquisition and reliability assessment built directly on pretrained models, dispensing with committees or auxiliary uncertainty heads altogether \cite{VargaUmbrichA2026,VargaUmbrichB2026,LiZ2026}, and reaction-aware sampling that couples active learning to transition-state search or metadynamics to capture chemistry far from equilibrium \cite{Achar2025,Vitartas2025,Martyka2025}.

The rise of universal, pretrained \enquote{foundation} MLIPs, trained on tens of millions of Materials-Project-derived configurations \cite{Deng2024,Jacobs2025}, has not eliminated the need for active learning, but it has changed the target.
Where early workflows built task-specific training sets for models trained from scratch, many current workflows use active learning to select additional data for fine-tuning a pretrained model.
Fine-tuning can also use a supplied dataset; active selection and the training regime are separate design choices.
Targeted refinement addresses failures such as potential energy surface \enquote{softening} at surfaces, defects and transition states \cite{Deng2024}, and distribution shift between a foundation model's training data and a downstream task \cite{Kreiman2025}.
Preserving prior knowledge during adaptation is a further challenge because fine-tuning can cause catastrophic forgetting \cite{Hwang2026,Liu2026}.
This shift is reflected in the growing share of MLIP publications that report some form of fine-tuning rather than from-scratch training, and in a parallel, more skeptical literature examining whether active learning, in either its classical or foundation-model-adapted form, is being applied correctly.
Recent critical assessments have documented bias arising from active data selection~\cite{Nair2026}.
They also emphasize that no single uncertainty quantification method dominates across scenarios and that reliable calibration in the unlabeled pool, where it matters most, remains largely underdeveloped~\cite{Nair2026,Park2026}.
The field also lacks a standardized way to demonstrate that active learning actually outperforms a well-chosen random or diversity-based baseline on a task-relevant metric \cite{Nair2026,Deng2024}.

Several substantial reviews have already been published on machine-learned interatomic potentials broadly, including data-generation strategies as one component among many \cite{Kulichenko2024,Kim2026,Jacobs2025,Zhang2025,Thiemann2025,Park2026,LiY2025}.
The contribution of our review is narrower and more focused.
Rather than cataloguing machine learning (ML) architectures or surveying the entire field, we focus on active learning as a specific methodological approach.
We track this approach across a wide range of primary research applications, with an emphasis on work from 2020 to 2026, and critically synthesize the evidence available on its reliability and limitations.

Sec.~\ref{sec2} introduces the active-learning loop and four complementary design choices: organization of the feedback loop, acquisition criterion, exploration strategy, and model training and updating regime.
It also discusses complementary data-efficiency strategies and limitations related to sampling bias, uncertainty calibration, and validation.
Sec.~\ref{sec3}--\ref{sec9} survey active learning and complementary approaches for on-the-fly, system-specific potentials (Sec.~\ref{sec3}), foundation-model fine-tuning and configuration generation (Sec.~\ref{sec4}), acquisition and reliability assessment beyond committees (Sec.~\ref{sec5}), reactive chemistry, rare events, and spectroscopy (Sec.~\ref{sec6}), mechanical, transport, and functional properties (Sec.~\ref{sec7}), high-entropy and compositionally disordered materials (Sec.~\ref{sec8}), and generative and autonomous materials discovery (Sec.~\ref{sec9}).
Sec.~\ref{sec10} examines benchmarks of foundation MLIPs and related system-specific models, emphasizing their implications for active learning, fine-tuning, and observable-specific validation.
We conclude by discussing the current state of the art in methodology and the future outlook for active learning-guided materials simulation.

\section{Active Learning. Theoretical Foundations}
\label{sec2}

\begin{figure*}[!t]
\begin{minipage}[h]{1\linewidth}
\center{ \includegraphics[width=1\linewidth]{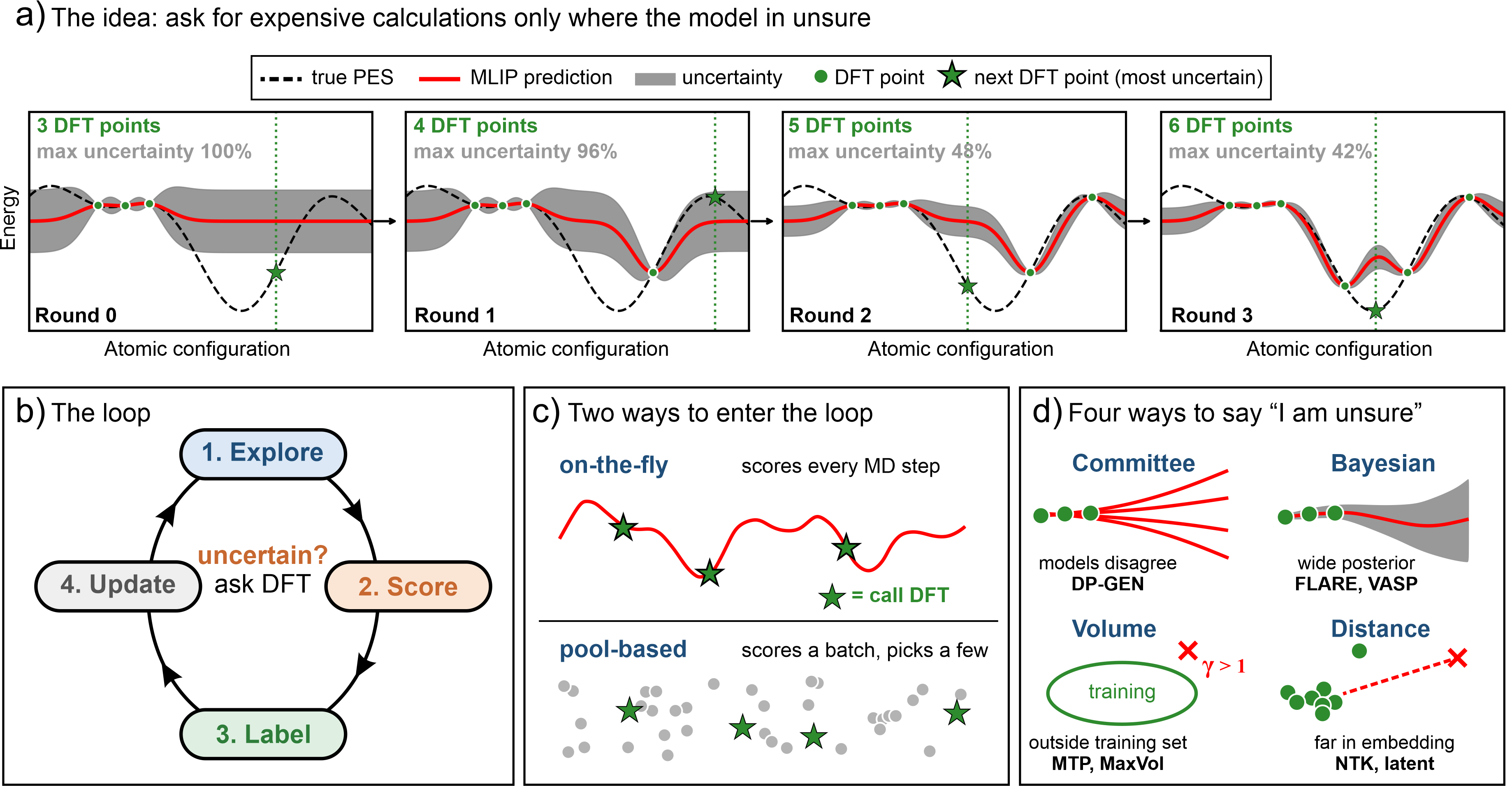}}  \\
\end{minipage}
\caption{\textbf{Active learning for interatomic potentials at a glance.}
\textbf{(a)} A schematic example of a one-dimensional energy surface with a Gaussian-process model. The MLIP model, shown in red, is trained on three DFT points (green) clustered in one region. Its uncertainty (gray band) is largest far from them, so the next DFT calculation (green star) is requested exactly there. After each new point, the band collapses locally, and the maximum uncertainty falls from 100\% to 42\% of its initial value within three rounds while the number of DFT data points increases to 6.
\textbf{(b)} The four-step loop for active learning of interatomic potentials. Explore configuration space with the current MLIP (MD, relaxation, metadynamics, or structure search), score each candidate, label only the selected ones with DFT, and update (retrain or fine-tune) the model.
\textbf{(c)} Candidates reach the loop in two ways: on-the-fly, where the uncertainty is checked at every MD step and the simulation pauses for a DFT call when it exceeds a threshold, or pool-based, where a large pre-generated pool is scored at once and a batch of the top-ranked candidates is labeled per round (Sec.~\ref{sec:feedback}).
\textbf{(d)} Four ways to quantify \enquote{uncertainty} (Sec.~\ref{sec:acquisition}, Sec.~\ref{sec5}), with representative frameworks: disagreement among models in an ensemble (committee; DP-GEN), a wide Bayesian posterior (FLARE, VASP on-the-fly), a configuration outside the volume spanned by the training set, measured by the D-optimality extrapolation grade $\gamma>1$ (moment tensor potentials, MaxVol), and a large distance in the model's own kernel or embedding space (neural tangent kernels, NTKs; latent-space distance).
}
\label{fig:theory}
\end{figure*}

\textit{The active-learning loop.}  
The basic idea of active learning is illustrated in Fig.~\ref{fig:theory}a, which shows several rounds of learning.
Initially, a MLIP is trained on a small set of DFT data (Round 0).
In this example, the uncertainty in the predicted energy is used to select the next configuration for a DFT calculation (green star).
The new reference data are added to the training set, the model is updated, and its uncertainty is reassessed before the next configuration is selected.

For MLIP training, this procedure can be expressed as the four-step cycle shown schematically in Fig.~\ref{fig:theory}b.
In step 1, candidate atomic configurations are generated, typically by running MD, relaxation, metadynamics, or a structural search.
In step 2, candidates are assessed using an acquisition criterion, which may be based on uncertainty, extrapolation, or structural coverage.
In step 3, the selected candidates are labeled with a reference method, usually DFT.
In step 4, the model is retrained or updated on the augmented dataset.
The cycle repeats until a chosen stopping criterion is met, for example, when uncertainty along the trajectories of interest falls below a threshold.

The approaches reviewed here differ in four complementary \emph{design choices}:
\begin{enumerate}
\item \emph{Organization of the feedback loop:} how exploration, labeling, and model updates are scheduled.
\item \emph{Acquisition criterion:} which candidates are selected for labeling.
\item \emph{Exploration strategy:} how candidate configurations are generated.
\item \emph{Model training and updating regime:} whether the model is trained from scratch or adapted from a pretrained model.
\end{enumerate}
These choices can be combined: the same acquisition criterion can be used in on-the-fly or pool-based workflows, and with training from scratch or fine-tuning.

\subsection{Organization of the feedback loop}
\label{sec:feedback}

This design choice concerns the scheduling of exploration, reference calculations, and model updates (Fig.~\ref{fig:theory}c).
On-the-fly active learning, exemplified by FLARE \cite{Vandermause2020}, the Bayesian on-the-fly scheme in VASP \cite{Jinnouchi2020}, and the MaxVol-based MLIP package, see Refs. \cite{Podryabinkin2017,Hodapp2020,Jalolov2024,LiJ2025}, interleaves uncertainty checking with the MD or relaxation trajectory itself.
In its simplest form, when the running uncertainty exceeds a threshold, the simulation pauses, a DFT calculation is launched, the model is updated, and the simulation resumes.
This tightly couples data generation to the dynamical process of interest and, as Sec.~\ref{sec3} shows, can converge with few reference calculations, sometimes fewer than 100, for well-behaved elemental and small-molecule systems \cite{Vandermause2020,Kulichenko2024}.

Pool-based active learning instead scores an already-generated collection of candidate configurations, often produced by MD with a cheaper or foundation potential, and selects a batch for labeling before the next model update.
The candidate pool may be fixed or regenerated in successive rounds using the updated model.
Workflows can also combine uncertainty checks during simulation with batch selection from the resulting trajectories~\cite{Zaverkin2024}.
Selecting from a pool allows comparisons across candidates, for example through clustering or diversity-aware batch selection such as largest-cluster-maximum-distance (LCMD)~\cite{Qi2024,VargaUmbrichB2026}.
Examples of pool-based selection for fine-tuning foundation MLIPs are discussed in Sec.~\ref{sec4}~\cite{Lee2026}.
The choice of loop organization is separate from whether the model is trained from scratch or fine-tuned.

\subsection{Acquisition criteria}
\label{sec:acquisition}

An acquisition criterion determines which candidate configurations are selected for reference calculations.
Selection may draw on uncertainty quantification (UQ), measures of extrapolation, or coverage of configuration space.
Below, we discuss \emph{committee disagreement}, \emph{Bayesian predictive uncertainty}, \emph{D-optimality and extrapolation}, and \emph{geometric coverage and diversity}, illustrated schematically in Fig.~\ref{fig:theory}d.
These approaches are not mutually exclusive and can be combined.

For uncertainty-based acquisition, \emph{calibration} describes how well the reported predictive uncertainty agrees quantitatively with observed prediction errors.
For example, nominal 90\% prediction intervals should contain the reference values in approximately 90\% of cases.

\subsubsection{Committee disagreement}

Query-by-committee (QBC) trains an ensemble of typically 3--8 independently initialized or independently subsampled models and uses their mutual disagreement, most often the standard deviation of predicted energies or forces, as the uncertainty signal.
An early application to interatomic potentials was demonstrated by Artrith and Behler in 2012~\cite{Artrith2012}, who used disagreement between neural-network potentials to identify configurations requiring additional reference calculations.
This strategy is architecture-agnostic and simple to implement.
It explains its ubiquity from the concurrent-learning DP-GEN platform of Zhang \emph{et al.}~\cite{Zhang2019}, which formalized exploration-labeling-training as a three-stage loop and demonstrated it on Al, Mg, and Al--Mg alloys.
The systematic committee-neural-network-potential framework of Schran \emph{et al.}~\cite{Schran2020} showed that averaging over committee members improves accuracy beyond that of any individual member.
However, its disagreement both drives active learning and can be biased during simulation to control the generalization error directly (a single committee model for water, trained on just 814 reference calculations, described liquid water, multiple ice phases, and the air-water interface, all including nuclear quantum effects), through modern concurrent learning pipelines in a broader sense \cite{Kulichenko2024,Achar2025,Ito2025,Gurlek2025}.
QBC multiplies training cost by the ensemble size and, as Sec.~\ref{sec5} details, its disagreement signal correlates only weakly and unreliably with true error in several recent comparisons \cite{Park2026,Nair2026}.

\subsubsection{Bayesian predictive uncertainty}

Bayesian methods instead derive uncertainty directly from a probabilistic model. 
Gaussian-process regression, as implemented in the FLARE framework, produces a closed-form posterior variance that combines an epistemic component (uncertainty in the inferred model) with a noise component optimized by maximizing the marginal likelihood \cite{Vandermause2020}.
In contrast, VASP \cite{kresse_ultrasoft_1999,kresse_ab_1993,kresse_efficient_1996} on-the-fly Bayesian linear regression, developed by Jinnouchi \emph{et al.}~\cite{Jinnouchi2020}, propagates a Bayesian error estimate through the same kernel ridge machinery used for force field fitting \cite{Jinnouchi2020,Ma2025,Volkmer2026,Angeletti2025}.
Partially Bayesian neural networks, in which only a chosen subset of layers carry probabilistic weights sampled via Hamiltonian Monte Carlo, approximate a fully Bayesian network's calibration at a fraction of the cost \cite{Allec2025}.

\subsubsection{D-optimality and extrapolation}

A third family, specific to the linear and moment-tensor potentials descended from compressed-sensing and cluster-expansion methods, uses the D-optimality or maximum-volume (MaxVol) criterion introduced for MTPs by Podryabinkin and Shapeev~\cite{Podryabinkin2017} (see Fig.~\ref{fig:theory}d).
Here, the \enquote{extrapolation grade} $\gamma$ of a configuration is the maximum ratio by which adding its local feature vector can increase the determinant of the training-set design matrix. A value of $\gamma > 1$ flags extrapolation, and a large value (typically $\gamma \gtrsim 10$) flags a candidate that is too extreme to be trusted, even temporarily \cite{Podryabinkin2017,Hodapp2020,Jalolov2024,Jalolov2026,LiJ2025}.

\subsubsection{Geometric coverage and diversity}
\label{sec:geometric}

Geometric acquisition criteria select configurations to improve coverage or diversity in a representation of atomic structure.
This representation can use explicit structural descriptors, such as cell parameters and bond geometry, or learned embeddings from a pretrained model.

Neural tangent kernels (NTKs) or pooled activation kernels derived from a pretrained model can define distances for selecting configurations that improve geometric coverage~\cite{VargaUmbrichA2026,VargaUmbrichB2026}.
Using frozen pretrained representations avoids the cost of training an auxiliary ensemble for acquisition.
We note that the same learned representation can support not only geometric coverage criteria but also Bayesian posterior-variance acquisition (although posterior variance is not itself a distance metric between configurations).
These kernel-based approaches, together with representation-based novelty measures~\cite{LiZ2026}, are discussed in Sec.~\ref{sec5}.

For example, Sharma and Sanvito~\cite{Sharma2024} replaced uncertainty-based acquisition with a purely geometric criterion, CBAD, that discretizes Cell parameters, Bonds, Angles, and Dihedrals into bins and greedily accepts only configurations that populate a previously unrepresented bin, combined with temperature-ramped MD generation.

Qi \emph{et al.}~\cite{Qi2024} proposed a related, explicitly non-iterative alternative, Dimensionality-Reduced Encoded Clusters with sTratified (DIRECT), which featurizes a candidate pool through embeddings of a pretrained M3GNet model, reduces dimensionality by principal component analysis (PCA), and stratified-samples across balanced iterative reducing and clustering using hierarchies (BIRCH) clusters in a single pass.

Wang \emph{et al.}~\cite{WangEntropy2026} addressed the problem that, while entropy-maximized biased molecular dynamics (MD) can produce a collectively redundant dataset by diversifying individual trajectories, it can still be improved.
They proposed a method that combines a local, per-configuration entropy bias with a global, dataset-aware filter.
This filter is based on the log-determinant of the fingerprint covariance matrix of the entire accumulated dataset.
The goal of this filter is to admit only genuinely new configurations.

\subsection{Exploration strategies}
\label{sec:exploration}

The exploration strategy determines how candidate configurations are generated for assessment by the acquisition criterion.
MD samples configurations along trajectories, while structural relaxation explores configurations on the way to local energy minima.
Metadynamics, uncertainty-biased dynamics, and explicit reaction searches can extend exploration to regions that are poorly sampled by ordinary MD, as illustrated in Sec.~\ref{sec3} and \ref{sec6}.
Structural searches and generative models can also propose new structures or compositions, as discussed in Sec.~\ref{sec4} and \ref{sec9}.
The examples below show how these exploration strategies are combined with the selection of configurations for reference calculations.

On accessible simulation timescales, unbiased MD may fail to sample events such as proton transfer or spontaneous defect formation, leaving relevant configurations absent from the training set.
Kulichenko \emph{et al.}~\cite{Kulichenko2023} addressed this directly with uncertainty-driven dynamics for active learning (UDD-AL), see Fig.~\ref{fig:sec3b}a.
A Gaussian bias potential, constructed from the uncertainty in the ensemble disagreement, was added to the molecular dynamics (MD) potential energy surface.
This actively steered the trajectory toward regions that were poorly sampled, in a manner that is conceptually similar to metadynamics, but using model uncertainty instead of a hand-chosen collective variable as the bias coordinate.

Zaverkin \emph{et al.}~\cite{Zaverkin2024} generalized the uncertainty-biasing idea further with an ensemble-free, gradient-based uncertainty estimate calibrated by conformal prediction and a novel \enquote{bias stress}, obtained by differentiating the uncertainty with respect to infinitesimal strain, see Fig.~\ref{fig:sec3b}c.
This additionally drives exploration of unit-cell and volume degrees of freedom in NpT simulations.

\begin{figure*}[!t]
\begin{minipage}[h]{1\linewidth}
\center{ \includegraphics[width=1\linewidth]{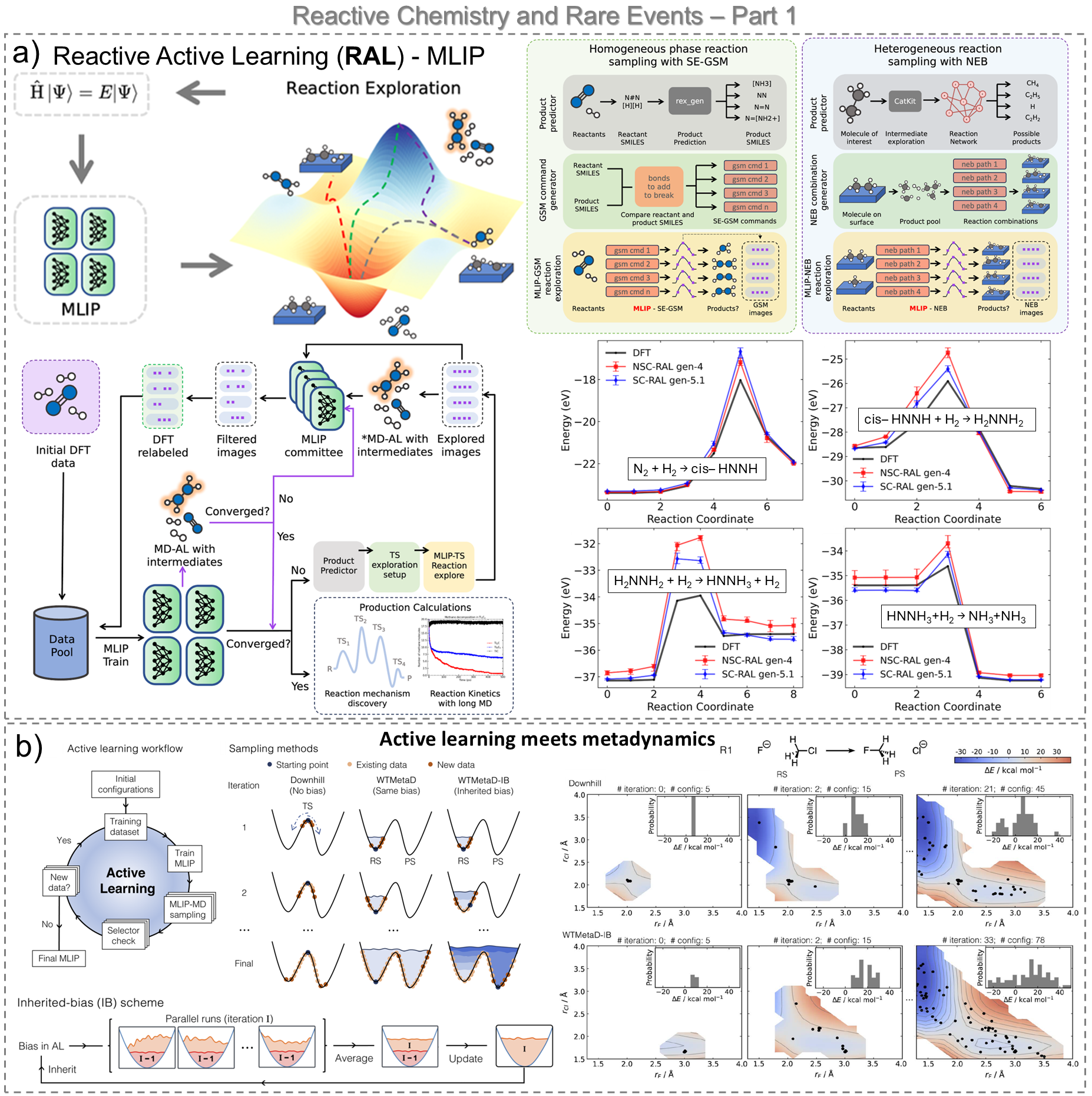}}  \\
\end{minipage}
\caption{\textbf{Active Learning for Reactive Chemistry with Reaction Discovery and Metadynamics.}
\textbf{(a)} The workflow of reactive active learning (RAL) (Ref.~\cite{Achar2025}) in which a committee of bootstrap MLIPs explores reactions by product prediction, transition-state search setup, and growing-string (homogeneous phase) or nudged-elastic-band (heterogeneous phase) path calculation (left).
Images with committee force deviations between user-defined thresholds are relabeled with DFT, optionally interleaved with MD-based active learning (bottom right).
Energy profiles of four uncatalyzed gas-phase ammonia-synthesis benchmark reactions from two RAL generations compared with DFT, with error bars spanning the committee predictions (bottom right).
\textbf{(b)} Active learning with metadynamics (Ref.~\cite{Vitartas2025}) showing the active-learning strategy and the three MLIP-MD sampling modes, downhill from the transition state, well-tempered metadynamics, and well-tempered metadynamics with inherited bias, in which the bias averaged over the parallel runs of one iteration becomes the starting bias of the next (left).
The training of MLIPs for the S$_\mathrm{N}$2 reaction of fluoride with chloromethane in implicit water by downhill and inherited-bias sampling, showing the collected training data on two-dimensional energy surfaces spanned by the forming and breaking bonds after successive iterations (right).
Panels adapted from the cited references.
}
\label{fig:sec6a}
\end{figure*}

Achar \emph{et al.}~\cite{Achar2025} addressed rare-event sampling with Reactive Active Learning (RAL), presented in Fig.~\ref{fig:sec6a}a, which alternates between a conventional configurational active learning step and an explicit reaction discovery step.
In the reaction discovery step, the model generates candidate products, identifies transition states using growing string or nudged elastic band (NEB) methods on the current potential energy surface, and relabels images based on DFT force deviation.

Vitartas \emph{et al.}~\cite{Vitartas2025} coupled active learning with well-tempered metadynamics in their work, see Fig.~\ref{fig:sec6a}b.
They used a new variant of metadynamics called metadynamics with inherited bias.
In this method, the accumulated bias potential is carried forward and averaged across parallel trajectories between active-learning iterations.

Unglert \emph{et al.}~\cite{Unglert2026} adopted a unique approach in which replica exchange nested sampling (RENS) serves as both an exploration engine and an acquisition mechanism, as shown in Fig.~\ref{fig:sec3c}c.
This approach selects thermodynamically representative configurations for density functional theory (DFT) labeling, preserving diversity and likelihood constraints in each round.

Zou \emph{et al.}~\cite{ZouSKMD2026} addressed the diversity-versus-representativeness tension in their work on Stein kernelized molecular dynamics (SKMD), see (Fig.~\ref{fig:sec3d}c).
They introduced a particle-based variational inference sampler that preserves the Boltzmann distribution of the machine learning interatomic potential (MLIP) as its asymptotic distribution.
This approach balances configurational diversity with physically meaningful likelihood, rather than simply trading one for the other, as done by other methods such as metadynamics and hyperactive learning.

Xue \emph{et al.}~\cite{Xue2025} used Bayesian optimization, a tree-structured Parzen estimator, over atom types, compositions, and symmetry-constrained crystal structures, with the energy standard deviation of three independently trained M3GNet-based models, as the uncertainty signal, to generate hypothetical crystals that the current model predicted worst and label them with DFT.

\subsection{Model training and updating}
\label{sec:training}

The training regime determines how the reference data are used to fit or update the model, while the acquisition criterion determines which configurations are selected for labeling.
A MLIP can be trained from scratch or adapted from a pretrained foundation model by fine-tuning its parameters on task-relevant data.
Both training regimes can be used within an iterative active-learning loop, as illustrated in Sec.~\ref{sec3} and \ref{sec4}.
Fine-tuning can also be performed on a supplied dataset without active data selection, as discussed in Sec.~\ref{sec4}.

Liu \emph{et al.}~\cite{Liu2026} presented a tutorial that placed fine-tuning explicitly within the active-learning workflow, see Fig.~\ref{fig:sec4c}a.
This study presented uncertainty-based filtering of a candidate set and iterative active learning as two ways to build the fine-tuning set.
They also caution that continuous application of active learning to fine-tune foundation models may cause catastrophic forgetting, unless safeguards are in place.
They name the automated and uncertainty-informed creation of fine-tuning datasets as a significant open challenge.

In the approach of Hwang \emph{et al.}~\cite{Hwang2026}, multi-head fine-tuning separates readout heads on a shared equivariant backbone trained on a weighted sum of the bulk and surface losses to resist catastrophic forgetting.
The study by Karimitari \emph{et al.}~\cite{karimitari2026systematic} presented the fine-tuning of the MACE-MH-1 foundation model using a multi-head replay strategy, where one head was trained on the new catalytic dataset and another retained 30000 OMAT configurations to prevent catastrophic forgetting.

The FIRE (fine-tuning with integrated replay and efficiency) framework~\cite{LiuFIRE2026} combined efficient configurational sampling with a replay-based continual-learning strategy to resist forgetting.

Radova \emph{et al.}~\cite{Radova2025} presented the \enquote{mace-freeze} scheme where the embedding and other lower layers of MACE-MP are kept fixed while the interaction, product, and readout parameters are retrained (their \enquote{f4} configuration).

Which architecture an active-learning practitioner should pair with a given query strategy is itself an accuracy-versus-cost trade-off rather than a settled choice.
A standardized head-to-head assessment of neural-network potentials, Gaussian approximation potentials (GAPs), spectral neighbor analysis potentials (SNAP), and moment tensor potentials on a common six-element DFT dataset found all four capable of near-DFT accuracy for energies, forces, and derived elastic and phonon properties \cite{Zuo2020}.
However, there are substantial differences in training data requirements and evaluation costs, and a general trade-off is required throughout the architecture choices made in Sec.~\ref{sec3}--\ref{sec9} \cite{Zuo2020}.

\subsection{Complementary strategies and limitations}
\label{sec:complementary}

\textit{Related and complementary data-efficiency strategies.} Active learning is not the only route to data efficiency and reliable predictions, and the literature surveyed here repeatedly pairs or contrasts it with complementary strategies.
These include symmetry- and physics-informed representations, data fusion with non-DFT sources, label-free test-time refinement, and regularization of local atomic energies.
We also discuss limitations of active learning related to sampling bias, uncertainty calibration, and validation.

Symmetry- and physics-informed representations can reduce the effective sample complexity directly, independent of any acquisition strategy.
The Bravais-inspired gradient-domain machine learning (BIGDML) framework \cite{Sauceda2022}, for example, exploits the full crystallographic symmetry group of a periodic solid to increase the effective number of training samples.
The authors achieved state-of-the-art accuracy in graphene force calculations with ten training geometries.
Atom-centered descriptors typically require orders of magnitude more training data, but with BIGDML, this problem is solved \cite{Sauceda2022}.

Data fusion with non-DFT sources, such as classical force fields used as a physically motivated prior \cite{YuanEC2025}, or macroscopic experimental observables like elastic constants and lattice parameters, can help correct systematic model deficiencies that additional DFT-only active learning cannot address \cite{Roecken2024}.
DFT itself can be the source of discrepancies, so it is important to use other sources of data to improve the accuracy of the model.

\begin{figure*}[!t]
\begin{minipage}[h]{1\linewidth}
\center{ \includegraphics[width=1\linewidth]{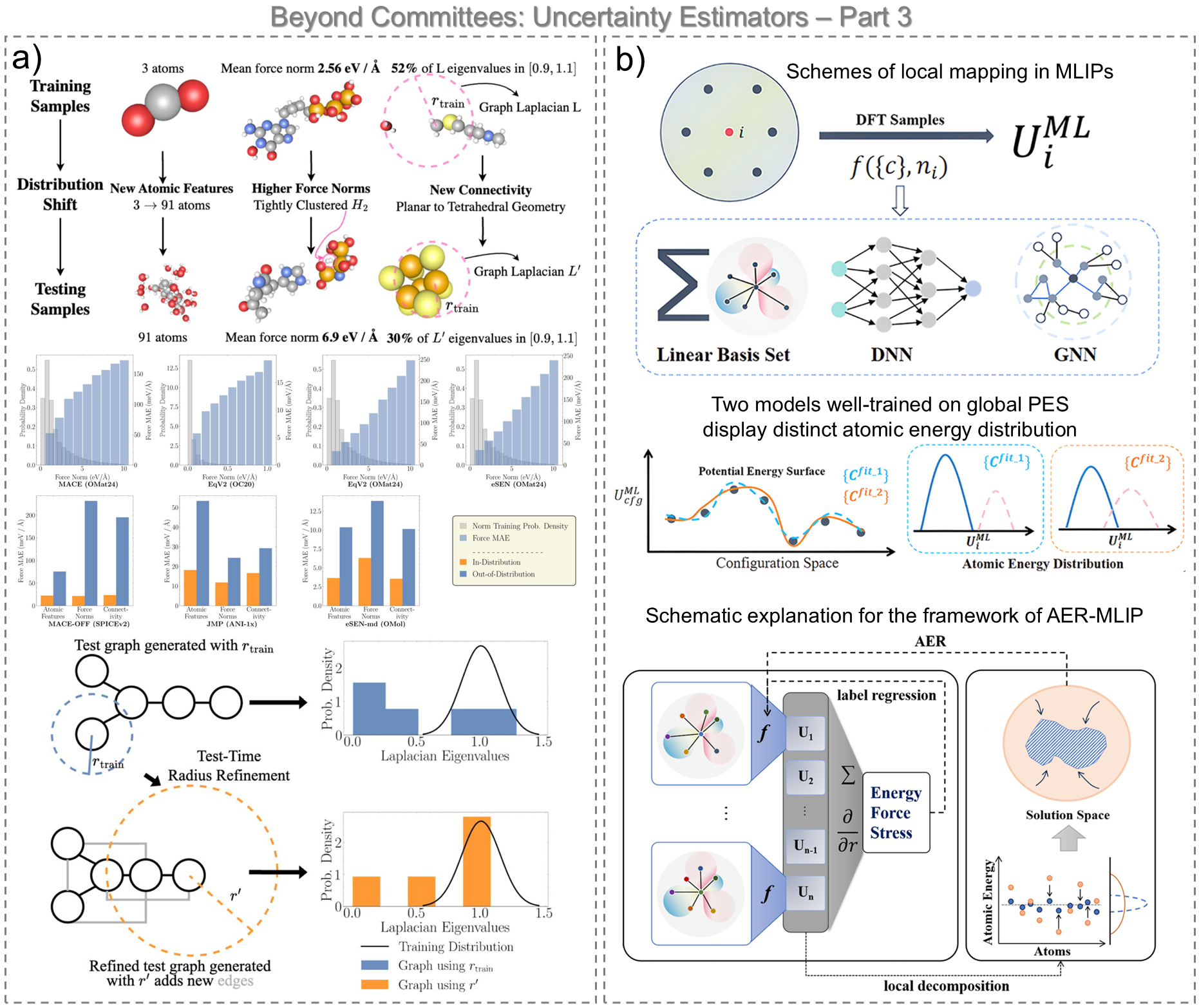}}  \\
\end{minipage}
\caption{\textbf{Label-Free Corrections with Distribution-Shift Diagnosis, Test-Time Refinement, and Local-Energy Regularization.}
\textbf{(a)} Distribution shifts in MLIPs (Ref.~\cite{Kreiman2025}) showing the three types of identified shifts, namely atomic-feature shift (much larger molecules than in training), force-norm shift (unusually close contacts with high force norms), and connectivity shift (graph geometries unlike those seen in training), illustrated with SPICE and SPICEv2 examples (top).
Force errors of large open-source models (MACE, EquiformerV2, eSEN, MACE-OFF, JMP, and eSEN-md) on in- and out-of-distribution structures, which are 2--10 times larger under shift (middle).
Test-time radius refinement, which chooses the graph cutoff for each test structure so that its Laplacian eigenvalues align with those of the training distribution (bottom).
\textbf{(b)} The local mapping of MLIPs within atomic energy regularization (Ref.~\cite{Hu2025}) and the non-uniqueness of the local atomic-energy projection, whereby two models trained equally well on the global potential-energy surface distribute the energy differently among atoms (top).
The AER-MLIP framework (bottom), which combines a data-driven label regression with a probability-based solution-space filter that suppresses spurious fluctuations of the atomic energies.
Panels adapted from the cited references.
}
\label{fig:sec5c}
\end{figure*}

Kreiman and Krishnapriyan~\cite{Kreiman2025} formally decompose MLIP generalization failure into three types, namely atomic-feature shift, force-norm shift, and graph connectivity shift, see Fig. \ref{fig:sec5c}a.
They propose two label-free test-time refinement strategies: radius refinement, which re-optimizes the graph cutoff for each test structure, and test-time training against a cheap physical prior as an alternative to costly retraining or active learning.

A distinct source of uncertainty, which is largely orthogonal to the acquisition process, is addressed by Hu \emph{et al.}~\cite{Hu2025}, see Fig.~\ref{fig:sec5c}b.
Because the atom-partitioned local energy of a MLIP is not a physically observable quantity in itself, independently trained models may agree closely on the total energy and forces, while disagreeing significantly on how this energy is distributed among individual atoms.
This ambiguity, which is usually harmless, can cause a prior gradient-based thermal-conductivity method to fail completely in systems with large atomic displacements.
Their atomic energy regularization (AER) framework uses a probabilistic prior based on long-range statistical correlations to constrain the local decomposition of the system.
This prior is built without the need for any new labeled data.

\textit{What active learning does not guarantee.} Many investigations state that an active learning cycle is not always beneficial.
Because it deliberately, and by design, selects atypical, high-uncertainty configurations rather than a representative sample of the target application distribution, actively curated training data violate the independent-and-identically-distributed assumption underlying standard empirical-risk-minimization validation.
This may cause an effect termed \enquote{active-learning bias} that can inflate apparent generalization \cite{Nair2026}.
Sampling strategies that focus solely on optimizing informativeness, without considering representativeness, physical plausibility, or relevance to the downstream task, have been shown to select statistically irrelevant, extreme, or distorted configurations. 
In at least one instance, this has caused random sampling to outperform active learning for MLIP training \cite{Nair2026}.
No single uncertainty quantification method dominates across different model architectures and applications, and each has documented failure modes. 
Bayesian methods, ensembles, and geometric criteria all have their own strengths and weaknesses.
The reliable calibration of uncertainty in the unlabeled pool, where acquisition decisions are made, remains largely unsolved \cite{Nair2026,Park2026}.
These caveats do not diminish the empirical successes catalogued in Sec.~\ref{sec3}--\ref{sec9}, but they do mean that a reported reduction in DFT calculations should be read as a demonstrated efficiency gain on the specific task and acquisition protocol tested, not as evidence that the resulting dataset is representative, complete, or transferable beyond it.
In addition to these issues, Maxson \emph{et al.}~\cite{Maxson2024} note that the literature on MLIPs has not yet reached a consensus on shared standards for how training sets, active learning protocols, and validation procedures are constructed.
They proposed a checklist for practitioners to ensure that the claimed data efficiency gains can be reproduced and evaluated independently. 
This checklist includes specifications for hardware and software, as well as guidelines for training data generation and validation practices.
This documentation gap compounds rather than resolves the active learning bias and calibration concerns mentioned above.

\section{On-the-Fly Active Learning for System-Specific Interatomic Potentials}
\label{sec3}

The clearest and most direct demonstration of the value of active learning is in the on-the-fly training of a customized, system-specific machine learning interatomic potential (MLIP) from a small or non-existent dataset, interleaving calls to the data collection function (DFT) with the simulation that the model is being trained for.

\subsection{Foundational on-the-fly learning frameworks}
\label{sec:on-the-fly-foundations}

Podryabinkin and Shapeev~\cite{Podryabinkin2017} developed the D-optimality/MaxVol acquisition criterion for moment tensor potentials (MTPs)~\cite{Shapeev2016} and compared it with a fixed-interval learning-on-the-fly baseline inspired by Li, Kermode, and De Vita~\cite{li2015molecular}.
The earlier \enquote{learning-and-remembering} scheme had demonstrated a factor-of-30 reduction in reference calculations.

For a 0.5~\textmu s parallel-replica molecular dynamics (MD) simulation of bulk lithium, a 128-atom system at the melting temperature, the D-optimality scheme performed with zero trajectory failures, using only about 50 additional density functional theory (DFT) calculations.
In the same study, the fixed-interval baseline added a configuration every 100 MD steps and required 1500 additional calculations to increase the trajectory's failure time by a factor of ten, from 15 ps to 150 ps, while still exhibiting failures.
In a separate threshold test, raising the extrapolation-grade threshold from $\gamma_{\mathrm{th}}=1$ to $11$ traded accuracy against cost, reducing the number of reference calculations from 3712 to 397 with force root-mean-square error (RMSE) staying near 0.015--0.016~eV/\AA{} throughout (the maximum error rose at $\gamma_{\mathrm{th}}=11$), and the authors recommend $\gamma_{\mathrm{th}}$ between 2 and 11 in practice.

Using the Bayesian on-the-fly framework implemented in VASP, Jinnouchi \emph{et al.}~\cite{Jinnouchi2020} reported that across a range of applications, alloy nanoparticle catalysis, hybrid-perovskite phase transitions, and nuclear-quantum-effect-corrected hydrogen thermodynamics in a Laves-phase metal hydride, more than 99\% of first-principles calculations could be bypassed during training while production runs evaluating thermodynamic and kinetic properties were accelerated by several orders of magnitude.
This approach has been tested using actively trained force fields for Rh$_{1-x}$Au$_x$ nanoparticles, which predicted an optimal catalyst size and composition of 2~nm and $x=0.3$ for NO decomposition, based on surface reactivity screening.
In another example, path-integral simulations using an actively trained potential for TiCr$_2$H$_{1.3}$ Laves hydride resulted in a hydrogenation enthalpy of -19 kJ/mol H$_2$, which is within the experimental range of -20 to -17 kJ/mol H$_2$.
The authors state that this level of agreement can only be achieved with a quantum simulation that takes zero-point vibrational contributions into account.

The D-optimality/MaxVol and VASP frameworks, together with FLARE~\cite{Vandermause2020}, were foundational developments in on-the-fly active learning, combining different acquisition criteria with the feedback loop described in Sec.~\ref{sec2}.
The examples below apply FLARE and local-environment MTP learning to phase-transition and defect simulations.

\begin{figure*}[!t]
\begin{minipage}[h]{1\linewidth}
\center{ \includegraphics[width=1\linewidth]{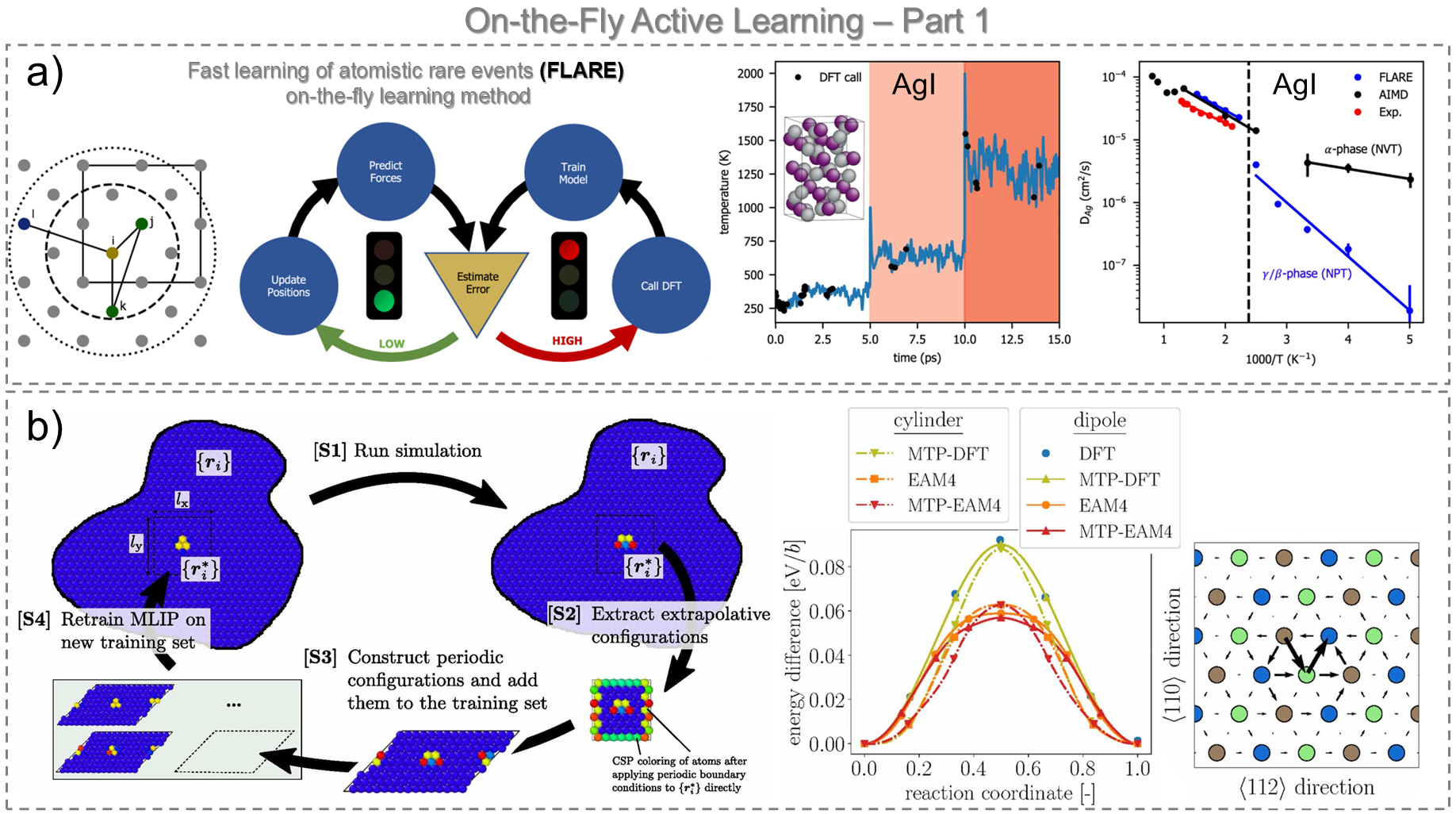}}  \\
\end{minipage}
\caption{\textbf{Foundational on-the-fly learning frameworks in action: Bayesian Gaussian processes and MaxVol extrapolation grades.}
\textbf{(a)} FLARE (Ref.~\cite{Vandermause2020}) framework representing the two- and three-body kernels that compare interatomic distances within cutoff spheres around each atom. 
It also includes a two-loop on-the-fly algorithm, which allows molecular dynamics to proceed on the Gaussian-process force field until the epistemic uncertainty of any force component exceeds a certain multiple of the noise uncertainty. 
When this happens, DFT (density functional theory) is called, and the local environments with the highest uncertainty are added to the training set (left).
The temperature profile from a 15~ps training run with DFT calls for AgI indicated by black dots (middle). 
The Ag diffusion coefficients calculated from 1~ns simulations of 10976 atoms using the trained force field, compared to AIMD (ab initio molecular dynamics) and experiment (right). This allows one to resolve the fast-ion phase transition at 420~K.
\textbf{(b)} Cut-and-periodize learning for a screw dislocation in bcc tungsten (Ref.~\cite{Hodapp2020}). 
The four-step loop of simulating until extrapolation is detected, running the simulation (S1), extracting the extrapolative core neighborhoods (S2), constructing periodic configuration for training that keep those neighborhoods while removing artificial boundary environments (S3), and retraining, S4 (left).
Energy curves along the dislocation-glide path for cylinder and periodic-dipole cells, on which the MTP trained on DFT and the EAM-trained MTP reproduce their references to within about 1~meV per Burgers vector together with the differential-displacement map of the intermediate core configuration (right).
Panels adapted from the cited references.
}
\label{fig:sec3a}
\end{figure*}

Vandermause \emph{et al.}~\cite{Vandermause2020} demonstrated Bayesian on-the-fly learning with FLARE, a Gaussian process force field whose inherent epistemic and noise uncertainties trigger new DFT calculations during molecular dynamics (MD), as shown in Fig.~\ref{fig:sec3a}a.
The most striking demonstration is the superionic conductor AgI, which undergoes a structural transition to a fast-ion-conducting phase near 420~K.
Starting from a 48-atom $\alpha$-phase cell of AgI, a 15~ps on-the-fly training run, in which the instantaneous temperature was raised at 5 and 10~ps, required only 39 DFT calculations in total.
The resulting force field, mapped to a tabulated form and run in 1~ns NpT simulations of 10976 atoms between 200 and 650~K, showed the sharp jump in silver diffusivity between 400 and 450~K (Fig.~\ref{fig:sec3a}a).
This marks the transition, and reproduced the $\alpha$-phase diffusion coefficients with a slight vertical offset but comparable activation energies of 0.107~eV, against 0.114~eV from \emph{ab initio} MD and 0.093~eV from experiment.
The same procedure needed fewer than 100 DFT calls to train an aluminum force field spanning melting and solidification, and made vacancy diffusion in aluminum more than 300 times faster to simulate than an equivalent \emph{ab initio} MD run (68.8~h on 32 cores).
The tabulated two-body force field ran at $5.6\times10^{-6}$~s/atom/timestep, on par with a classical embedded-atom method (EAM) potential ($2.2\times10^{-6}$~s/atom/timestep) and roughly four orders of magnitude below the $\sim$0.1~s/atom/timestep reported in the literature for a Gaussian approximation potential for silicon.

Hodapp and Shapeev~\cite{Hodapp2020} extended the complementary D-optimality/MaxVol on-the-fly framework, developed for moment tensor potentials, to simulations far too large for direct DFT, see Fig.~\ref{fig:sec3a}b.
By detecting locally extrapolated $\sim100$-atom neighborhoods within multi-thousand-atom cells and cutting, symmetrizing, and periodizing them into small DFT-tractable training configurations, the researchers computed the Peierls barrier and Peierls stress of a $\langle111\rangle$ screw dislocation in bcc tungsten using training sets of only 20-40 configurations.
The obtained Peierls stress of $2.29\pm0.05$~GPa lies within the range of 1.71--2.8~GPa from earlier DFT estimates. 
Meanwhile, the same protocol applied to a classical EAM potential reproduced a large-scale reference calculation with an error of 6--8\%.
This same cut-and-periodize strategy for extracting local training fragments from otherwise intractable large-scale structures recurs throughout the mechanical-properties investigations surveyed in Sec.~\ref{sec7} \cite{Podryabinkin2022,Jalolov2024,LiJ2025,Jalolov2026}.

\subsection{Rare-event sampling and configurational coverage}

\begin{figure*}[!t]
\begin{minipage}[h]{1\linewidth}
\center{ \includegraphics[width=1\linewidth]{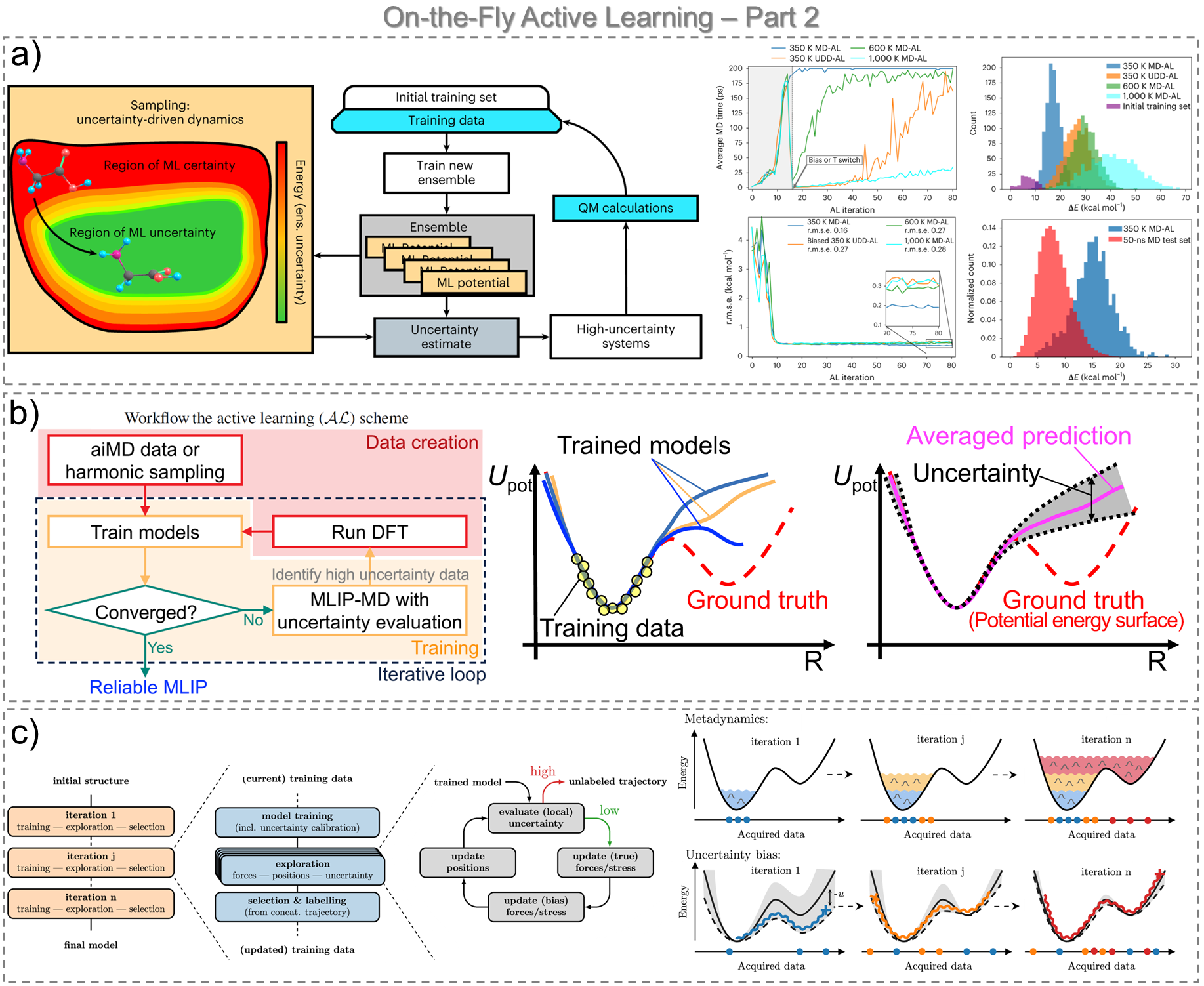}}  \\
\end{minipage}
\caption{\textbf{Exploring where the model is blind: Uncertainty-biased dynamics and ensemble uncertainty.}
\textbf{(a)} Uncertainty-driven dynamics for active learning (UDD-AL) for glycine (Ref.~\cite{Kulichenko2023}) in which a bias derived from ensemble uncertainty drives the dynamics toward regions of ML uncertainty whose configurations are labeled by QM calculations and added to the training set (left). 
Comparison of 350, 600, and 1000~K MD-AL with 350~K UDD-AL in terms of the average MD time needed to meet the uncertainty criterion (right), the energy distributions of the sampled data sets, the test-set RMSE versus AL iteration, and the energy distribution of the 350~K MD-AL data against the 50~ns MD test set.
\textbf{(b)} Active learning for strongly anharmonic materials (Ref.~\cite{Kang2025}) in which an initial AIMD or harmonic-sampling dataset trains a MLIP whose MLIP-MD trajectories are screened for high-uncertainty configurations that are labeled with DFT until convergence (left), and the ensemble uncertainty, defined as the spread of predictions of several models trained on different data, which diverge beyond the trained region (right).
\textbf{(c)} Uncertainty-biased MD (Ref.~\cite{Zaverkin2024}) showing the AL loop with parallel MD explorations terminated at an uncertainty threshold, followed by selection, labeling, and retraining (left). 
Metadynamics, which fills the energy surface with Gaussian biases along hand-picked collective variables (right top), and uncertainty bias, which subtracts the predicted energy uncertainty from the energy surface to push trajectories toward poorly known regions without collective variables (right bottom).
Panels adapted from the cited references.
}
\label{fig:sec3b}
\end{figure*}

The following studies address limitations of naive on-the-fly sampling, discussed in Sec.~\ref{sec:exploration}.
For acetylacetone proton transfer at 350~K, uncertainty-driven dynamics (UDD) sampling~\cite{Kulichenko2023} produced 90 reactive events in 0.5~ns of simulation, compared to zero for unbiased MD at the same temperature, and only 48 events at 620~K.
Kang \emph{et al.}~\cite{Kang2025} identified a related and more insidious failure mode in equivariant graph-neural-network potentials for strongly anharmonic materials, see Fig.~\ref{fig:sec3b}b. 
Despite sub-meV/atom and sub-0.01-eV/\AA{} test errors on 112 bulk materials from the NOMAD database, roughly 9\% of these materials showed qualitatively wrong rare-event behavior.
This behavior can be classified into two categories: a \enquote{missing minimum} (CuI) where the model fails to learn a spontaneous defect state, and a \enquote{fictitious minimum} (AgGaSe$_2$) where the model creates an unphysical minimum.
An acquisition function that combines a soft preference for high uncertainty with a penalty for unphysical high predicted energies, applied after approximately 300-330 additional density functional theory (DFT) calculations, corrected both of these issues and restored phonon lifetime predictions that were in agreement with \textit{ab initio} molecular dynamics (AIMD).

The uncertainty-biased MD scheme of Zaverkin \emph{et al.}~\cite{Zaverkin2024} is described in Sec.~\ref{sec:exploration}.
For alanine dipeptide (Fig.~\ref{fig:sec3b}c) this beat unbiased 300~K and 600~K MD by factors of 13$\times$ and 2$\times$ in energy and force RMSE respectively, and for the flexible metal--organic framework MIL-53(Al) it outperformed metadynamics with a hand-chosen collective variable past roughly 200 sampled configurations.

When applied to silicon, germanium, and titanium at the r2SCAN level, replica exchange nested sampling (RENS)~\cite{Unglert2026} (Sec.~\ref{sec:exploration}) converged to full pressure-temperature phase diagrams within 10-15 iterations. For silicon, it reproduced the accuracy of a large, manually curated database of 2475 structures from an initial pool of only 480 configurations, right panel on Fig.~\ref{fig:sec3c}c.
For the metal--organic frameworks ZIF-8 and MOF-5, selection using cell parameters, bonds, angles, and dihedrals (CBAD)~\cite{Sharma2024} (Sec.~\ref{sec:geometric}) reached full 1000~K MD stability with only a few hundred training configurations and a 392-parameter linear SNAP potential, reproducing experimental thermal-expansion coefficients and a phenylene-ring rotation free-energy barrier within 0.03~eV of experiment.

Wang \emph{et al.}~\cite{WangEntropy2026} tested their combined entropy-biased sampling and dataset-aware filtering (Sec.~\ref{sec:geometric}) on carbon, silicon, and NaCl.
They found that the entropy-driven sampling produced 3-10 times lower energy mean absolute error (MAE) compared to random-MD sampling.
However, the broader fingerprint-space coverage led to an overrepresentation of distorted configurations for carbon.
To resolve this issue, the authors blended 80\% of entropy-selected configurations with 20\% near-equilibrium random-MD configurations, achieving a balance between force/stress accuracy and computational efficiency.

Stein kernelized molecular dynamics (SKMD)~\cite{ZouSKMD2026} (Sec.~\ref{sec:exploration}) was demonstrated on two examples: active learning of a neural network model of the M\"uller--Brown potential and fine-tuning of a MACE potential for alanine dipeptide.
In both cases, SKMD achieved higher model accuracy using the same number of training samples as the active-learning baselines.

\subsection{Automated and scalable workflows}

\begin{figure*}[!t]
\begin{minipage}[h]{1\linewidth}
\center{ \includegraphics[width=0.92\linewidth]{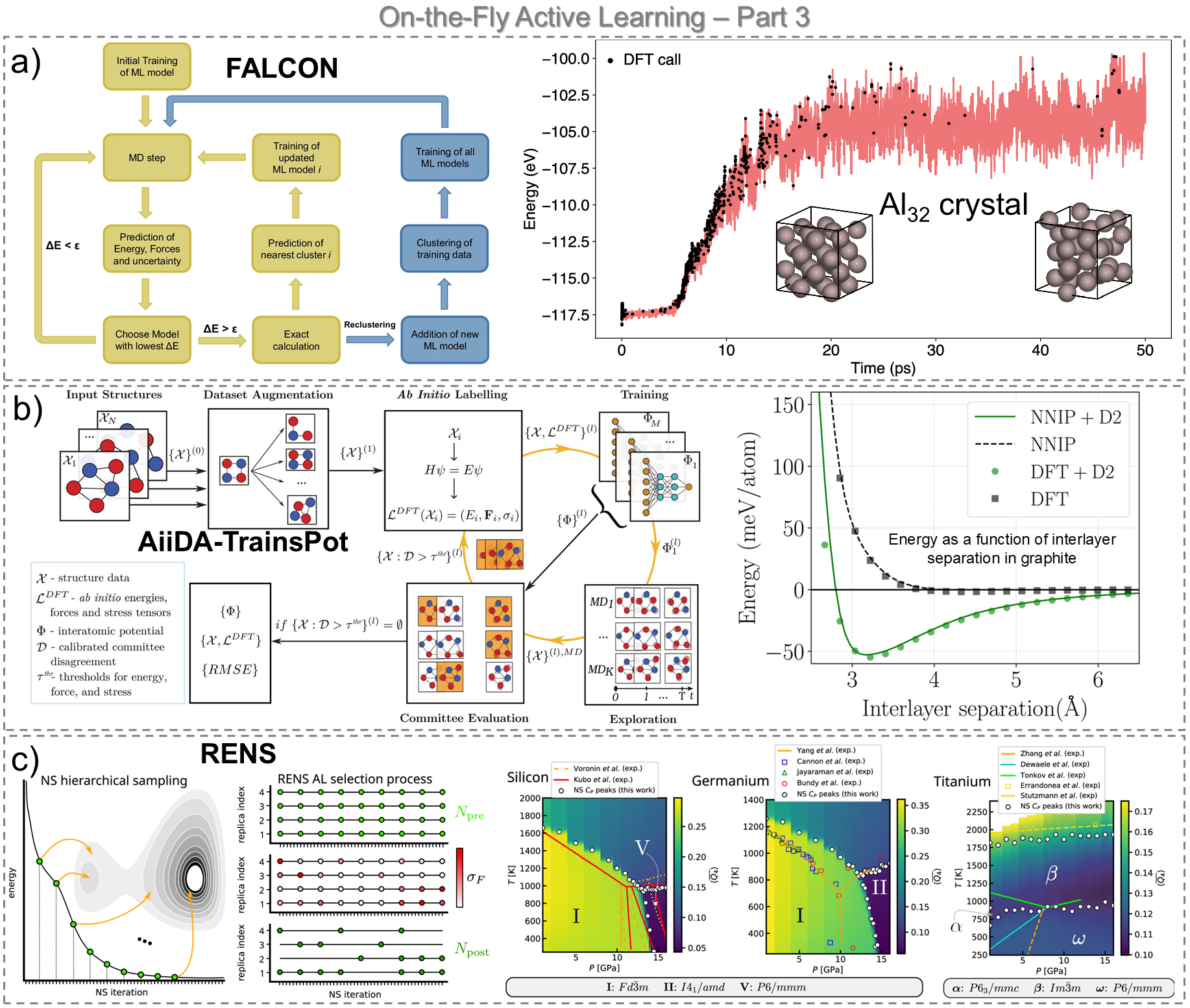}}  \\
\end{minipage}
\caption{\textbf{Automated and scalable on-the-fly workflows: clustered GPR, committee-driven pipelines, and nested sampling.}
\textbf{(a)} FALCON (Ref.~\cite{Felis2025}) represents the on-the-fly loop extended by $k$-means clustering of the training structures, in which the model with the lowest predicted uncertainty is chosen, an exact calculation is triggered only when the uncertainty exceeds the accuracy threshold.
Here only the nearest cluster model is retrained, and all data are reclustered once the average model size is reached (left). 
Total energy during the first 50~ps of an on-the-fly simulation of an Al$_{32}$ crystal heated to 3000~K, with DFT calls marked as black dots (right).
\textbf{(b)} AiiDA-TrainsPot (Ref.~\cite{bidoggia2026aiida}) representing the automated workflow of structure augmentation, \emph{ab initio} labeling, training of a committee of potentials, MD exploration, and committee evaluation, in which structures whose calibrated disagreement exceeds a threshold are added to the reference set (left). Binding energy of graphite as a function of interlayer spacing for the trained NNIP with and without D2 dispersion corrections compared with DFT (right).
\textbf{(c)} Replica-exchange nested-sampling active learning (Ref.~\cite{Unglert2026}) including the nested-sampling hierarchy and the RENS selection process, in which trajectories from several replicas are sliced into $N_\mathrm{pre}$ candidate configurations, colored by force uncertainty, and subsampled to $N_\mathrm{post}$ configurations for labeling (left).
The resulting phase diagrams of silicon, germanium, and titanium, with energy--volume and enthalpy curves of the competing phases (right).
Panels adapted from the cited references.
}
\label{fig:sec3c}
\end{figure*}

The following studies address the software-engineering and computational-scaling side of on-the-fly training.
Felis and Dononelli's FALCON~\cite{Felis2025} tackles the $O(N^3)$ retraining cost that plagues a single, ever-growing Gaussian-process model by clustering accumulated training data with $k$-means into multiple smaller Gaussian process regression (GPR) sub-models, retraining only the nearest sub-model per new point, see Fig.~\ref{fig:sec3c}a.
For a 115-atom water-in-carbon-nanotube DFT system this reduced the total active-learning cost to 38.4~h against an estimated 500-plus days for full AIMD.
At the same time the melting of bulk aluminum required only 500 DFT calculations in the first 25~ps of heating and just 69 more over the following 225~ps while reproducing AIMD-level radial distribution functions and diffusion coefficients.
Bidoggia \emph{et al.}~\cite{bidoggia2026aiida} packaged committee-disagreement-calibrated on-the-fly active learning into AiiDA-TrainsPot, see Fig.~\ref{fig:sec3c}b.
AiiDA-TrainsPot is an automated, provenance-tracked workflow orchestrating DFT, data augmentation, and classical MD, demonstrated on pristine and defective carbon allotropes and monolayer W$_x$Mo$_{1-x}$Te$_2$ alloys. 
The architecture of the workflow and iterative active-learning strategy are summarized in Fig.~\ref{fig:sec3c}b, which schematically depicts how initial structures are augmented, labeled by DFT, used to train a committee of potentials, and then evaluated on MD-sampled configurations to identify the most uncertain structures for subsequent labeling.
A distinguishing feature of this approach is the on-the-fly calibration of committee disagreement against \textit{ab initio} reference errors, which ensures reliable uncertainty estimates and minimizes both false positives and false negatives in the selection of configurations for first-principles calculations\cite{bidoggia2026aiida}.

\subsection{Reducing the cost of reference calculations}

\begin{figure*}[!t]
\begin{minipage}[h]{1\linewidth}
\center{ \includegraphics[width=1\linewidth]{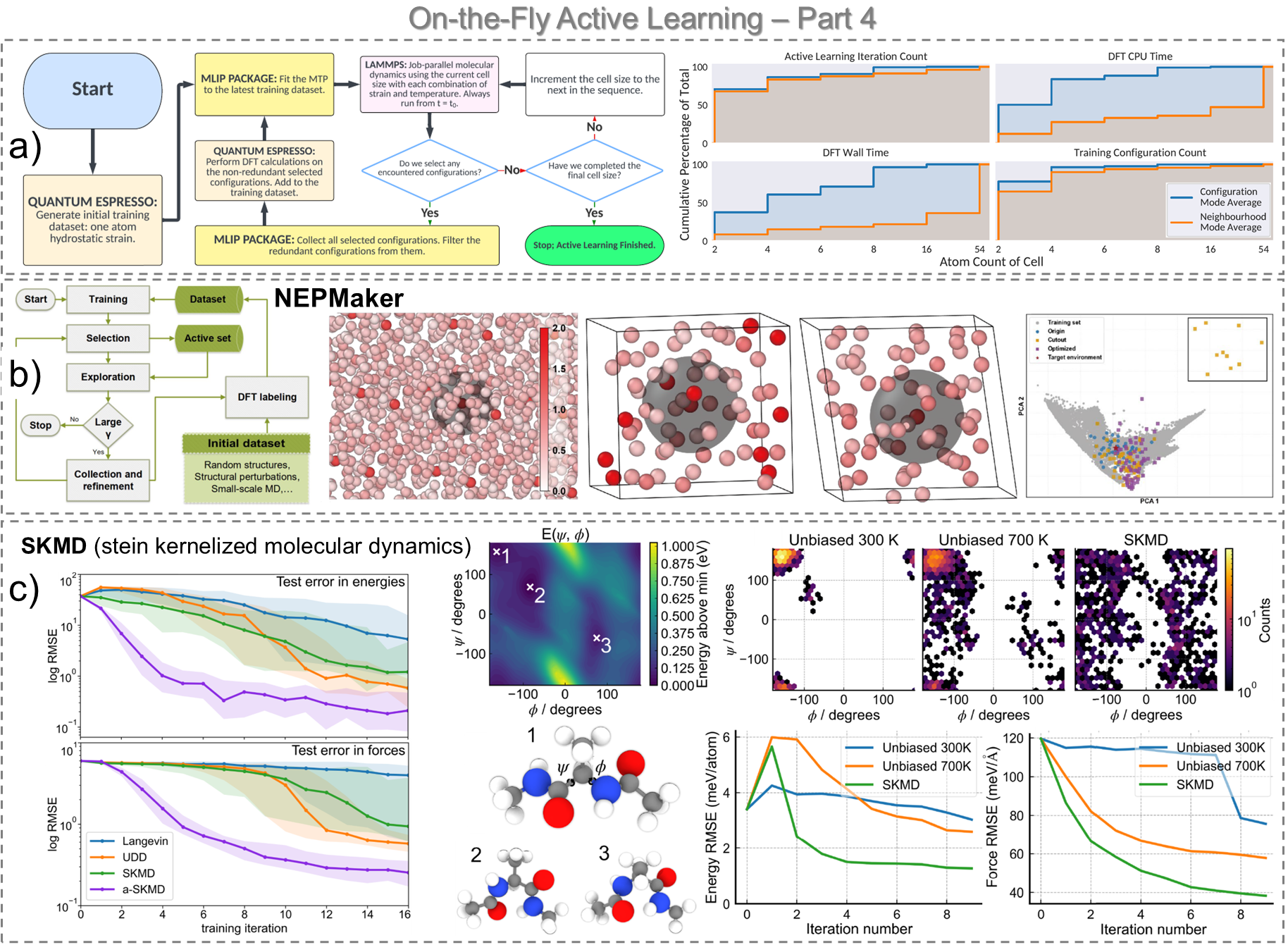}}  \\
\end{minipage}
\caption{\textbf{Cheaper labels and smarter sampling: Small cells, local fragments, and kernel-steered dynamics.}
\textbf{(a)} The computational protocol of small-cell active learning for potassium (Ref.~\cite{Meng2025}) in which an initial one-atom hydrostatic-strain dataset is used to train an MTP, LAMMPS MD is run at each cell size, non-redundant encountered configurations are labeled with DFT and added, and the cell size is incremented until the final size is reached (left).
Cumulative distributions of AL iteration count, DFT CPU time, DFT wall time, and training-configuration count by cell size for configuration-mode and neighborhood-mode extraction (right).
\textbf{(b)} NEPMaker (Ref.~\cite{WangNEP2026}) which is the NEP active-learning framework of training, MaxVol active-set selection, exploration, collection and refinement of large-$\gamma$ configurations, and DFT labeling (left).
Uncertainty-based optimization of boundary atomic environments, in which a target environment in a liquid-silicon supercell (colored by extrapolation grade) is cut out.
The cut introduces spurious extrapolative boundary environments visible in a principal-component map of the descriptors, and the atoms outside the sphere and the lattice are optimized to remove them while preserving the target environment (right).
\textbf{(c)} Stein kernelized molecular dynamics (Ref.~\cite{ZouSKMD2026}) which is the Ramachandran energy surface and three minima of alanine dipeptide (left).
The coverage of the $(\psi,\phi)$ surface by samples from unbiased MD at 300 and 700~K and from SKMD, and the energy and force RMSE of the fine-tuned MACE model versus iteration (left).
Test error in energies and forces on the M\"uller--Brown benchmark for Langevin, UDD, SKMD, and adaptive-SKMD sampling, with the median and $25^{th}$--$75^{th}$ percentile band over 50 trials (right).
Panels adapted from the cited references.
}
\label{fig:sec3d}
\end{figure*}

The following studies reduce the cost of labeling by using small cells or local fragments.
Meng \emph{et al.}~\cite{Meng2025} have systematically characterized small-cell active learning, using the MTP within MLIP package \cite{novikov_mlip_2021, podryabinkin_mlip-3_2023} (see Fig.~\ref{fig:sec3d}a), for potassium and sodium-potassium systems with cells containing only 1-8 atoms, rather than the conventional 54-atom cells used for on-the-fly training. 
They showed that this small-cell approach achieves up to two orders of magnitude cost savings, with some training runs finishing in less than 120 core hours, while still yielding accurate \textit{ab initio} agreement on static and thermodynamic properties. 
The small training cells are too small to include solid-liquid interfaces or critical exponents, but they extend the lineage of small-cell studies for boron, aluminum, magnesium, zirconium, and NaCl.
Wang \emph{et al.}~\cite{WangNEP2026} applied the same local-fragment and MaxVol-based approach (Sec.~\ref{sec:on-the-fly-foundations}) to the GPU-accelerated neuroevolution potential (NEP) framework using their NEPMaker tool.
The general procedure is shown in Fig.~\ref{fig:sec3d}b.
This method helps to identify extrapolated local chemical configurations within large-scale simulations on the fly and embed them into locally periodic structures with optimized boundary atoms that remain close to the training distribution, see Fig.~\ref{fig:sec3d}b.
This extends the application of large-cell active learning to widely used high-throughput machine learning-based interatomic potential (MLIP) architectures beyond the moment tensor potential ecosystem where it was originally developed.

\subsection{Comparison and limitations}

Beiersdorfer \emph{et al.}~\cite{Beiersdorfer2026} used an iterative, committee-error-driven protocol to directly compare kernel-based GAP and equivariant MACE architectures for a liquid electrolyte.
It was found that correcting a specific chemically significant failure mode, which caused spurious Li-Li ion pairing artifacts, required several additional rounds of active learning beyond what the initial training of either architecture had deemed sufficient.
This highlights the fact that convergence of overall error metrics does not necessarily guarantee convergence for every chemically relevant interaction.

\begin{table*}[htbp]
\centering
\caption{Representative studies of on-the-fly active learning for system-specific interatomic potentials (Sec.~\ref{sec3}).}
\label{tab:sec3}
\small
\setlength{\tabcolsep}{3pt}
\resizebox{1\textwidth}{!}{%
\begin{tabular}{|c|l|c|l|l|c|}
\hline
\multicolumn{1}{|c|}{\textbf{\#}} & \multicolumn{1}{|c|}{\textbf{Authors}} & \multicolumn{1}{|c|}{\textbf{Year}} & \multicolumn{1}{|c|}{\textbf{System / Method}} & \multicolumn{1}{|c|}{\textbf{Key Active-Learning Result}} & \multicolumn{1}{|c|}{\textbf{Fig.}} \\
\hline
1 & Vandermause et al.~\cite{Vandermause2020} & 2020 & AgI, Al; on-the-fly Bayesian GP (FLARE) & AgI 420~K transition from 39 DFT calls; Al $<$100 calls & Fig.~\ref{fig:sec3a}a \\
2 & Hodapp \& Shapeev~\cite{Hodapp2020} & 2020 & bcc-W screw dislocation; MaxVol + local fragments & 20--40 configs; Peierls stress 2.29~GPa, inside the earlier DFT range & Fig.~\ref{fig:sec3a}b \\
3 & Kulichenko et al.~\cite{Kulichenko2023} & 2023 & Glycine, acetylacetone; uncertainty-biased MD & 90 vs.\ 0 reactive events (350~K, biased vs.\ unbiased) & Fig.~\ref{fig:sec3b}a \\
4 & Zaverkin et al.~\cite{Zaverkin2024} & 2024 & Alanine dipeptide, MIL-53(Al); gradient UQ + bias stress & 13$\times$ lower energy RMSE vs.\ unbiased 300~K MD & Fig.~\ref{fig:sec3b}c \\
5 & Sharma \& Sanvito~\cite{Sharma2024} & 2024 & ZIF-8, MOF-5; geometric CBAD selection & Few hundred configs; 392-parameter linear SNAP potential & -- \\
6 & Kang et al.~\cite{Kang2025} & 2025 & CuI, AgGaSe$_2$; ensemble UQ + energy penalty & 9\% of 112 materials show rare-event PES pathologies & Fig.~\ref{fig:sec3b}b \\
7 & Felis \& Dononelli~\cite{Felis2025} & 2026 & Pt clusters, bulk Al, water/CNT; clustered GPR (FALCON) & $>$300$\times$ speedup vs.\ full AIMD (water/CNT) & Fig.~\ref{fig:sec3c}a \\
8 & Beiersdorfer et al.~\cite{Beiersdorfer2026} & 2026 & Liquid electrolyte; GAP vs.\ MACE, committee AL & Li--Li artifact needs extra AL generations in both models & -- \\
9 & Bidoggia et al.~\cite{bidoggia2026aiida} & 2026 & Carbon allotropes, W$_x$Mo$_{1-x}$Te$_2$; AiiDA-TrainsPot & Automated, provenance-tracked on-the-fly workflow & Fig.~\ref{fig:sec3c}b \\
10 & Unglert et al.~\cite{Unglert2026} & 2026 & Si, Ge, Ti; replica-exchange nested sampling & Full $P$--$T$ phase diagrams converge in 10--15 iterations & Fig.~\ref{fig:sec3c}c \\
11 & Meng et al.~\cite{Meng2025} & 2025 & K, Na--K; small-cell (1--8 atom) MTP training & Up to 100$\times$ cost savings vs.\ large-cell training & Fig.~\ref{fig:sec3d}a \\
12 & Wang et al.~\cite{WangNEP2026} & 2026 & Large-cell NEP; NEPMaker local-fragment MaxVol & Extends cut-and-periodize AL to GPU-based NEP/GPUMD & Fig.~\ref{fig:sec3d}b \\
13 & Wang, Rao, Zhu~\cite{WangEntropy2026} & 2026 & C, Si, NaCl; entropy-maximized dataset-aware AL & 3--10$\times$ lower energy MAE vs.\ random MD at matched $N$ & -- \\
14 & Zou, Birks, et al.~\cite{ZouSKMD2026} & 2026 & M\"uller--Brown, alanine dipeptide; SKMD sampler & Boltzmann-preserving sampler beats AL baselines at fixed $N$ & Fig.~\ref{fig:sec3d}c \\
\hline
\end{tabular}%
}
\end{table*}

Taken together, the studies summarized in Table~\ref{tab:sec3} demonstrate that active learning can reliably and reproducibly reduce the number of calculations needed to train a MLIP system to a target accuracy.
This reduction is often one to several orders of magnitude compared to naive AIMD sampling.
The studies also show that this trend holds across different model classes (Gaussian processes, moment tensor potentials, linear SNAP, and neural-network potentials) and acquisition criteria (Bayesian posterior variance, D-optimality/MaxVol, ensemble disagreement, and geometric coverage).
Exploration strategies include conventional MD, uncertainty- or entropy-biased dynamics, and replica exchange nested sampling (Sec.~\ref{sec:exploration}).

What these studies do not generally establish is whether the resulting potential is reliable beyond the specific trajectory, defect, or property it was trained for.
The recurring need for uncertainty-\emph{biasing} illustrates that an acquisition criterion can only assess configurations reached by the exploration strategy.
The finding of Beiersdorfer \emph{et al.} that a chemically important interaction can remain incorrect even after bulk error metrics converge also shows that aggregate validation errors do not guarantee reliability for every chemically relevant interaction.
Where the methods described in the section clearly outperform brute-force alternatives, it is precisely in situations where a large, computationally intractable structure (such as a dislocation core or a nanotube pore) can be reduced to a smaller, locally representative fragment for labeling~\cite{Hodapp2020,Felis2025}.
Where they are most vulnerable is in rare, qualitatively different events that the exploration strategy may fail to sample adequately.
These limitations motivate the exploration and acquisition strategies discussed above~\cite{Kulichenko2023,Zaverkin2024,Kang2025}.

\section{Active Learning Meets Foundation Models. Fine-Tuning and Configuration-Space Generation}
\label{sec4}

The arrival of universal pretrained foundation models, such as MACE-MP\cite{batatia2025foundation}, CHGNet\cite{deng2023chgnet}, M3GNet\cite{chen2022universal}, SevenNet\cite{bendaoud2026sevennet}, MatterSim\cite{yang2024mattersim}, which were trained on tens of millions of material-project derived relaxation trajectory configurations, has not solved the data generation problem that active learning was designed to address. 
Instead, it has simply relocated the problem.

\subsection{Fine-tuning on supplied datasets}

\begin{figure*}[!t]
\begin{minipage}[h]{1\linewidth}
\center{ \includegraphics[width=0.97\linewidth]{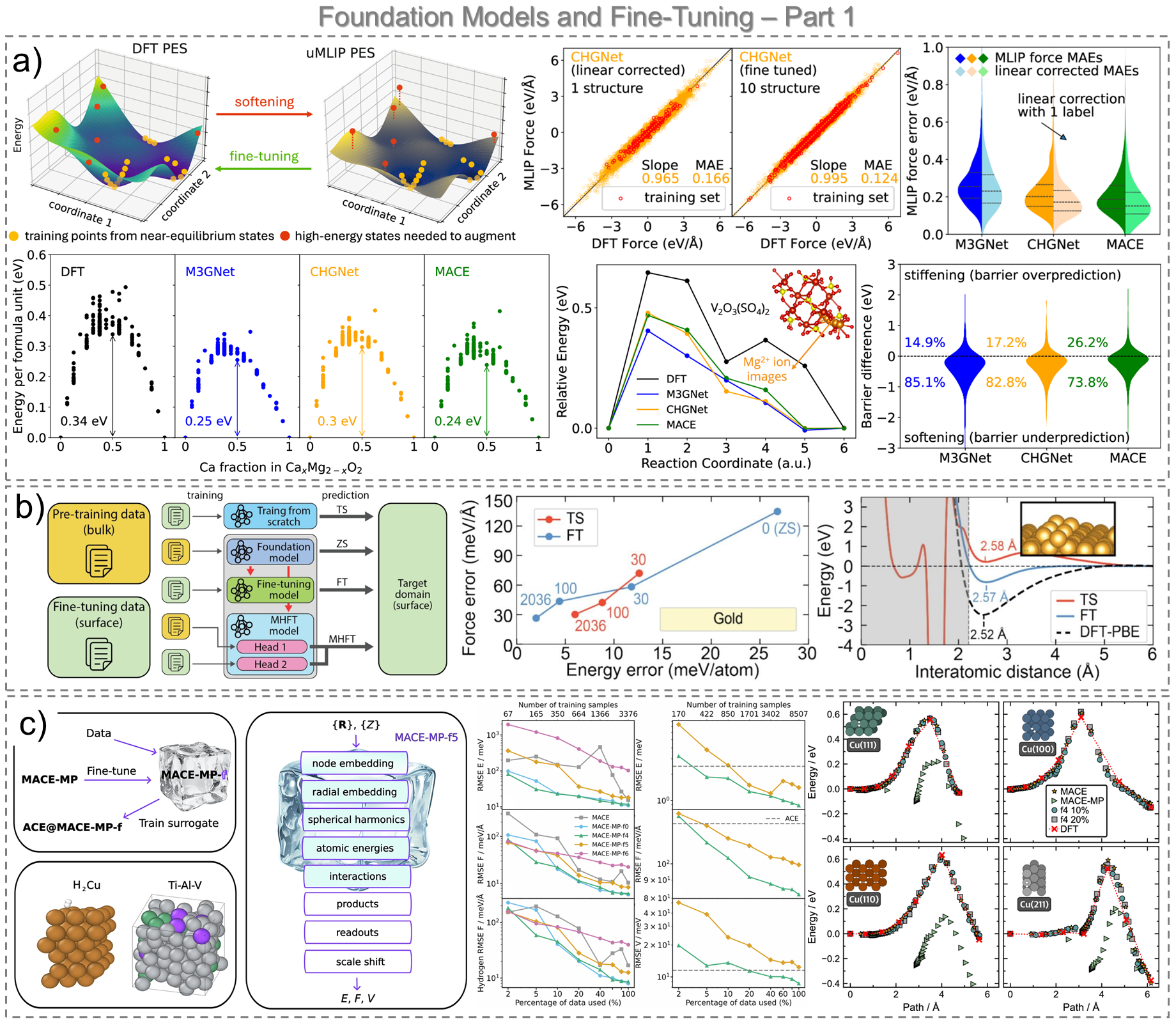}} \\
\end{minipage}
\caption{\textbf{Why and how foundation potential models are fine-tuned. Review of systematic softening, data-efficient fine-tuning, and frozen transfer learning.}
\textbf{(a)} Systematic softening of universal MLIPs (Ref.~\cite{Deng2024}) is shown by schematic of a DFT potential-energy surface and its softened universal-MLIP counterpart, trained mainly on near-equilibrium states (orange points) and underestimating the high-energy states (red points) that fine-tuning must supply.
Formation energies of Ca$_x$Mg$_{2-x}$O$_2$ solid solutions from DFT and universal MLIPs, collectively underestimated.
An example Mg-ion migration path in V$_2$O$_3$(SO$_4$)$_2$ and the distribution of 477 barrier differences between universal MLIPs and DFT, showing systematically underestimated barriers.
Parity plots for CHGNet on mp-973966 after a linear correction fitted to a single DFT-labeled structure (force MAE 0.190$\to$0.166~eV/\AA) and after fine-tuning all parameters on 10 structures (0.124~eV/\AA), together with the distributions of force MAEs and linearly corrected MAEs of M3GNet, CHGNet, and MACE for 1000 of compounds.
\textbf{(b)} Fine-tuning bulk-trained MACE-MP-0a for surfaces (Ref.~\cite{Hwang2026}) with the comparison of four training schemes training from scratch (TS) on surface data, zero-shot use of the foundation model (ZS), conventional fine-tuning (FT), and multi-head fine-tuning (MHFT) with separate readout heads for the bulk and surface domains (left).
Energy and force RMSE on the gold-surface test set versus training-set size (40, 200, and 2000 configurations, \enquote{0} denotes zero-shot) for TS and FT models, and the Au--Au diatomic curves of TS and FT models trained on 2000 configurations compared with DFT-PBE.
\textbf{(c)} Frozen transfer learning of MACE-MP (Ref.~\cite{Radova2025}) with shown workflow in which the foundation model is fine-tuned with partially frozen layers and then used to label data for a fast atomic cluster expansion (ACE) surrogate, the parameter blocks of MACE that are frozen, and the two benchmark systems (H$_2$ on Cu and Ti--Al--V).
Learning curves for the f6, f5, f4, and f0 freezing levels against a from-scratch MACE model (H$_2$/Cu: energy, force, and H-atom force RMSE) or a custom ACE model trained on the full database (Ti--Al--V: energy, force, and virial RMSE) as a function of the fraction of training data used.
Minimum-energy paths for H$_2$ dissociation on Cu(111), Cu(100), Cu(110), and Cu(211) from DFT, the from-scratch MACE model, the untuned MACE-MP, and MACE-MP-f4 fine-tuned on 10\% and 20\% of the data, in which the spurious minima of the untuned model disappear (right).
Panels adapted from the cited references.
}
\label{fig:sec4a}
\end{figure*}

Deng \emph{et al.}~\cite{Deng2024} demonstrated why a targeted correction is still necessary even when using M3GNet, CHGNet, and MACE-MP-0. 
These models systematically underpredict energies and forces in out-of-distribution high-energy configurations such as surfaces, defects, and ion migration transition states, see Fig.~\ref{fig:sec4a}a.
The mean absolute error for Mg-ion migration barriers is between 0.34 and 0.49 eV, and the bias is negative (underestimated) in more than 90\% of the 1000 tested compounds.
Since much of this error is systematic rather than random, it can be corrected significantly, by an average of 15\% across a diverse test set. 
This correction can be achieved with a single DFT-labeled structure and one-parameter linear adjustment.
Further reduction can be obtained with fine-tuning on only ten structures.
This explains why fine-tuning of foundation MLIPs often yields striking data efficiency.

This data efficiency has been documented at scale and across architectures. 
Hwang \emph{et al.}~\cite{Hwang2026} compared zero-shot inference, training from scratch, conventional fine-tuning, and multi-head fine-tuning for adapting MACE-MP-0a-small to the surfaces of Au, RuO$_2$, and BiVO$_4$, using fixed, sequentially generated training sets rather than active selection.
See Fig.~\ref{fig:sec4a}b and Sec.~\ref{sec:training} for details.
For gold, fine-tuning on just 40 configurations reduces the energy error from 14.8 to 4.3~meV/atom and the force error from 74.7 to 37.4~meV/\AA, matching or beating a from-scratch model trained on 2000 configurations.
What is often overlooked is that fine-tuning on a single element carries a risk of catastrophic forgetting, improving performance on that element while potentially degrading performance on others.
The energy error on the original bulk data rose from 11 to 302~meV/atom, and the surface-energy RMSE on Ag, Cu, Pt, and Ir rose from 27 to 88~meV/\AA$^2$. 
Multi-head fine-tuning limited the bulk error to 27~meV/atom and reduced the out-of-domain RMSE to 16~meV/\AA$^2$, at about 100 times the training time of plain fine-tuning and with the gold energy error increasing from 1.0 to 2.8~meV/atom. 
Karimitari \emph{et al.}~\cite{karimitari2026systematic} applied multi-head replay (Sec.~\ref{sec:training}) to fine-tune MACE-MH-1 for catalytic applications.
A key feature of this fine-tuning approach is that it is much less sensitive to training-set sampling than training from scratch, and it can achieve strong generalization even with small, chemically different datasets.
The most important result is that the FT-All model~\cite{karimitari2026systematic}, fine-tuned on 49860 configurations, performed best across both metallic and metal-oxide catalysts, screening 90781 bimetallic alloy adsorption energies with an MAE of 0.15 eV, including on unseen high-index surfaces such as (532).
It also achieved an MAE of 0.30 eV for the oxygen evolution reaction on iridium oxide polymorphs, outperforming out-of-box MACE-MH-1 by 0.08 eV and the best from-scratch model by 0.14 eV.
This approach offers a scalable and data-efficient perspective for catalytic screening, but its accuracy is fundamentally bounded by the chemical coverage of the pretrained model and the fine-tuning set, meaning that out-of-distribution intermediates like OOH$^*$ can still fail and require targeted DFT sampling despite the method's overall promise.
Radova \emph{et al.}~\cite{Radova2025} showed that the degree of freezing matters, see Fig.~\ref{fig:sec4a}c.
Their \enquote{f4} configuration of \enquote{mace-freeze}, with frozen lower layers of MACE-MP (Sec.~\ref{sec:training}), performed as well as the fully unfrozen model at lower training cost, whereas a readout-only model (\enquote{f6}) did not beat a from-scratch potential at low data.
For H$_2$ dissociative chemisorption on Cu, the f4 model, trained on 20\% of the reference data (664 out of 3376 configurations), matched a MACE model trained from scratch on all of the data. 
10\% of this data was sufficient to reproduce the hydrogen sticking probabilities of the full database model and to eliminate the spurious minima produced by the unmodified foundation model along the reaction pathways.
For a Ti--Al--V alloy, five models, each trained on a different 10\% subset, matched or exceeded the performance of a linear atomic cluster expansion (ACE) model trained on the full database in validation, elasticity, and vibrational tests, indicating that the basis set genuinely \enquote{fills gaps} that a small, bespoke dataset would otherwise leave.
The reduced training sets were subsets of existing databases, rather than actively selected.
Using the fine-tuned model as a labeler for a fast ACE surrogate (2200 structures from 600 model-driven trajectories) resulted in errors below those of the original ACE model fitted to the full DFT database, and a 17-fold reduction in evaluation time (22.6~ms vs. 390.5~ms), and on Cu-H data, delta-learning (fitting only the residuals to the basis set) was worse than training from scratch for all but the smallest training sets, and therefore worse than frozen transfer learning.

Hänseroth and coworkers~\cite{HanserothB2026} approached the problem from two complementary angles.
One study~\cite{HanserothB2026}, schematically presented in the Fig.~\ref{fig:sec4b}a benchmarked naive fine-tuning of five foundation architectures (MACE-MP-0, GRACE, SevenNet, MatterSim, ORB) across seven chemically diverse systems.
No active-learning-based configuration selection was applied at all.
A fixed, equidistantly sub-sampled 2000-configuration AIMD dataset was used. 
For this setup, force-error reduction of 81--89\% and energy-error reduction by two to four orders of magnitude were found in 34 out of 35 architecture/system combinations. 
This suggests that the results are essentially architecture-independent.
Main results presented in the top panel of Fig.~\ref{fig:sec4b}a.

\subsection{Configuration generation and coverage expansion}

\begin{figure*}[!t]
\begin{minipage}[h]{1\linewidth}
\center{ \includegraphics[width=1\linewidth]{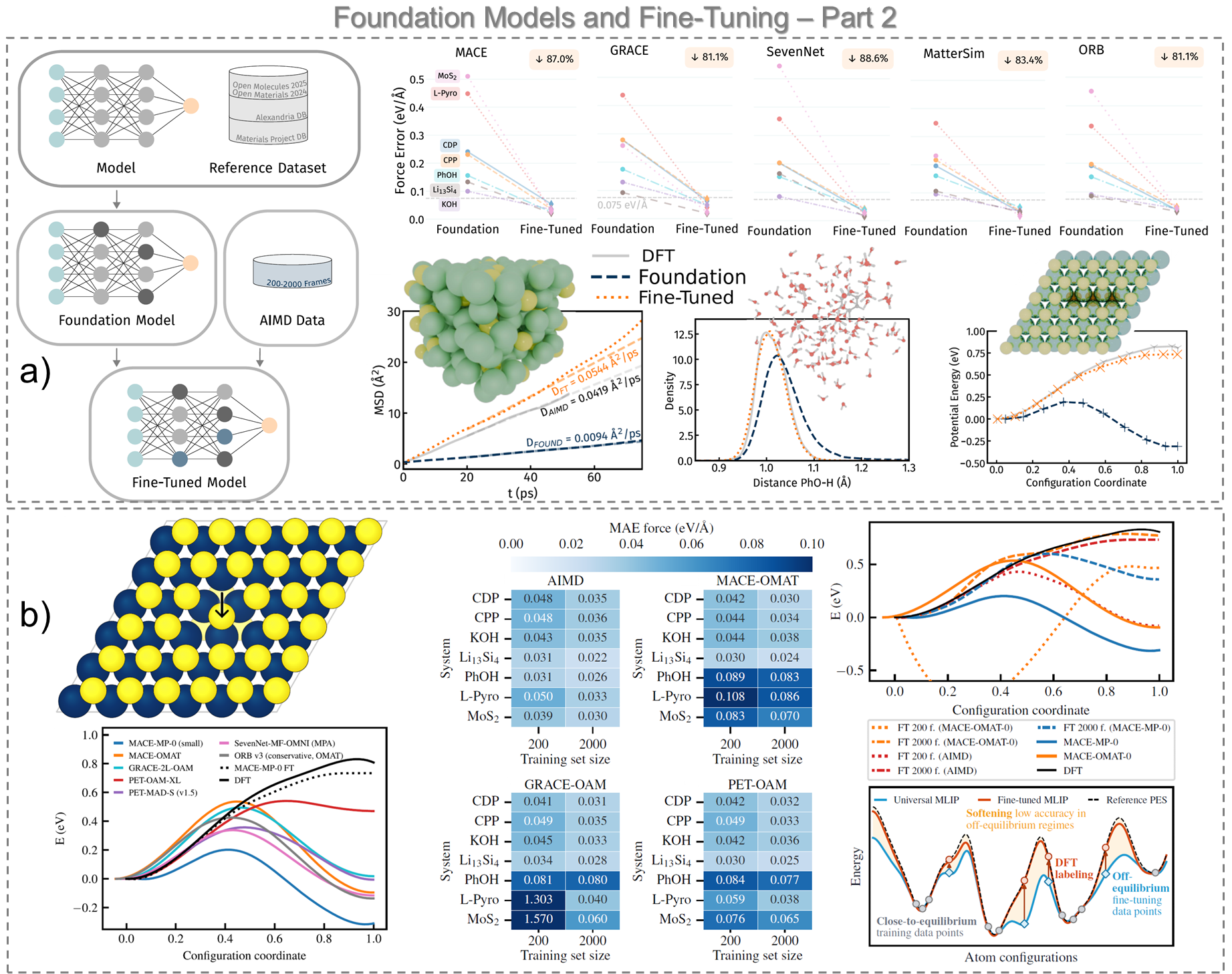}}  \\
\end{minipage}
\caption{\textbf{Fine-Tuning Across Architectures and Universal Potentials as Configuration-Space Generators.}
\textbf{(a)} Fine-tuning five foundation architectures on the same data (Ref.~\cite{HanserothB2026}) showing the fine-tuning scheme, in which a foundation model pretrained on a large reference dataset is adapted using 200--2000 frames of an AIMD trajectory. RMSE of forces of the foundation and fine-tuned MACE-MP-0, GRACE-1L-OAM, SevenNet-0, MatterSim-Large, and ORB-v2 models for seven systems, with average error reductions of 81--89\%. 
Physical properties from DFT, foundation, and fine-tuned models, namely Li-ion mean-squared displacements and diffusion coefficients in Li$_{13}$Si$_4$ (ORB), the O--H bond-length distribution of phenol in water (SevenNet), and the energy profile of a sulfur jump into a line of sulfur vacancies in MoS$_2$ (MACE).
\textbf{(b)} Universal potentials as configuration-space generators (Ref.~\cite{HanserothA2026}) for the casae of the sulfur-vacancy jump in MoS$_2$ and its nudged-elastic-band energy profile from several universal models, a MACE-MP-0 model fine-tuned on 2000 AIMD frames, and DFT (left). 
Force MAEs of MACE-MP-0 fine-tuned on 200 or 2000 configurations taken from AIMD or from DFT-relabeled MD trajectories of MACE-OMAT-0, GRACE-2L-OAM, and PET-OAM-XL for seven systems (middle).
MoS$_2$ energy profiles of models fine-tuned on 200 or 2000 configurations sampled by AIMD, MACE-MP-0, or MACE-OMAT-0, of which only the 2000-configuration AIMD and MACE-OMAT-0 sets recover the DFT curve (right) and a schematic of how molecular dynamics with a softened universal model samples high-energy configurations whose DFT relabeling corrects the fine-tuned potential-energy surface.
Panels adapted from the cited references.
}
\label{fig:sec4b}
\end{figure*}

The companion study~\cite{HanserothA2026} shown in the Fig.~\ref{fig:sec4b}b instead used universal MLIPs purely as cheap configuration-space \emph{generators}, exploiting their known softening bias to oversample rare, high-energy regions.
Then, DFT-relabeling of the sampled trajectory for fine-tuning was performed and showed that, in the case of the hardest test case (the sulfur-vacancy jump barrier in MoS$_2$, Fig.~\ref{fig:sec4b}b) 2000 DFT-relabeled configurations sampled with the newer MACE-OMAT-0 model recovered the DFT energy profile, whereas 2000 sampled with the older MACE-MP-0 model did not. 
An iterative, three-generation self-training loop also recovered the profile with only 600 DFT points, whereas single-pass fine-tuning on either 600 AIMD configurations or 600 MACE-MP-0 generated configurations failed.

A related strategy targets not a narrow downstream task but a foundation model's elemental coverage itself.
Using an uncertainty-guided crystal search (Sec.~\ref{sec:exploration}), Xue \emph{et al.}~\cite{Xue2025} expanded a universal potential from 5 to 25 main-group elements.
Details are presented in Fig.~\ref{fig:sec4e}b.
For the five-element seed, 3373 such structures lowered the energy MAE from about 0.6 to 0.045~eV/atom, and newly added elements such as O or F appeared more than twice as often in the sampled structures as the other elements.
Reaching 50~meV/atom needed roughly 1000--1400 labeled structures per element, so the DFT budget per element was not reduced, but the time to train to 100~meV/atom fell from 73--203~h from scratch to under 5~h (14 vs.\ 1~h for 25 elements).
The elements from the same periodic-table group as existing ones (Na, Si) were learned faster than dissimilar ones (O, F).

\subsection{Data selection and training regime}

The choice of training data is separate from the choice between training from scratch and fine-tuning (Sec.~\ref{sec:training}).
A training set can be selected once, for example using uncertainty~\cite{Liu2026} or geometric coverage~\cite{Qi2024,Lee2026}, and then used for either training regime~\cite{Lee2026}.
Alternatively, selection, reference calculations, and model updating can be repeated within an iterative active-learning loop, again using either training regime.

\begin{figure*}[!t]
\begin{minipage}[h]{1\linewidth}
\center{ \includegraphics[width=1\linewidth]{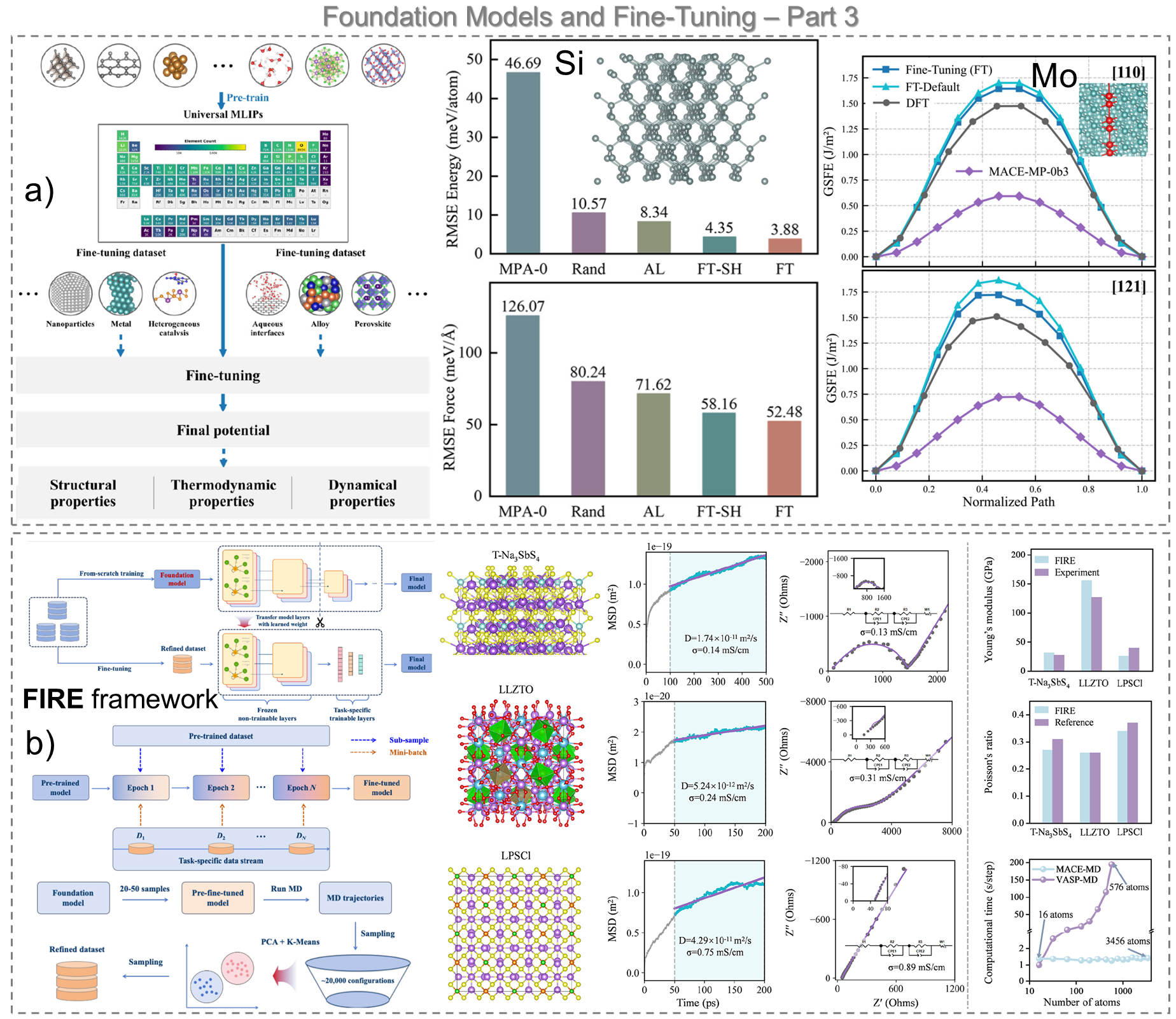}}  \\
\end{minipage}
\caption{\textbf{Fine-Tuning Workflows in Practice. A Tutorial Benchmark and Replay-Based Continual Fine-Tuning for Battery Interfaces.}
\textbf{(a)} Fine-tuning tutorial for MACE-MP-0 (Ref.~\cite{Liu2026}) presenting the overall workflow from pretrained universal MLIPs through task-specific fine-tuning datasets to the final potential and its structural, thermodynamic, and dynamical properties (left).
The silicon out-of-distribution test (grain boundaries, di-interstitials, stacking-fault paths, amorphous structure), with energy and force RMSE for the original MACE-MPA-0, models trained on randomly chosen (Rand) and ensemble-uncertainty-filtered (AL) 5\% subsets, single-head (FT-SH) and multi-head (FT) fine-tuned models (middle).
Generalized stacking-fault energy profiles of Mo along $\langle110\rangle$ and $\langle112\rangle$ for models fine-tuned with default loss weights and with a force weight of 100, compared with DFT (right).
\textbf{(b)} The FIRE framework for solid--solid battery interfaces (Ref.~\cite{LiuFIRE2026}) training from scratch versus fine-tuning with frozen layers and task-specific trainable layers. Replay-augmented continual fine-tuning that mixes sub-sampled pretraining data into every epoch of the task-specific data stream, and the sampling workflow in which a model pre-fine-tuned on 20--50 samples runs MD to generate $\sim$20\,000 configurations that are reduced by PCA and $k$-means clustering (left).
Property predictions of the resulting potentials, namely mean-squared displacements and room-temperature impedance spectra for T-Na$_3$SbS$_4$, LLZTO, and LPSCl, Young's moduli and Poisson ratios compared with experiment, and the MD wall time per step, nearly constant up to 3456 atoms in contrast to the cubic scaling of VASP (right).
Panels adapted from the cited references.
}
\label{fig:sec4c}
\end{figure*}

Liu \emph{et al.}~\cite{Liu2026}, whose fine-tuning workflow is described in Sec.~\ref{sec:training}, used an ensemble-uncertainty filter to cut the training set to 5\% of the original data on a silicon out-of-distribution benchmark.
MACE models trained on that subset beat those trained on randomly chosen subsets (energy RMSE 8.3 vs.\ 10.6~meV/atom), but fine-tuning the foundation model was clearly better still (3.9~meV/atom), so the authors conclude that fine-tuning is more data-efficient than training from scratch on uncertainty-driven subsets.
Elsewhere, fine-tuned MACE reduced the C$_{11}$ error of bcc molybdenum from 45\% to 1.5\% with respect to DFT.
On heterointerfaces it reached lower errors on forces than the long-range (latent-Ewald) CACE-LR model (e.g., 18 vs.\ 68~meV/\AA{} for LiCl/GaF$_3$), which the authors tentatively attribute to long-range physics implicitly encoded during pretraining while stressing that this is case dependent.
The same research group subsequently applied this fine-tuning philosophy to a genuinely difficult target, namely solid-solid battery interfaces, with their FIRE (fine-tuning with integrated replay and efficiency) framework~\cite{LiuFIRE2026}, see Fig.~\ref{fig:sec4c}b and Sec.~\ref{sec:training}.
FIRE reached energy RMSE below 1~meV/atom and force RMSE near 20~meV/\AA{} across six solid-solid interface systems.
FIRE showed an order-of-magnitude improvement over existing models, while requiring only 10\% of the original training datasets, and the fine-tuned potentials reproduced key mechanical and electrochemical interfacial properties in close agreement with experiment.
This directly addressed the configurational diversity and multi-phase heterogeneity that make solid-solid interfaces particularly hard to sample compared to the bulk or solid-liquid systems that dominate Sec.~\ref{sec7}.

Zhou and Komsa~\cite{Zhou2026} examined fine-tuning and training-set construction for the two-dimensional, five-component transition-metal-dichalcogenide alloy (Mo,Ta,Nb,W,V)S$_2$.
None of the five foundation models they tested produced reliable zero-shot mixing energies.
The mean absolute error was 16~meV/unit for MACE-small, 25~meV/unit for MACE-small-0b2, and significantly larger for MatterSim-1M and CHGNet when tested on multicomponent databases.
At the same time fine-tuning MACE-small-0b2 reduced the error to about 2~meV/unit (1.5~meV/unit on average) at the largest training sets.
Training structures generated by exhaustive cluster-expansion enumeration generalize more evenly than randomly-sized samples, which reach lower errors on randomly generated test structures at larger scales but fail badly on ordered structures.
In a five-component system, the enumerated models outperformed their random counterparts, and Monte Carlo annealed subsets of the enumerated pool performed similarly well.
Training sets containing two- and three-component alloys transferred to the full five-component system more effectively than binary-only sets, and fine-tuned models remained stable in structural relaxation at small data sizes where models trained from scratch failed.

Lee \emph{et al.}~\cite{Lee2026} directly compared sampling strategy against training regime for augmenting SevenNet0 on sodium-oxide solid electrolytes, see Fig.~\ref{fig:sec4d}b.
Under training-from-scratch, the largest-cluster-maximum-distance (LCMD) selector targeting diversity raised the diffusivity agreement with AIMD ($R^2$) from 0.48 to 0.70, compared to 0.59--0.60 for random or Dimensionality-Reduced Encoded Clusters with sTratified (DIRECT) selection.
However, under fine-tuning, uniform coverage dominated, and naive random selection actively degraded performance below the untouched baseline through catastrophic forgetting. 
This is a cautionary example of the interaction between sampling strategy and training regime, as neither factor alone can predict the outcome.

\begin{figure*}[!t]
\begin{minipage}[h]{1\linewidth}
\center{ \includegraphics[width=0.92\linewidth]{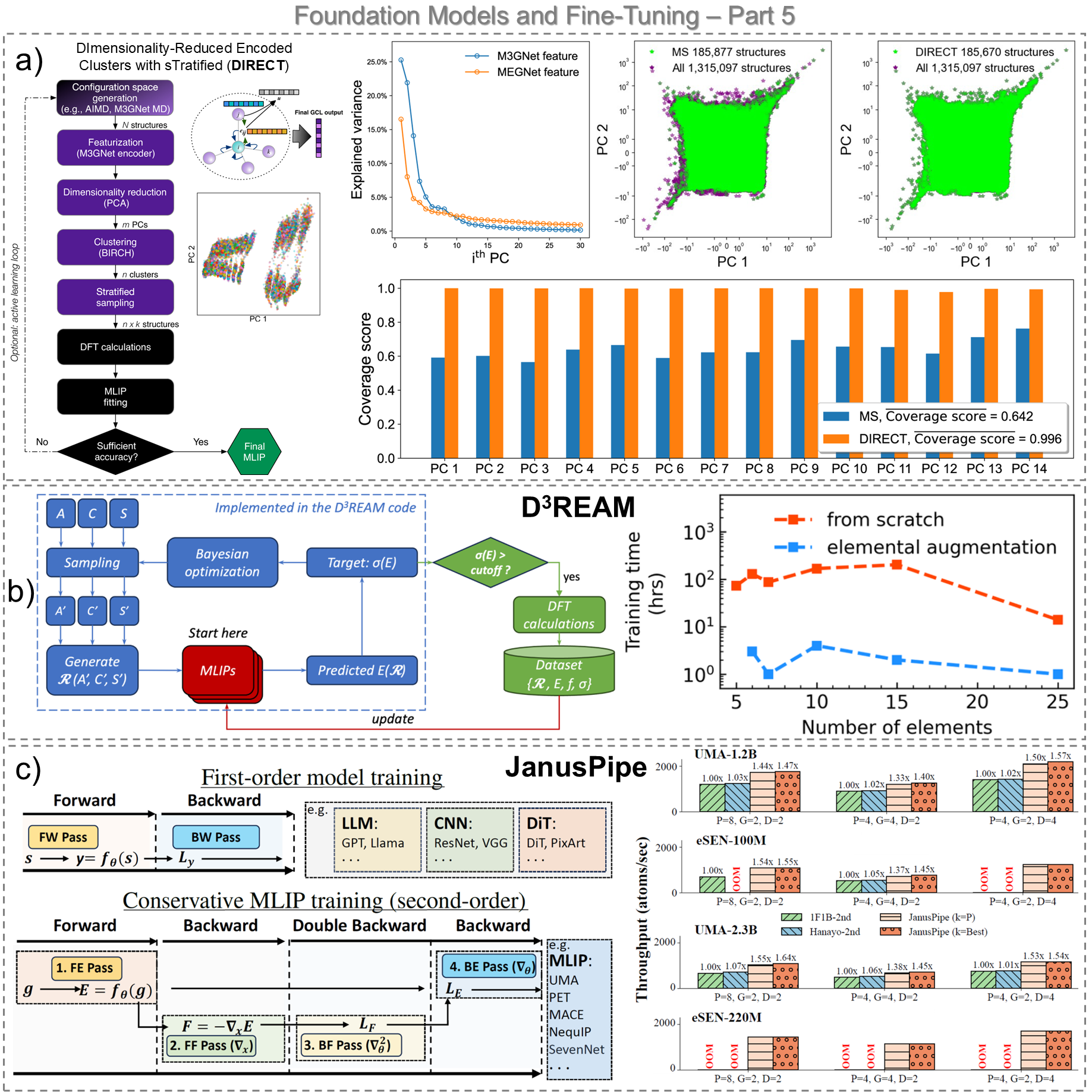}} \\
\end{minipage}
\caption{\textbf{Selecting and Scaling Training Data for Foundation Potentials with DIRECT Sampling, Elemental Augmentation, and Pipeline-Parallel Training.}
\textbf{(a)} DIRECT sampling (Ref.~\cite{Qi2024}) representing the workflow of configuration-space generation, M3GNet featurization, PCA, BIRCH clustering, and stratified sampling before DFT calculations and MLIP fitting, with an optional active-learning loop (left).
The explained variance of the first 30 principal components of M3GNet and MEGNet encodings (top right), the coverage of the first two principal components of the Materials Project relaxation data by the manually sampled and DIRECT-sampled sets of similar size, and the coverage scores of both sets over the first 14 principal components (bottom right).
\textbf{(b)} Elemental augmentation of a universal potential (Ref.~\cite{Xue2025}) using the D$^3$REAM workflow, in which Bayesian optimization over atom, composition, and structure pools generates crystal structures whose ensemble-predicted energy spread $\sigma(E)$ is largest, and structures above a cutoff are labeled with DFT and added to the training set (left)
The number of training structures per element needed to reach an energy MAE of 50~meV/atom, together with the training time to reach 100~meV/atom from scratch and by elemental augmentation, as a function of the number of elements (right).
\textbf{(c)} Pipeline-parallel training of conservative MLIPs (Ref.~\cite{WangHY2026}) with the single forward and backward pass of first-order models contrasted with the four phases of conservative MLIP training (forward energy, forward force, backward force, backward energy), which involve a double-backward pass (left).
The end-to-end throughput of JanusPipe compared with the 1F1B-2nd and Hanayo-2nd schedules for UMA-1.2B, UMA-2.3B, eSEN-100M, and eSEN-220M across pipeline, graph, and data parallel settings, with OOM marking configurations that ran out of memory (right).
Panels adapted from the cited references.
}
\label{fig:sec4e}
\end{figure*}

The DIRECT sampling protocol of Qi \emph{et al.}~\cite{Qi2024} (Sec.~\ref{sec:geometric}) is presented in Fig.~\ref{fig:sec4e}a.
On the Materials Project relaxation data, DIRECT selection of about 186\,000 structures covered the feature space almost completely (coverage score 0.996 vs. 0.642 for the manually curated set of the same size used for the original M3GNet potential), see right panel of Fig.~\ref{fig:sec4e}a.
The resulting potential was markedly more reliable on the hardest $\sim$10\% of test structures and on hypothetical S-containing materials (energy-above-hull error below 0.091 vs. 0.158~eV/atom for half of them).
The MD simulation using the M3GNet potential generated 274\,000 candidate structures for titanium hydrides.
From these, 954 clusters were identified, yielding 947 converged DFT (density functional theory) structures using a moment tensor potential (MTP).
Its test-set energy MAE was only about 1~meV/atom above that of an MTP built with 28 active-learning iterations and 2077 structures (both below 7~meV/atom), that is, comparable accuracy with roughly half the DFT calculations.
This potential reproduced hydrogen-diffusion barriers of 0.42, 0.14, and 0.91~eV in HCP, BCC, and FCC Ti against experimental values of 0.45, 0.15, and 0.92~eV. 
Layering active learning on top of the DIRECT-trained model then needed no iterations at 300 and 1000~K and five for the hardest cases (3000~K and H$_{72}$ at 300~K), versus roughly 10--28 from scratch.
A 75\% reduction in iterations indicating that the two strategies are complementary rather than substitutes. 

\subsection{Distillation and computational cost}

\begin{figure*}[!t]
\begin{minipage}[h]{1\linewidth}
\center{ \includegraphics[width=1\linewidth]{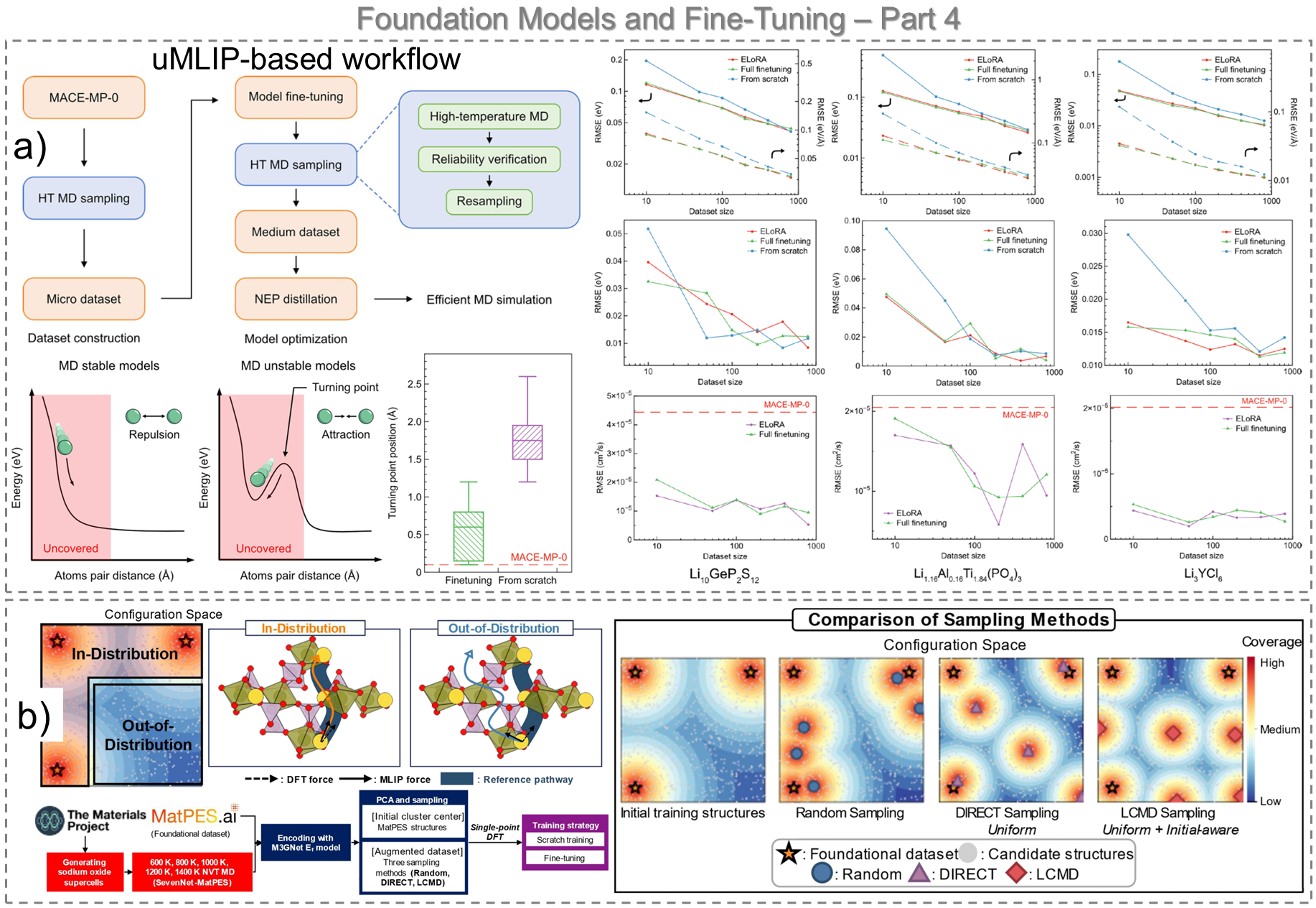}}  \\
\end{minipage}
\caption{\textbf{Single-Shot Data Generation and the Choice of Sampling Strategy for Fine-Tuning on Solid Electrolytes.}
\textbf{(a)} Single-shot fine-tuning and distillation workflow (Ref.~\cite{ZhangW2026}) in which high-temperature MD simulation with MACE-MP-0 and a reliability check supplies a micro dataset for DFT where the fine-tuned MACE model generates a medium-sized dataset by further high-temperature MD, and a lightweight NEP model is distilled from it for efficient MD (left)
Analysis of short-range repulsion through a schematic of the diatomic energy curves of stable and unstable models and the distribution of the distance at which the diatomic force turns from repulsive to attractive, for fine-tuned models with stable MD and for unstable models trained from scratch, with MACE-MP-0 as reference (left).
The energy and force RMSE, the error of nudged-elastic-band relative energies, and the error of Li diffusion coefficients in Li$_{10}$GeP$_2$S$_{12}$, Li$_(1+x)$Al$_x$Ti$_{2-x}$(PO$_4$)$_3$ (LATP), and Li$_3$YCl$_6$ versus training-set size for ELoRA fine-tuning, full fine-tuning, and training from scratch (right).
The untuned MACE-MP-0 in shown as a dashed line.
\textbf{(b)} Sampling strategies for augmenting a universal MLIP on sodium-oxide solid electrolytes (Ref.~\cite{Lee2026}) which conceptualize configuration space in which training structures define in-distribution regions of accurate forces and out-of-distribution regions cause force errors and trajectory deviations of mobile ions.
The augmentation workflow (left), from high-temperature SevenNet-MatPES MD (600--1400~K) through M3GNet encoding, PCA, and random, DIRECT, or LCMD selection to single-point DFT and either training from scratch or fine-tuning.
A schematic of the coverage achieved by each strategy, uneven for random selection, uniform for DIRECT, and uniform but aware of the initial dataset for LCMD (right).
Panels adapted from the cited references.
}
\label{fig:sec4d}
\end{figure*}

Fine-tuned foundation models can also serve as teachers for cheaper surrogates.
Zhang \emph{et al.}~\cite{ZhangW2026} replaced the iterative process with a single workflow for solid electrolytes (Li$_{10}$GeP$_2$S$_{12}$,
Li$_{1+x}$Al$_x$Ti$_{2-x}$(PO$_4$)$_3$ (LATP), Li$_3$YCl$_6$), see Fig.~\ref{fig:sec4d}a.
A high-temperature molecular dynamics simulations with MACE-MP-0 were used to keep the non-lithium framework rigid.
The trajectories were then used to select only about 200 configurations for density functional theory (DFT), compared to the tens of thousands that would typically be used in conventional active learning.
The zero-shot foundation model showed a 22.8\% relative error in nudged-elastic-band migration-barrier energies and 60.8\% in the diffusion coefficient, Fig.~\ref{fig:sec4d}a.
Fine-tuning on the 200 configurations (full or with the parameter-efficient ELoRA) lowered these to 10.7\% and 13.2\%, matched a from-scratch model trained on twice as many configurations, and kept high-temperature MD stable even with 10 configurations, whereas from-scratch MACE and NEP models remained unstable with 800.
The fine-tuned model was then distilled into a lightweight NEP, trained on 2000 model-labeled configurations (with a ZBL repulsion).
This new model was stable and 20 times faster in ASE (hundreds of times faster in optimized GPU engines, according to the authors), but only modestly more accurate than a NEP fitted directly to the 200 DFT points.
For LATP, the error obtained by NEP was 17\% lower and its force RMSE was 11\% lower than the original model.
The NEP deviated from its teacher's diffusion coefficient by only 54\%.
The workflow was validated only for bulk phases, with charged systems and heterogeneous interfaces left open.

Baghishov \emph{et al.}~\cite{Baghishov2026} explored a complementary aspect of the same trade-off, namely not which pre-trained model to fine-tune, but how much DFT convergence and model expressivity a custom-built potential actually needs.
Using an entropy-maximized beryllium dataset and a leverage-score-based subsampling strategy, an active learning-adjacent D-optimality selection method, they demonstrated that energy errors with respect to the highest convergence reference dropped from 497~meV/atom to 0.54~meV/atom across six different DFT convergence levels.
Leverage sampling achieved within 1~meV/atom accuracy of the full dataset result with roughly 10 times fewer configurations compared to random sampling.
Matching model complexity to the DFT convergence level rather than maximizing each independently can reduce total training compute by one to two orders of magnitude, according to the authors.

Finally, as fine-tuning and configuration generation workflows increasingly train larger and larger conservative (force-as-energy-gradient) foundation architectures on larger and larger actively or passively curated datasets, the distributed training infrastructure itself has become a bottleneck that the field is beginning to explicitly address.
Wang \emph{et al.}~\cite{WangHY2026} presented a JanusPipe that has improved the pipeline parallel training throughput of conservative machine learning models on 32 GPUs by 1.51$\times$ and 1.45$\times$ on average over the 1F1B and Hanayo schedules adapted for second-order training. 
This work was motivated in part by the trends in data scale documented in this section.

\subsection{Comparison and limitations}

\begin{table*}[htbp]
\centering
\caption{Representative studies of fine-tuning and configuration-space generation from foundation MLIPs (Sec.~\ref{sec4}).}
\label{tab:sec4}
\small
\setlength{\tabcolsep}{3pt}
\resizebox{1\textwidth}{!}{%
\begin{tabular}{|c|l|c|l|l|c|}
\hline
\multicolumn{1}{|c|}{\textbf{\#}} & \multicolumn{1}{|c|}{\textbf{Authors}} & \multicolumn{1}{|c|}{\textbf{Year}} & \multicolumn{1}{|c|}{\textbf{System / Method}} & \multicolumn{1}{|c|}{\textbf{Key Active-Learning / Fine-Tuning Result}} & \multicolumn{1}{|c|}{\textbf{Fig.}} \\
\hline
1 & Deng et al.~\cite{Deng2024} & 2024 & Surfaces, defects, Mg-ion migration; softening diagnosis & Single structure + linear correction cuts error $\sim$15\% & Fig.~\ref{fig:sec4a}a \\
2 & Qi et al.~\cite{Qi2024} & 2024 & MP relaxation set, TiH$_n$; DIRECT stratified sampling & Within $\sim$1~meV/atom of 28-iteration AL with 947 vs.\ 2077 DFT structures & Fig.~\ref{fig:sec4e}a \\
3 & Radova et al.~\cite{Radova2025} & 2025 & Cu--H, Ti--Al--V; frozen-layer transfer learning & Cu--H: 20\% of the data matches from-scratch MACE & Fig.~\ref{fig:sec4a}c \\
4 & Xue et al.~\cite{Xue2025} & 2025 & 5$\to$25 main-group elements; BO on ensemble deviation & Train time to 100~meV/atom: $<$5~h vs.\ 73--203~h from scratch & Fig.~\ref{fig:sec4e}b \\
5 & Baghishov et al.~\cite{Baghishov2026} & 2026 & Beryllium; leverage-score sub-sampling & 10$\times$ fewer configs than random for 1~meV/atom target & -- \\
6 & Hwang et al.~\cite{Hwang2026} & 2026 & Au, RuO$_2$, BiVO$_4$ surfaces; fine-tuning vs.\ from scratch, multi-head variant & 40 configs match 2000-config from-scratch model & Fig.~\ref{fig:sec4a}b \\
7 & H\"anseroth et al.~\cite{HanserothB2026} & 2026 & 7 systems, 5 architectures; naive fine-tuning & 81--89\% force-error reduction, arch.-independent & Fig.~\ref{fig:sec4b}a \\
8 & H\"anseroth et al.~\cite{HanserothA2026} & 2026 & MoS$_2$ vacancy jump; config-space generation & MoS$_2$ barrier recovered with 600 DFT points (iterative self-training) & Fig.~\ref{fig:sec4b}b \\
9 & Liu et al.~\cite{Liu2026} & 2026 & Si, Mo, Li$_{10}$GeP$_2$S$_{12}$, heterointerfaces; FT tutorial & Si: fine-tuning beats 5\%-subset AL (3.9 vs.\ 8.3~meV/atom); Mo C$_{11}$ error 45\%$\to$1.5\% & Fig.~\ref{fig:sec4c}a \\
10 & Lee et al.~\cite{Lee2026} & 2026 & Na-oxide solid electrolytes; DIRECT/LCMD $\times$ FT & LCMD raises $R^2$ 0.48$\to$0.70 (scratch); FT favors coverage & Fig.~\ref{fig:sec4d}b \\
11 & Zhang et al.~\cite{ZhangW2026} & 2026 & Li$_{10}$GeP$_2$S$_{12}$, LATP, Li$_3$YCl$_6$; FT + distillation & $\sim$200 DFT points in one shot vs.\ tens of thousands for conventional AL & Fig.~\ref{fig:sec4d}a \\
12 & Zhou \& Komsa~\cite{Zhou2026} & 2026 & (Mo,Ta,Nb,W,V)S$_2$ 2D HEA; enumerated FT & Mixing-energy MAE 25$\to\sim$2~meV/unit (MACE-small-0b2 base) & -- \\
13 & Liu et al.~\cite{LiuFIRE2026} & 2026 & Solid-solid battery interfaces; FIRE replay FT & Order-of-magnitude error cut with 10\% of the data & Fig.~\ref{fig:sec4c}b \\
14 & Wang et al.~\cite{WangHY2026} & 2026 & UMA, eSEN; pipeline-parallel training (JanusPipe) & 1.51$\times$/1.45$\times$ throughput vs.\ 1F1B/Hanayo on 32 GPUs & Fig.~\ref{fig:sec4e}c \\
\hline
\end{tabular}%
}
\end{table*}

Several studies surveyed in this section show that fine-tuning a pretrained MLIP can match or exceed the accuracy of a model trained from scratch, while using only a small fraction of the task-specific training data~\cite{Radova2025,Hwang2026,Liu2026}.
For example, Radova \emph{et al.}~\cite{Radova2025} achieved comparable accuracy with 10--20\% of the training data for Cu--H$_2$ and Ti--Al--V systems, while Hwang \emph{et al.}~\cite{Hwang2026} matched or exceeded the accuracy of training from scratch for Au surfaces using 2--10\% of the data.
The studies summarized in Table~\ref{tab:sec4} cover different architectures (MACE, GRACE, SevenNet, MatterSim, ORB) and chemistries (surfaces, alloys, solid electrolytes, solid-solid battery interfaces, and 2D materials).
The examples of Radova and Hwang also show that such gains can arise from reusing pretrained knowledge without active data selection.

There are three qualifications.
First, the success of fine-tuning is dependent on the prior knowledge of the foundation model being relevant to the target domain.
Fine-tuning narrows the model's transferability, as shown by Hwang \emph{et al.}~\cite{Hwang2026}, and can lead to \enquote{negative transfer} when the target system differs significantly from the pre-training distribution, as warned by Radova \emph{et al.}~\cite{Radova2025}. 
This caveat is further emphasized by the evidence from benchmarking in Sec.~\ref{sec10}, which suggests that fine-tuning may sometimes be neutral or even harmful.
Second, the layers that are frozen, the selection of training data (random, DIRECT, LCMD, or enumeration), and the starting point of training (from scratch or fine-tuning) all interact in complex ways that are difficult to predict from first principles. 
Lee \emph{et al.}~\cite{Lee2026} compared sampling strategies within each training regime, allowing the benefit of data selection to be assessed separately for training from scratch and fine-tuning; the optimal sampling strategy varied between the two regimes.
Third, distillation and configuration-generation strategies that exploit a foundation model's own systematic errors, such as its softening bias, to oversample rare configurations, are elegant but self-referential.
The resulting fine-tuned or distilled model may inherit the artifacts of the generating model, unless independently corrected.
This has been documented for MoS$_2$ configurations generated from an already-poor foundation prior~\cite{HanserothA2026}.
Similarly, a distilled student may also fall short of its teacher. For example, there is a 54\% discrepancy in the LATP diffusion coefficient of a distilled NEP model~\cite{ZhangW2026}. 

Fine-tuning is a model-updating choice that can be combined with iterative active learning or applied to a fixed training set (Sec.~\ref{sec:training}).
It converts a large, expensive, general-purpose prior into a targeted correction, provided that the prior is trustworthy for the chemistry in question.
When combined with active data selection, this reuse of prior knowledge offers a powerful route to reducing the reference calculations needed for a target accuracy.

\section{Beyond Committees. Bayesian, Kernel, and Representation-Based Uncertainty}
\label{sec5}

Query-by-committee is a simple and architecture-agnostic approach, but it increases training cost proportionally to the number of models in the ensemble.
As several studies have shown, the disagreement signal produced by this method is not always a reliable indicator of true predictive error.
A growing body of recent research instead derives acquisition signals from the Bayesian structure of a single model, gradient geometry, or learned latent representation. 
In some cases, this is done by exploiting the embeddings of a pre-trained foundation model without any additional training.

\subsection{Bayesian and kernel-based acquisition}

\begin{figure*}[!t]
\begin{minipage}[h]{1\linewidth}
\center{ \includegraphics[width=1\linewidth]{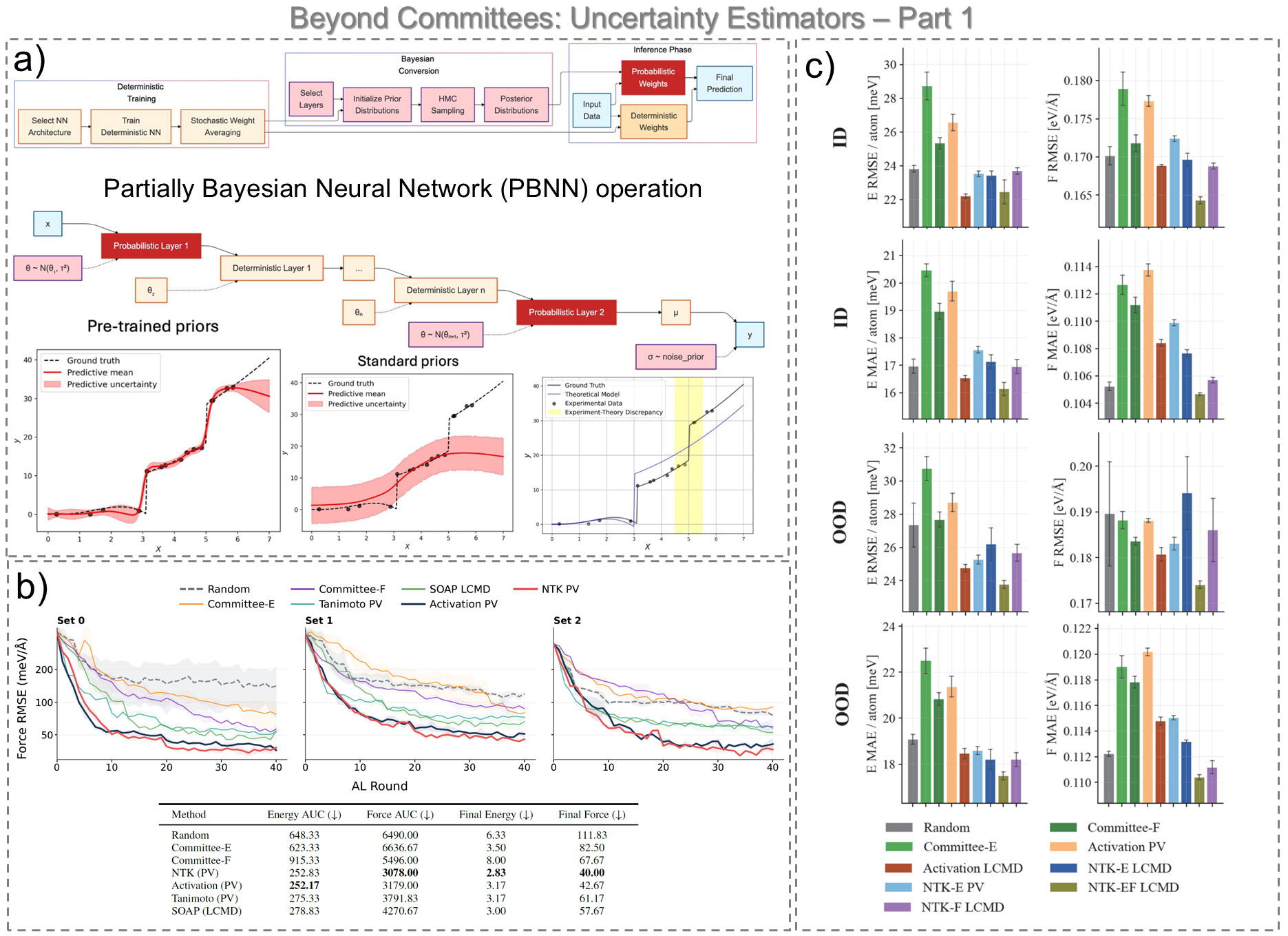}}  \\
\end{minipage}
\caption{\textbf{Uncertainty Beyond Committees with Partially Bayesian Networks and Kernels Built on Pretrained Models.}
\textbf{(a)} The training scheme of Partially Bayesian neural networks (Ref.~\cite{Allec2025}) in which a deterministic network is first trained with stochastic weight averaging, a selected subset of layers is then made probabilistic with priors initialized from the pretrained weights while all other weights stay frozen, and posteriors are sampled with Hamiltonian Monte Carlo (NUTS), together with the alternation of probabilistic and deterministic stages in the resulting model (top).
One-dimensional example predictions in which a Bayesian network with priors pretrained on \enquote{theoretical} data follows the theory where it holds and adapts to \enquote{experimental} data where it does not, whereas standard priors miss the second transition and give overly conservative uncertainty.
\textbf{(b)} Pretrained-model representations as acquisition signals (Ref.~\cite{VargaUmbrichA2026}) showing obtained force RMSE versus active-learning round (five structures added per round) on three T1x reaction pools for random selection, energy- and force-disagreement committees, Tanimoto and SOAP descriptor kernels, and posterior-variance acquisition with pooled-activation and neural-tangent kernels of a pretrained MACE model, with the areas under the learning curves and final errors tabulated (top).
The model-based kernels are the strongest acquisition signals (bottom).
\textbf{(c)} Force-aware neural tangent kernels on a 200\,000-structure OC20 pool (Ref.~\cite{VargaUmbrichB2026}) with shown final-round energy and force RMSE and MAE on the in-distribution split and on the mean of three out-of-distribution splits (held-out adsorbates, bulks, or both) for random selection, committees, activation-kernel and energy-, force-, and joint energy--force NTK acquisition with posterior-variance or LCMD selection. The joint energy--force NTK gives the lowest errors on every metric.
Panels adapted from the cited references.
}
\label{fig:sec5a}
\end{figure*}

Allec and Ziatdinov~\cite{Allec2025} propose a hybrid approach that combines partially Bayesian neural networks with traditional neural networks, see Fig.~\ref{fig:sec5a}a.
In this approach, only a small number of layers, typically the first hidden layer and the output layer, use probabilistic weights sampled from a Hamiltonian Monte Carlo algorithm, while the rest of the layers are deterministic. 
This allows the model to track the performance of a fully Bayesian neural network on molecular (FreeSolv, ESOL) and material (steel fatigue strength, oxide electrical conductivity) datasets, while cutting computational costs by nearly a factor of four.
Additionally, initializing the weights with prior information from a theory-informed pre-training pass, such as DFT-computed band gaps, improves the calibration of the model during the early stages of active learning.
The most methodologically advanced recent work in this area creates acquisition signals directly from the parameter or activation space of a pre-trained MACE foundation model, without the use of any auxiliary uncertainty heads, Bayesian training, or ensembles.
Varga-Umbrich \emph{et al.}~\cite{VargaUmbrichA2026} introduced two such signals, namely a finite-width neural tangent kernel (NTK) posterior variance and a pooled activation kernel.
The methods were evaluated on reactive chemistry benchmarks (T1x, RGD, PMechDB), with substantial gains over energy- and force-disagreement committees on T1x, see Fig.~\ref{fig:sec5a}b.
They showed that both methods reduce the number of rounds required to reach a target error by an average of 38\% (energy RMSE) and 28\% (force RMSE), compared to random acquisition. 
Committee disagreement, on the other hand, exhibited an unstable, poorly calibrated energy-force trade-off and a weak, round-to-round unstable correlation with true, held-out error.

A direct follow-up~\cite{VargaUmbrichB2026} extended this framework in three directions simultaneously, see Fig.~\ref{fig:sec5a}c.
It introduced a pipeline for generating posterior-variance that scales linearly with the number of features and can handle a much larger number of candidate structures, up to 200\,000. 
The pipeline uses a force-aware neural tangent kernel (NTK) that is built from the derivative of the predicted forces with respect to the parameters. 
This kernel is combined with a joint energy-force kernel to achieve the best performance. 
The robustness of the model was tested under deliberately biased candidate pools. 
The model-based kernel performed better than the committee approach, which required training M+1 full models per round, while achieving similar performance with half the wall-clock time per round.

\subsection{Novelty and reliability from learned representations}

\begin{figure*}[!t]
\begin{minipage}[h]{1\linewidth}
\center{ \includegraphics[width=1\linewidth]{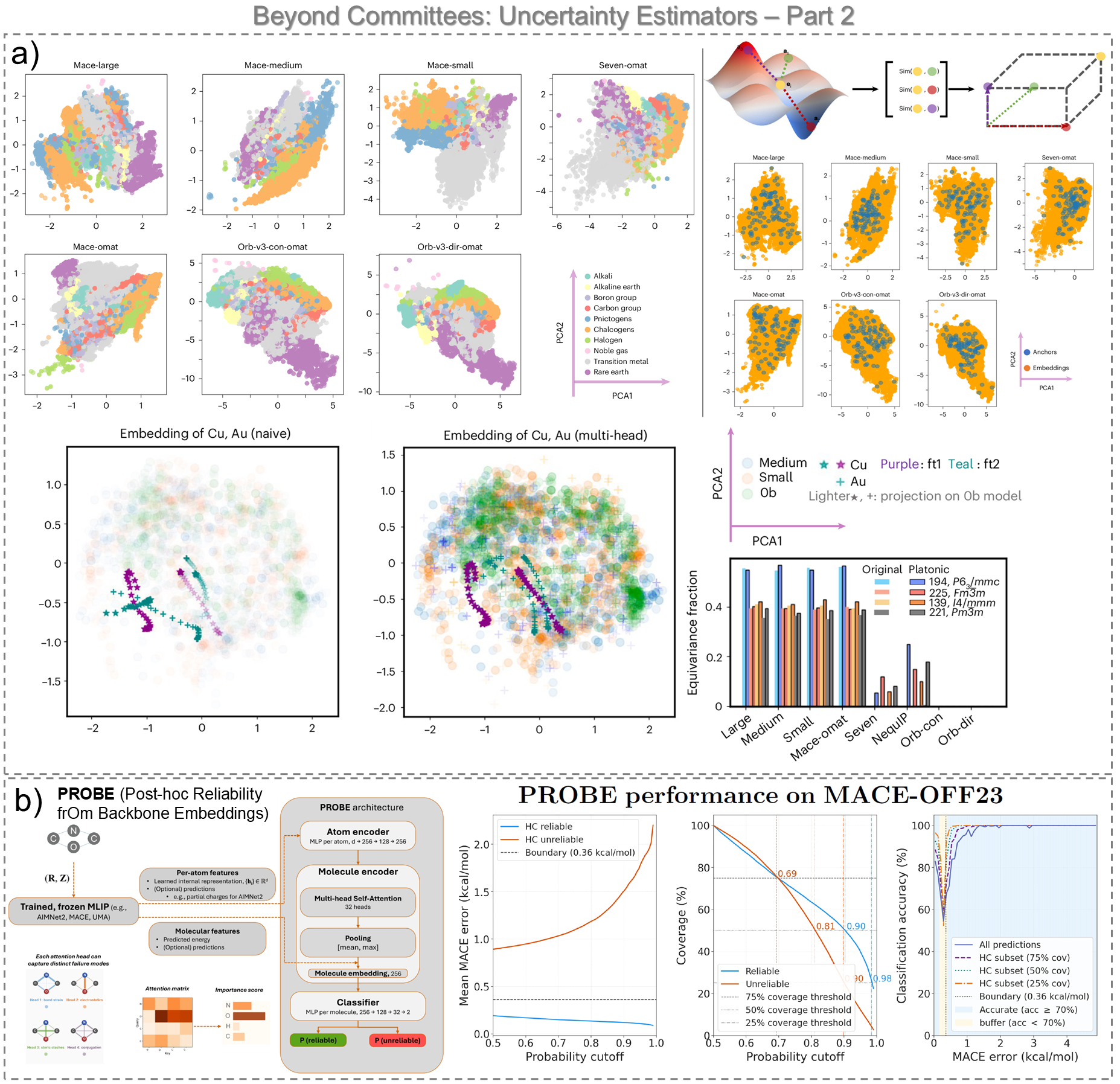}}  \\
\end{minipage}
\caption{\textbf{Reading Reliability from a Single Pretrained Model with Platonic Embeddings and Post-Hoc Reliability Probes.}
\textbf{(a)} Platonic representation of foundation MLIPs (Ref.~\cite{LiZ2026}) demonstrating two-dimensional PCA projections of the atomic embeddings of seven foundation MLIPs, whose variance directions and element clustering differ before transformation (top left).
Construction of the unified coordinate system, in which each embedding is projected onto a set of anchor vectors by cosine similarity, and the 100 DIRECT-sampled anchors overlaid on the original embedding manifold (top right) 
Diagnostic applications in the unified space, where naive fine-tuning on Cu--Cu data collapses the embeddings of unseen Au atoms (catastrophic forgetting) while multi-head fine-tuning keeps them near their pretrained positions, and the fraction of equivariant embeddings in each model (bottom).
\textbf{(b)} The architecture of PROBE (Ref.~\cite{Mehdi2026}) in which a compact classifier with an atom encoder, multi-head self-attention, and pooling reads the frozen per-atom representations of a pretrained MLIP and outputs the probability that a prediction is reliable, with per-atom importance scores from the attention weights identifying the atoms responsible for an unreliable prediction (left).
Performance on 50\,195 held-out molecules for MACE-OFF23, showing the mean model error of the reliable and unreliable classes and the high-confidence coverage as functions of the probability cutoff, and the classification accuracy as a function of the magnitude of the model error (right).
Panels adapted from the cited references.
}
\label{fig:sec5b}
\end{figure*}

In addition to this acquisition-focused work, Li and Walsh~\cite{LiZ2026}, motivated by the \enquote{platonic representation} hypothesis that sufficiently capable models converge on a shared statistical geometry of the data they are trained on, showed (Fig.~\ref{fig:sec5b}a), that seven architecturally diverse foundation models (MACE, SevenNet, Orb variants) project onto a common, largely model-independent latent space.
The distance a new configuration deviates from this shared space is a ground-truth-free, ensemble-free proxy for structural novelty, which tracks known generative model \enquote{novelty} scores and detects out-of-distribution configurations.
It also detects catastrophic forgetting-induced embedding collapse after naive single-task fine-tuning and broken rotational/permutation symmetry in non-equivariant architectures.
The authors explicitly state~\cite{LiZ2026} that a rigorous and calibrated correspondence between the manifold distance and the physical prediction error has not yet been established.
Therefore, they offer it as a promising heuristic screening tool rather than a proven quantifier of uncertainty.
Mehdi, Cho, and Isayev~\cite{Mehdi2026} follow a closely related strategy but with a more directly calibrated approach called PROBE (Post-hoc Reliability from Backbone Embeddings), see Fig.~\ref{fig:sec5b}b.
Instead of regressing a single uncertainty value, they cast the reliability of MLIP as a selective classification problem.
They train a compact, discriminative post-hoc classifier on the frozen per-atom representations of a pretrained model to output a probability of reliability that does not require any modification to the original MLIP.
Across large held-out evaluation sets and two structurally distinct architectures, PROBE outperforms ensemble disagreement as a binary reliability signal.
Its reliability probability tracks the actual error monotonically and its advantage grows with the expressiveness of the backbone representation.
This is an explicitly favorable scaling trend, as machine learning models move toward ever-larger foundation-scale backbones, which make full ensembles increasingly impractical to train.
Proposed by Li and Walsh~\cite{LiZ2026} Platonic manifold and by Mehdi \emph{et al.}~\cite{Mehdi2026} PROBE classifier, when read together, suggest a broader pattern.
As foundation models mature, uncertainty quantification is shifting from training auxiliary ensembles to post-hoc, architecture-agnostic probes of a single pre-trained model's internal representations.

\subsection{Complementary data-efficiency strategies}

A related, diagnostic rather than acquisition-oriented body of work asks not how to select new training data, but why a model fails on data it was never shown. 
Several of its methods are complementary to or substitutable for active learning itself.
In the study of Kreiman and Krishnapriyan~\cite{Kreiman2025}, test-time training before fine-tuning (Sec.~\ref{sec:complementary}) significantly reduces the amount of fine-tuning data required to achieve in-distribution accuracy, by a factor of ten.
This is demonstrated by a 25\% reduction in final error when using all available data. 
This shows that label-free, architecture-level corrections can effectively substitute for a significant portion of what would otherwise be needed through an active learning loop, as demonstrated by DFT.

When applied to nine independently trained models of vacancy-doped $\beta$-Cu$_{1.95}$Se, the atomic energy regularization (AER) framework of Hu \emph{et al.}~\cite{Hu2025} (Sec.~\ref{sec:complementary}) reduced the spread in the predicted lattice thermal conductivity to a standard deviation of only 0.027~W/m$\cdot$K around a mean value of 0.552~W/m$\cdot$K, which agreed well with experimental results.
The authors reported that the same constraint transfer successfully to a structurally distinct machine learning architecture on the same system. 
This suggests that posterior-based regularization of local quantities is a promising complement to model-level uncertainty quantification, which is discussed throughout the rest of this paper. 
Two further strategies expand the meaning of \enquote{data efficiency beyond active learning} in this context. 
One strategy is to exploit the full symmetry group of a periodic crystal to multiply the effective size of a small training set.
This allows the BIGDML framework (Bravais-Inspired Gradient-Domain Machine Learning) to match state-of-the-art graphene force accuracy with only ten training geometries, as demonstrated by Sauceda \emph{et al.}~\cite{Sauceda2022}.
Another strategy is to fuse a small amount of density functional theory (DFT) training data with macroscopic experimental observables through differentiable simulation.
This reduces titanium elastic property errors from 6-24\% to below 3\% across a 4-973~K range, while only modestly degrading DFT metric accuracy, as reported by Roecken \emph{et al.}~\cite{Roecken2024}.

\subsection{Comparison with committees and limitations}

Shuang \emph{et al.}~\cite{Shuang2025} showed, in a broader benchmark discussed in Sec.~\ref{sec10}, that a six-model ensemble of equivariant foundation potentials provides a useful and well-correlated uncertainty signal for defect energetics.
Across the evaluated datasets, the Spearman correlations of ensemble disagreement with energy and force RMSE are 0.88 and 0.70, respectively.
This suggests that committee-based uncertainty quantification (UQ), while less effective for acquisition than kernel-based methods in specific reactive-chemistry benchmarks~\cite{VargaUmbrichA2026,VargaUmbrichB2026}, remains a practical option, particularly when a model-specific kernel is unavailable, provided its disagreement signal has been validated for the target system.

\begin{table*}[htbp]
\centering
\caption{Representative studies of acquisition, reliability assessment, and complementary data-efficiency strategies (Sec.~\ref{sec5}).}
\label{tab:sec5}
\small
\setlength{\tabcolsep}{3pt}
\resizebox{1\textwidth}{!}{%
\begin{tabular}{|c|l|c|l|l|c|}
\hline
\multicolumn{1}{|c|}{\textbf{\#}} & \multicolumn{1}{|c|}{\textbf{Authors}} & \multicolumn{1}{|c|}{\textbf{Year}} & \multicolumn{1}{|c|}{\textbf{UQ / Data-Efficiency Strategy}} & \multicolumn{1}{|c|}{\textbf{Key Result}} & \multicolumn{1}{|c|}{\textbf{Fig.}} \\
\hline
1 & Sauceda et al.~\cite{Sauceda2022} & 2022 & Symmetry-multiplied kernel regression (BIGDML) & Graphene force accuracy matched with 10 geometries & -- \\
2 & R\"ocken \& Zavadlav~\cite{Roecken2024} & 2024 & DFT + experimental-observable data fusion & Ti elastic-property error 6--24\%$\to$$<$3\% & -- \\
3 & Allec \& Ziatdinov~\cite{Allec2025} & 2025 & Partially Bayesian neural networks & Matches full-BNN accuracy at $\sim$4$\times$ lower cost & Fig.~\ref{fig:sec5a}a \\
4 & Kreiman \& Krishnapriyan~\cite{Kreiman2025} & 2026 & Label-free test-time training/refinement & 10$\times$ less fine-tuning data for target accuracy & Fig.~\ref{fig:sec5c}a \\
5 & Shuang et al.~\cite{Shuang2025} & 2025 & 6-model foundation-MLIP committee & Spearman $\rho$=0.88 (energy) with true RMSE & -- \\
6 & Hu et al.~\cite{Hu2025} & 2025 & $\beta$-Cu$_{1.95}$Se; posterior local-energy regularization & Thermal-conductivity std.\ dev.\ 0.027~W/m$\cdot$K across models & Fig.~\ref{fig:sec5c}b \\
7 & Varga-Umbrich et al.~\cite{VargaUmbrichA2026} & 2026 & Pretrained NTK / activation-kernel acquisition & 38\%/28\% fewer rounds (energy/force) vs.\ random & Fig.~\ref{fig:sec5a}b \\
8 & Varga-Umbrich et al.~\cite{VargaUmbrichB2026} & 2026 & Force-aware, chunked NTK acquisition & Scales to 200k candidates; beats committees on OC20 & Fig.~\ref{fig:sec5a}c \\
9 & Li \& Walsh~\cite{LiZ2026} & 2026 & Shared ``platonic'' latent-manifold distance & Ground-truth-free novelty/OOD proxy across 7 models & Fig.~\ref{fig:sec5b}a \\
10 & Mehdi et al.~\cite{Mehdi2026} & 2026 & Post-hoc classifier on frozen embeddings (PROBE) & Beats ensemble disagreement as reliability classifier & Fig.~\ref{fig:sec5b}b \\
\hline
\end{tabular}%
}
\end{table*}

The studies of Varga-Umbrich \emph{et al.}~\cite{VargaUmbrichA2026,VargaUmbrichB2026} show substantial acquisition gains from pretrained NTK and activation kernels over committee disagreement on T1x and OC20, with the broader comparison depending on the dataset and error metric.
The first study documented weak and unstable correlations between committee disagreement and prediction error~\cite{VargaUmbrichA2026}; the follow-up found that representation-based acquisition was more robust to biased candidate pools~\cite{VargaUmbrichB2026}.
On OC20, the joint energy--force NTK also combined lower prediction errors with lower per-round computational cost than the committee baselines~\cite{VargaUmbrichB2026}.
Together, these results make pretrained representations a compelling basis for efficient active learning, while the positive committee result above supports retaining validated committees as a practical option.
The two kernel studies used MACE, while PROBE was evaluated on two structurally distinct architectures~\cite{Mehdi2026}.
The optimal balance between energy- and force-based signals remains heuristic rather than based on a clear principle \cite{VargaUmbrichB2026}.
The platonic manifold distance metric, although conceptually elegant and free from ensembles, is explicitly described by its authors as a screening tool that needs further research to link it to calibrated predictive uncertainty \cite{LiZ2026}.

Alongside acquisition and reliability assessment, this section discusses complementary strategies such as test-time training, symmetry-informed data efficiency, and experiment-fused training (Sec.~\ref{sec:complementary}).
These strategies can be combined with active data selection to make better use of a limited reference calculation budget.

\section{Active Learning for Reactive Chemistry, Rare Events, and Spectroscopy}
\label{sec6}

Reactive chemistry poses a more challenging version of the sampling problem compared to equilibrium molecular dynamics (MD).
The most representative examples are the transition states, bond breaking events, and excited state crossings that determine the outcome of a reaction which are, by definition, rare, have high energy, and are structurally different from any equilibrium configuration that a modeler might want to sample in advance.
However, these are exactly the configurations that a reactive machine learning (ML) model must get right.

\subsection{Reaction discovery and biased sampling}

For ammonia synthesis, Reactive Active Learning (RAL)~\cite{Achar2025} (Sec.~\ref{sec:exploration}) converged after five generations, with 91.7\% of the transition-state images within a 0.5~eV/\AA{} force-deviation tolerance.
Benchmarked against the prior nanoreactor active-learning approach on the methanimine system at a matched compute budget, the nanoreactor baseline had not yet converged after 47 generations. 
It showed potential-energy-surface offsets of several eV and generated intermediates with a convex-hull volume that was about two orders of magnitude smaller than the initial bootstrap data in a uniform manifold approximation and projection (UMAP) map.
For the S$_{\mathrm{N}}$2 reaction of fluoride with chloromethane, metadynamics with inherited bias~\cite{Vitartas2025} (Sec.~\ref{sec:exploration}) achieved lower energy and force errors than either plain downhill-from-transition-state sampling or metadynamics without inherited bias, at approximately half the total number of configurations and computations of the latter.
For a glycosylation reaction in explicit dichloromethane, it resolved a stepwise S$_{\mathrm{N}}$1/S$_{\mathrm{N}}i$ mechanism at a computational cost that the authors estimate to be less than 0.001\% of the equivalent \textit{ab initio} MD free-energy calculation.

\subsection{Excited-state dynamics}

Martyka \emph{et al.}~\cite{Martyka2025} extended active learning to non-adiabatic and excited-state dynamics using a multi-state neural network architecture that shares a single model across an arbitrary number of electronic states, as well as a novel \enquote{gap-driven} sampling scheme shown in Fig.~\ref{fig:sec6b}a.
This biases trajectories towards small energy-gap regions prone to conical intersections.
For fulvene, the full protocol was built from scratch and converged in three days, using 19 active learning iterations and 5\,950 training points.
This is in contrast to 104 days and 160 iterations for a simplified single-state protocol, which still did not converge after 28 days.
For the photoisomerization of azobenzene, a challenging reactive photochemistry target, predicted quantum yields agreed with high-level reference calculations and experiments.
Additionally, a previously unreported oscillatory re-equilibration was revealed in the photoisomerization yield on a long timescale ($>$1~ps).

\subsection{Vibrational spectroscopy}

\begin{figure*}[!t]
\begin{minipage}[h]{1\linewidth}
\center{ \includegraphics[width=0.85\linewidth]{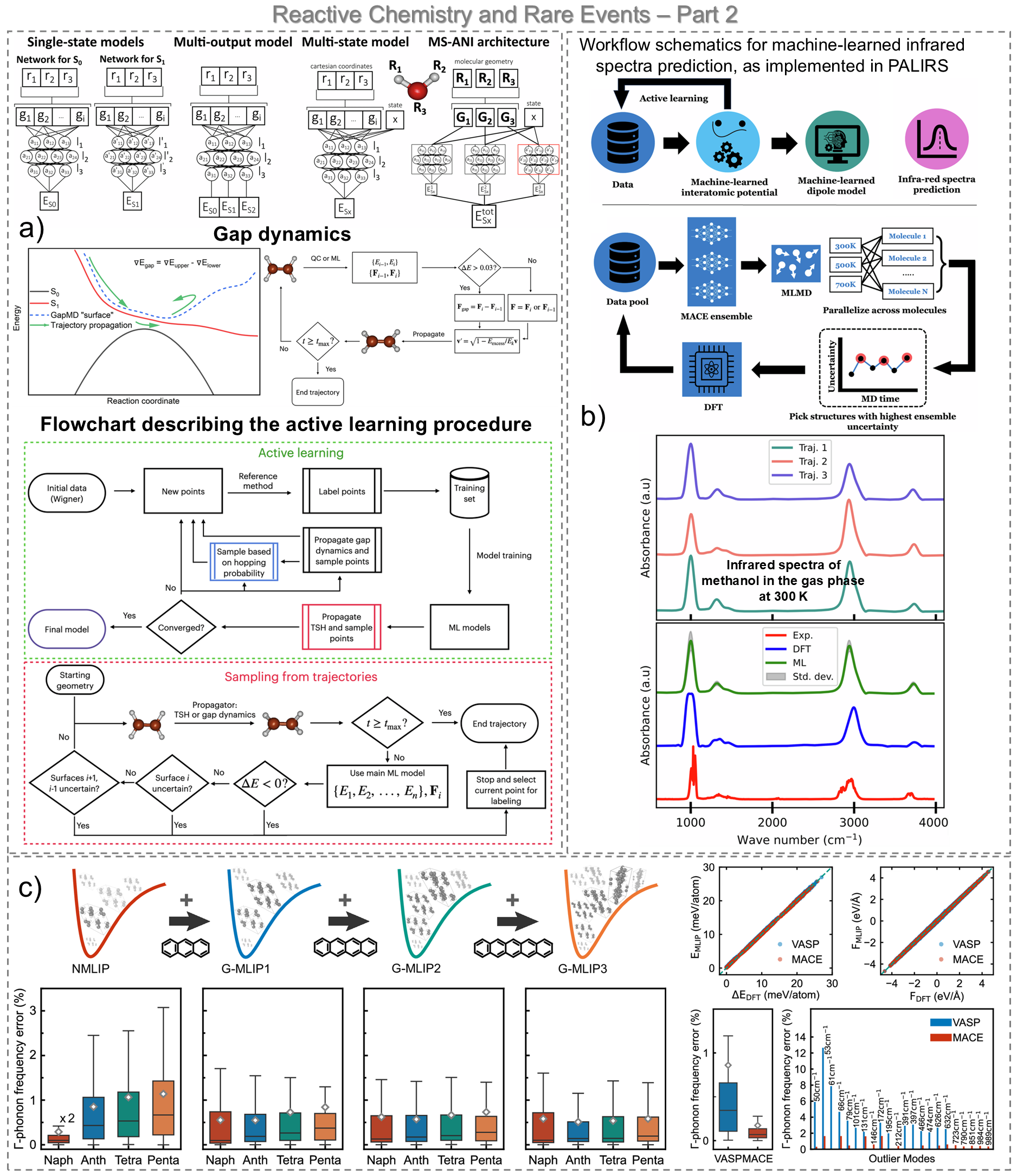}} \\
\end{minipage}
\caption{\textbf{Active Learning for Excited States and Vibrational Spectroscopy.}
\textbf{(a)} Multi-state learning with gap-driven dynamics (Ref.~\cite{Martyka2025}) representing single-output, multi-output, and multi-state atomistic networks for excited-state energies and the architecture of the multi-state MS-ANI model (top).
Gap-driven dynamics, which propagates trajectories toward small energy gaps between adjacent states and its algorithm (middle).
The active-learning loop, in which surface-hopping and gap-driven trajectories are sampled by negative energy gaps and by the uncertainty of the current and adjacent surfaces, relabeled, and used to retrain the models (bottom).
\textbf{(b)} The machine-learning workflow of Python-based Active Learning Code for Infrared Spectroscopy (PALIRS) (Ref.~\cite{Bhatia2025}) for predicting infrared spectra from MD with a MLIP and a dipole model, and the active-learning framework that builds the training data for both the potential-energy and the dipole-moment surfaces (top).
Gas-phase methanol spectra at 300~K from three ML trajectories and their average compared with DFT-based AIMD and the experimental NIST spectrum (bottom).
\textbf{(c)} The active-learning scheme for generalized MLIPs for polyacene molecular crystals (Ref.~\cite{Gurlek2025}) used to build potentials for naphthalene, anthracene, tetracene, and pentacene crystals, with error box plots of their $\Gamma$-point phonon frequencies for the naphthalene-specific and three generalized MLIPs (left).
The comparison of VASP MLIPs and a MACE committee for crystalline naphthalene in terms of relative energies, forces, harmonic phonon frequencies, and their outliers (right).
Panels adapted from the cited references.
}
\label{fig:sec6b}
\end{figure*}

A related application uses active learning not to resolve a reaction mechanism directly but to make a downstream spectroscopic observable computationally tractable.
Bhatia \emph{et al.}~\cite{Bhatia2025} built Python-based Active Learning Code for Infrared Spectroscopy (PALIRS), see Fig.~\ref{fig:sec6b}b.
The code actively samples configurations with the highest force-ensemble disagreement across multiple molecular dynamics (MD) runs at different temperatures to train machine learning (ML) models for anharmonic infrared (IR) spectra of small organic molecules. 
This replaces the need for more than 100\,000 single-point DFT calculations, which would be required for a direct \textit{ab initio}-MD-based IR calculation.
PALIRS reduces this number to fewer than 1\,000 calculations per molecule, representing a reduction of more than two orders of magnitude.
The ML spectra reached a mean Pearson correlation of 0.81 with experiment over 24 molecules, and for methanol the ML spectrum agreed with experiment even more closely than the spectrum from a single 50~ps DFT-based \emph{ab initio} MD run ($r=0.73$), which the authors attributed to the broader sampling of three independent ML trajectories. 
Gurlek \emph{et al.}~\cite{Gurlek2025} applied an eight-model committee active-learning scheme to polyacene molecular crystals (naphthalene through pentacene) targeting harmonic and anharmonic vibrational properties, see Fig.~\ref{fig:sec6b}c.
It was shown that a single machine learning interatomic potential (MLIP) actively trained on all four acene molecules generalized, without further training, to an unseen pentacene-in-naphthalene system with an average vibrational frequency error under 1\% and an insertion energy error of only 0.1~meV/atom.
This allowed the authors to resolve the host/guest mode hybridization and directly extract guest-mode vibrational lifetimes from molecular dynamics simulations.

\subsection{Comparison and limitations}

A broader electronic-structure-learning review by Hindges \emph{et al.}~\cite{Hindges2026} situates gap-driven and Bayesian active learning for non-adiabatic dynamics within the wider landscape of machine-learned electronic-structure methods.
They separately reported that Bayesian active-learning-guided training-set construction for alloy-composition/property prediction reduced the required labeled data by factors of roughly 2.5$\times$ and 1.7$\times$ for two different target systems with respect to passive, random data collection.
At the same time they achieved comparable predictive accuracy, a smaller but structurally identical gain to those reported throughout this section.
Both the reactive-chemistry and spectroscopy studies discussed above, along with the broader reviews of Kulichenko \emph{et al.}~\cite{Kulichenko2024} and Kim \emph{et al.}~\cite{Kim2026} on reactive-MLIP, which provide a more detailed description of the on-the-fly Bayesian, D-optimality, and query-by-committee methods that underlie this section, situate spectroscopy and vibrational property prediction as a natural and increasingly mainstream extension of reaction-oriented active learning.
The observable of interest, such as a spectrum, lifetime, or quantum yield, is only as reliable as the dynamical trajectory used to calculate it, which is precisely the coverage problem that active learning was designed to solve.

\begin{table*}[htbp]
\centering
\caption{Representative studies of active learning for reactive chemistry, rare events, and spectroscopy (Sec.~\ref{sec6}).}
\label{tab:sec6}
\small
\setlength{\tabcolsep}{3pt}
\resizebox{1\textwidth}{!}{%
\begin{tabular}{|c|l|c|l|l|c|}
\hline
\multicolumn{1}{|c|}{\textbf{\#}} & \multicolumn{1}{|c|}{\textbf{Authors}} & \multicolumn{1}{|c|}{\textbf{Year}} & \multicolumn{1}{|c|}{\textbf{System / Method}} & \multicolumn{1}{|c|}{\textbf{Key Active-Learning Result}} & \multicolumn{1}{|c|}{\textbf{Fig.}} \\
\hline
1 & Achar et al.~\cite{Achar2025} & 2025 & NH$_3$ synthesis, methanimine, TiC methane activation; RAL & Converges where nanoreactor-AL fails at matched budget & Fig.~\ref{fig:sec6a}a \\
2 & Bhatia et al.~\cite{Bhatia2025} & 2025 & 24 organic molecules; ensemble AL for IR spectra (PALIRS) & $>$100$\times$ fewer DFT calls than AIMD-based IR & Fig.~\ref{fig:sec6b}b \\
3 & Gurlek et al.~\cite{Gurlek2025} & 2025 & Polyacene crystals, host--guest system; committee AL & Generalizes to unseen host--guest with $<$1\% freq.\ error & Fig.~\ref{fig:sec6b}c \\
4 & Martyka et al.~\cite{Martyka2025} & 2025 & Azobenzene, fulvene; multi-state gap-driven AL & Fulvene: converges in 3 days vs.\ 104 days for simplified protocol & Fig.~\ref{fig:sec6b}a \\
5 & Vitartas et al.~\cite{Vitartas2025} & 2026 & S$_{\mathrm N}$2, sigmatropic shift, glycosylation; AL+metadynamics & Half the data/compute of plain metadynamics & Fig.~\ref{fig:sec6a}b \\
\hline
\end{tabular}%
}
\end{table*}

The five primary studies surveyed in this section, along with the reactive-MLIP literature that contextualizes them, demonstrate that active learning tailored to reaction discovery, excited-state dynamics, and vibrational spectroscopy is now capable of revealing reaction mechanisms, quantum yields, and spectroscopic observables that were previously computationally inaccessible using direct \emph{ab initio} MD or AIMD calculations.
For reaction discovery and excited-state dynamics, exploration is enhanced through methods such as explicit transition-state search, biased or inherited-bias metadynamics, and gap-driven excited-state sampling.
These methods are able to resolve these quantities at a fraction of the cost of reference calculations.
The clearest evidence that this represents genuine mechanistic progress, rather than just a faster route to the same static picture, comes from cases where actively trained dynamics reveal physics that a static or low-level calculation misses entirely.
For example, free-energy barrier for a sigmatropic rearrangement obtained by Vitartas \emph{et al.}~\cite{Vitartas2025} is in excellent agreement with experimental data, where a static harmonic approximation DFT treatment underestimates it by 2.6~kcal/mol. 
Martyka \emph{et al.}~\cite{Martyka2025} also discovered a long-term oscillatory re-equilibration in azobenzene photoisomerization that is invisible to single-snapshot analyses. 

The reaction-discovery and excited-state studies illustrate why uncertainty thresholding on unbiased trajectories can miss rare configurations that determine the target observables.
These studies enhance exploration through explicit transition-state search~\cite{Achar2025}, inherited metadynamics bias~\cite{Vitartas2025}, or gap-driven biasing~\cite{Martyka2025} to reach the configurations that are important.
For spectroscopy, the workflows of Bhatia \emph{et al.}~\cite{Bhatia2025} and Gurlek \emph{et al.}~\cite{Gurlek2025} combine MD sampling with committee-based acquisition.
In terms of the design choices in Sec.~\ref{sec2}, MD, reaction searches, and biasing determine how candidate configurations are generated, while committee disagreement provides a criterion for selecting configurations for reference calculations.
This reinforces the general caution expressed in Sec.~\ref{sec3}, which states that on-the-fly acquisition alone is not guaranteed to cover qualitatively distinct and low-probability events.

\section{Active Learning for Mechanical, Transport, and Functional Properties}
\label{sec7}

Mechanical strength, ion and defect diffusion, and functional responses such as ferroelectric switching all depend on simulations that reach length and time scales, thousands to tens of thousands of atoms, nanoseconds to microseconds, that direct DFT cannot address, which makes this domain a natural proving ground for the cut-and-periodize on-the-fly strategies introduced in Sec.~\ref{sec3}.

\subsection{Mechanical properties}

\begin{figure*}[!t]
\begin{minipage}[h]{1\linewidth}
\center{ \includegraphics[width=1\linewidth]{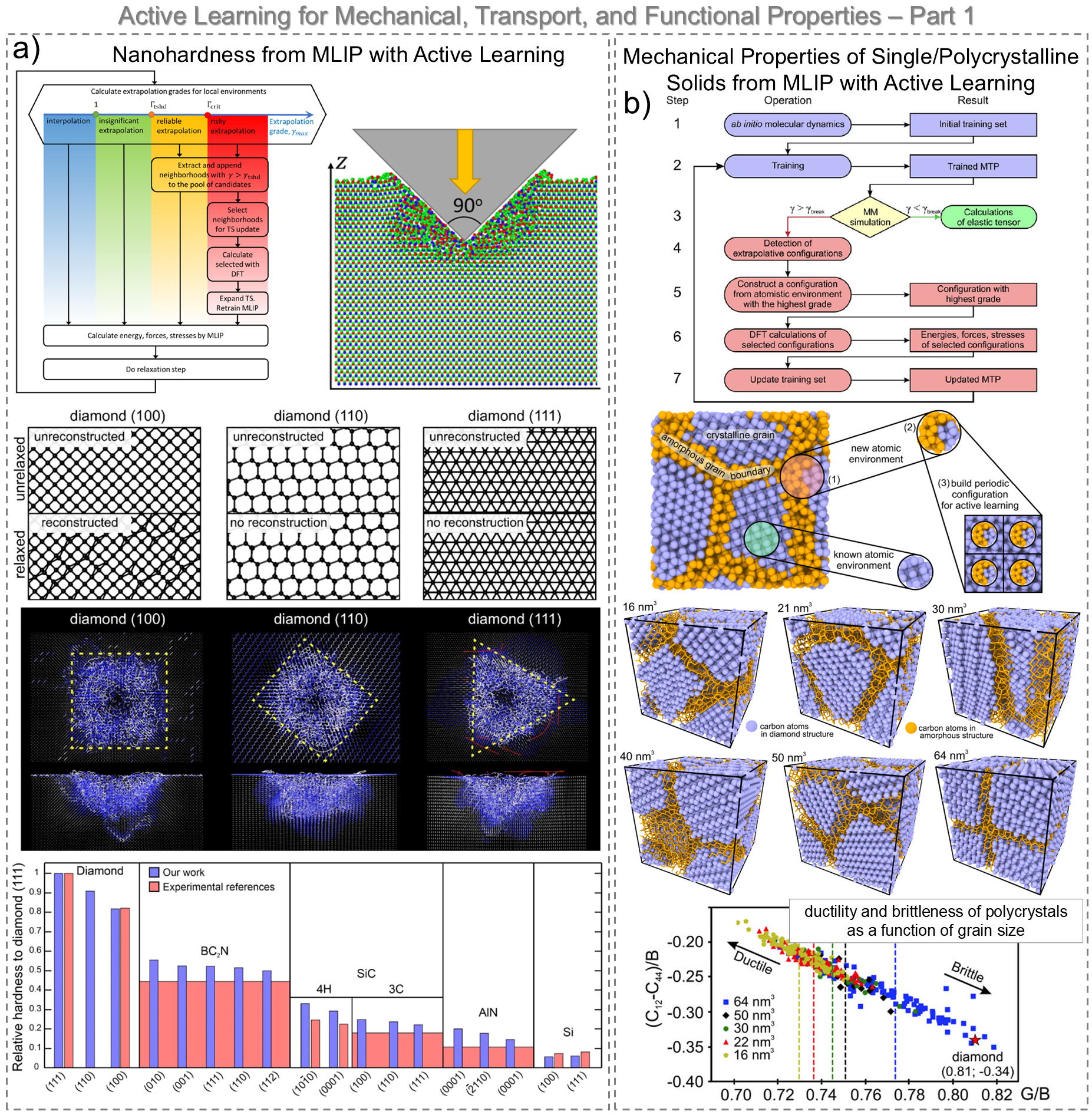}}  \\
\end{minipage}
\caption{\textbf{Learning on Local Atomic Environments for Mechanical Properties to Compute Nanohardness and Polycrystalline Elasticity.}
\textbf{(a)} The nanoindentation simulation setup used in Ref.~\cite{Podryabinkin2022} together with the on-the-fly workflow, in which the extrapolation grade is evaluated for every atomic neighborhood and the MTP is retrained whenever the largest grade exceeds a critical value, the extrapolative neighborhoods being computed with DFT as non-periodic clusters (top).
Unrelaxed, relaxed, and indented (100), (110), and (111) surfaces of diamond at a load of 3.5~$\mu$N, with bonds colored by length (middle).
Calculated versus experimental Vickers hardness of the studied compounds normalized to the diamond (111) surface (bottom).
\textbf{(b)} Scheme of the active learning on local atomic environments used for simulations of elastic properties of polycrystalline diamond (Ref.~\cite{Jalolov2024}) in which the region with the highest extrapolation grade is cut from the large structure and converted into a periodic configuration for DFT energies, forces, and stresses (top).
Atomistic models of polycrystals with average grain volumes from 16 to 64~nm$^3$ (amorphous and diamond structures shown in different colors)(middle).
The ductility--brittleness correlation between $(C_{12}-C_{44})/B$ and $G/B$ as a function of grain size, with single-crystal diamond marked by a star (bottom).
Panels adapted from the cited references.
}
\label{fig:sec7a}
\end{figure*}

\begin{figure*}[!t]
\begin{minipage}[h]{1\linewidth}
\center{ \includegraphics[width=0.93\linewidth]{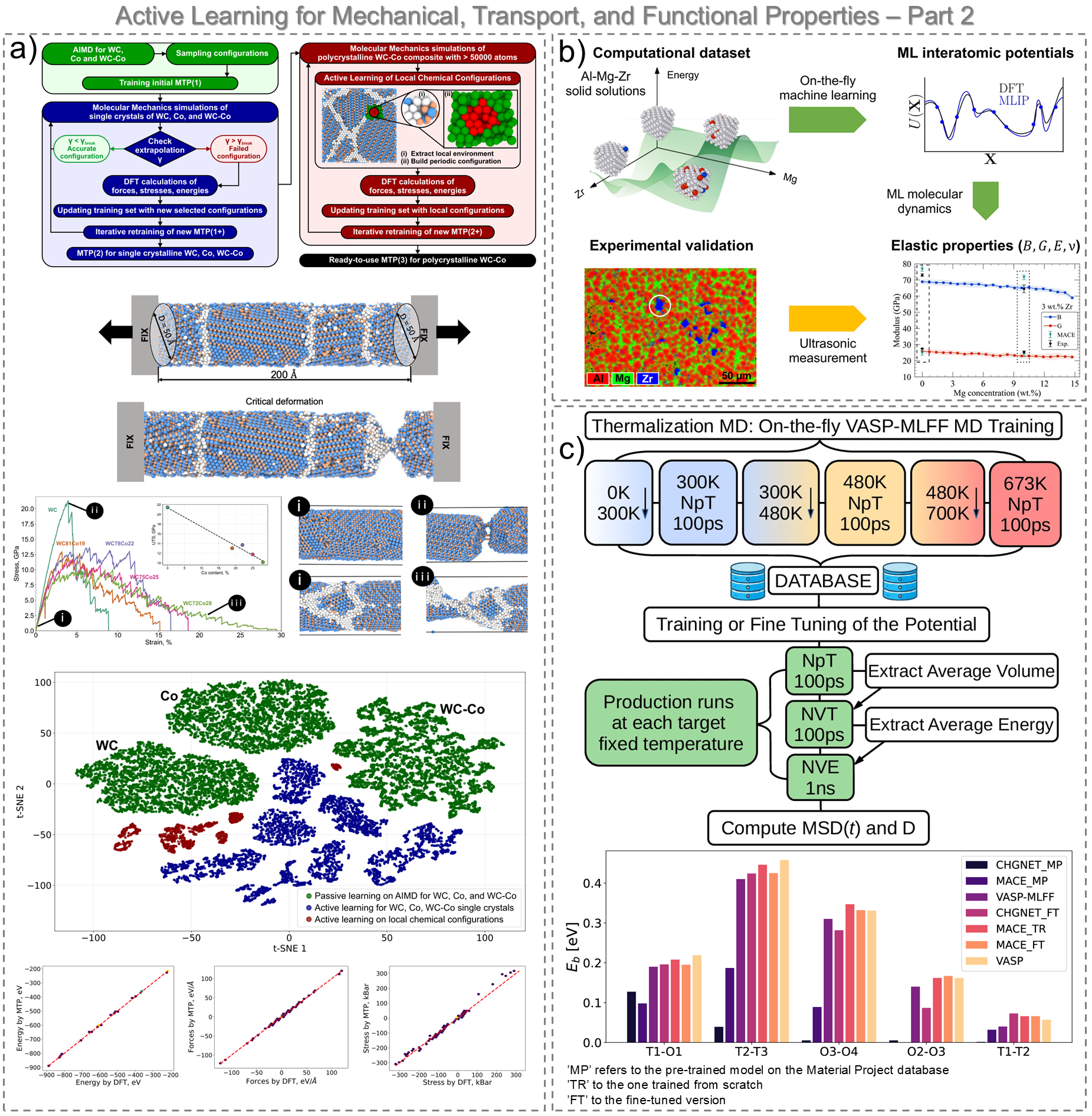}} \\
\end{minipage}
\caption{\textbf{Active Learning for Mechanical Properties of Composites and Alloys and for Hydrogen Diffusion in Metals.}
\textbf{(a)} The general procedure of active learning on local chemical configurations for WC--Co composites (Ref.~\cite{Jalolov2026}). Subsequent improvements of MLIP from passive learning on AIMD data through active learning on WC and Co single crystals and small WC--Co heterostructures to active learning on local chemical configurations (ALLCC) cut from composite structures of more than 50\,000 atoms and completed by an outer shell into periodic cells for DFT (top).
Tensile simulations of WC polycrystals and WC--Co cemented carbide with different Co contents, with stress--strain curves, atomic structures at the critical strain, and star plots of ultimate tensile strength, yield strength, Young's modulus, and critical strain (middle).
A t-SNE map of the training set colored by learning stage together with parity plots of energies, forces, and stresses for the final MTP (bottom).
\textbf{(b)} Elastic properties of Al--Mg--Zr solid solutions (Ref.~\cite{Volkmer2026}): the strategy of sampling the solid-solution phase space by DFT structural optimizations, training a MLIP on the most favorable structures by on-the-fly learning with Bayesian linear regression, and deriving elastic properties from stress--strain curves in ML-assisted MD.
\textbf{(c)} Hydrogen diffusion in magnesium (Ref.~\cite{Angeletti2025}): the protocol, in which configurations collected by on-the-fly active learning of a VASP force field during NpT heating from 0 to 700~K and at the target temperatures are used to fine-tune or train MLIPs, whose mean-squared displacements give the diffusion coefficients; and hydrogen transition barriers between octahedral and tetrahedral sites from climbing-image nudged elastic band (NEB) calculations with the pretrained, from-scratch, and fine-tuned models compared with DFT.
Panels adapted from the cited references.
}
\label{fig:sec7b}
\end{figure*}

Mortazavi \emph{et al.}~\cite{Mortazavi2023}, in a widely cited minireview of MLIPs for mechanical-property simulation, make the case explicitly that active learning \enquote{cannot be completely omitted} even after a careful two-step passive-training protocol, since classical empirical potentials and even static DFT itself can predict qualitatively wrong failure mechanisms for 2D materials under strain.
For graphene this methods overestimated tensile strength by up to 200~GPa against an experimental $130\pm10$~GPa.
The local-fragment, cut-and-periodize strategy underlying much of this section was itself first demonstrated for a such mechanical property as hardness.
Podryabinkin  \emph{et al.}~\cite{Podryabinkin2022} computed nanoindentation-derived nanohardness from active learning-trained machine learning potentials (MLPs) for diamond, aluminum nitride (AlN), silicon carbide (SiC), boron carbide nitride (BC2N), and silicon (Si), detecting extrapolative \enquote{local atomic environments} within simulation cells that were too large for direct density functional theory (DFT) calculations and relabeled them as finite, non-periodic clusters. 
The general idea, scheme and results are presented in Fig.~\ref{fig:sec7a}a.
This allowed the on-the-fly framework to apply to bulk hardness simulations. 
The resulting orientation-resolved hardness was determined for diamond (100)/(110)/(111) surfaces, while for silicon the values of 11.7--12.5 GPa agreed far better with experimental Vickers and Berkovich measurements (11.1 GPa and 12--14 GPa for silicon, respectively), than a purely empirical model based on elastic constants (7.7 GPa). 
This demonstrates that active learning-trained MLPs can resolve mechanical properties, such as crystallographic hardness anisotropy, that are not captured by empirical models or bulk elastic constants alone.
Jalolov \emph{et al.}~\cite{Jalolov2024} adapted the same local-fragment idea for polycrystalline diamond, extending the D-optimality on-the-fly framework with a bootstrapping scheme that extracts and periodizes locally extrapolative $\sim$100-atom fragments from Voronoi-tessellated polycrystal cells, see Fig.~\ref{fig:sec7a}b.
Computing elastic moduli and Pugh--Pettifor ductility ratios as a function of grain size (16--64~nm$^3$) were evaluated without ever performing DFT on the full polycrystalline structure.
The bulk modulus rose monotonically from $\sim$400~GPa at the smallest grain size, close to that of amorphous carbon, to about 480--500~GPa at the largest, still below the 550~GPa single-crystal value. 
A closely related follow-up, Jalolov \emph{et al.}~\cite{Jalolov2026} applied the same local-fragment active-learning strategy, there named active learning on local chemical configurations (ALLCC), to WC--Co cemented-carbides across the full 6--28\% cobalt-content range, see Fig.~\ref{fig:sec7b}a.
This allowed reducing energy MAE from 286~eV for a passively trained potential to 1.5~meV over three progressive training stages while adding only $\sim$20 new local configurations per pass, and quantitatively reproduced the composite's brittle-to-ductile transition, with ultimate tensile strength falling 53\% (21.4 to 10.1~GPa) and critical tensile strain rising sixfold as cobalt content increased. 
Volkmer \emph{et al.}~\cite{Volkmer2026} used on-the-fly Bayesian active learning as implemented in VASP code directly, bypassing 98\% of candidate DFT calculations, to model elastic properties of Al--Mg--Zr solid solutions up to 2048-atom supercells.
Bulk modulus decreasing from 72.8 to 59.0~GPa was found as Mg and Zr content increased and validating the predictions within a few GPa of ultrasonic experimental measurements on cast alloys.
Except for the one composition (AlZr$_3$) containing a secondary Al$_3$Zr precipitate phase the homogeneous-solid-solution training set could not represent, a clean illustration of active learning's fidelity being bounded by the physical scope of its training domain rather than by sampling density within that domain.

\subsection{Diffusion and ionic transport}

Angeletti \emph{et al.}~\cite{Angeletti2025} directly compared Bayesian active learning on-the-fly against pretrained and pretrained-then-fine-tuned universal MLIPs (CHGNet and MACE) for hydrogen diffusion in magnesium, see Fig.~\ref{fig:sec7b}c.
The pretrained foundation models alone were more than an order of magnitude worse than the on-the-fly potential. 
However, fine-tuning these models on the actively generated dataset allowed MACE to surpass even the accuracy of the on-the-fly potential and correctly reproduce the experimental Arrhenius diffusion curve. 
The activation energy of MACE was 0.28~eV, which is close to reported experimental value of 0.25~eV, whereas the on-the-fly model had an activation energy of 0.40~eV.
This result echoes the theme of foundation-model plus targeted correction from Sec.~\ref{sec4} in a transport property setting.

\begin{figure*}[!t]
\begin{minipage}[h]{1\linewidth}
\center{ \includegraphics[width=1\linewidth]{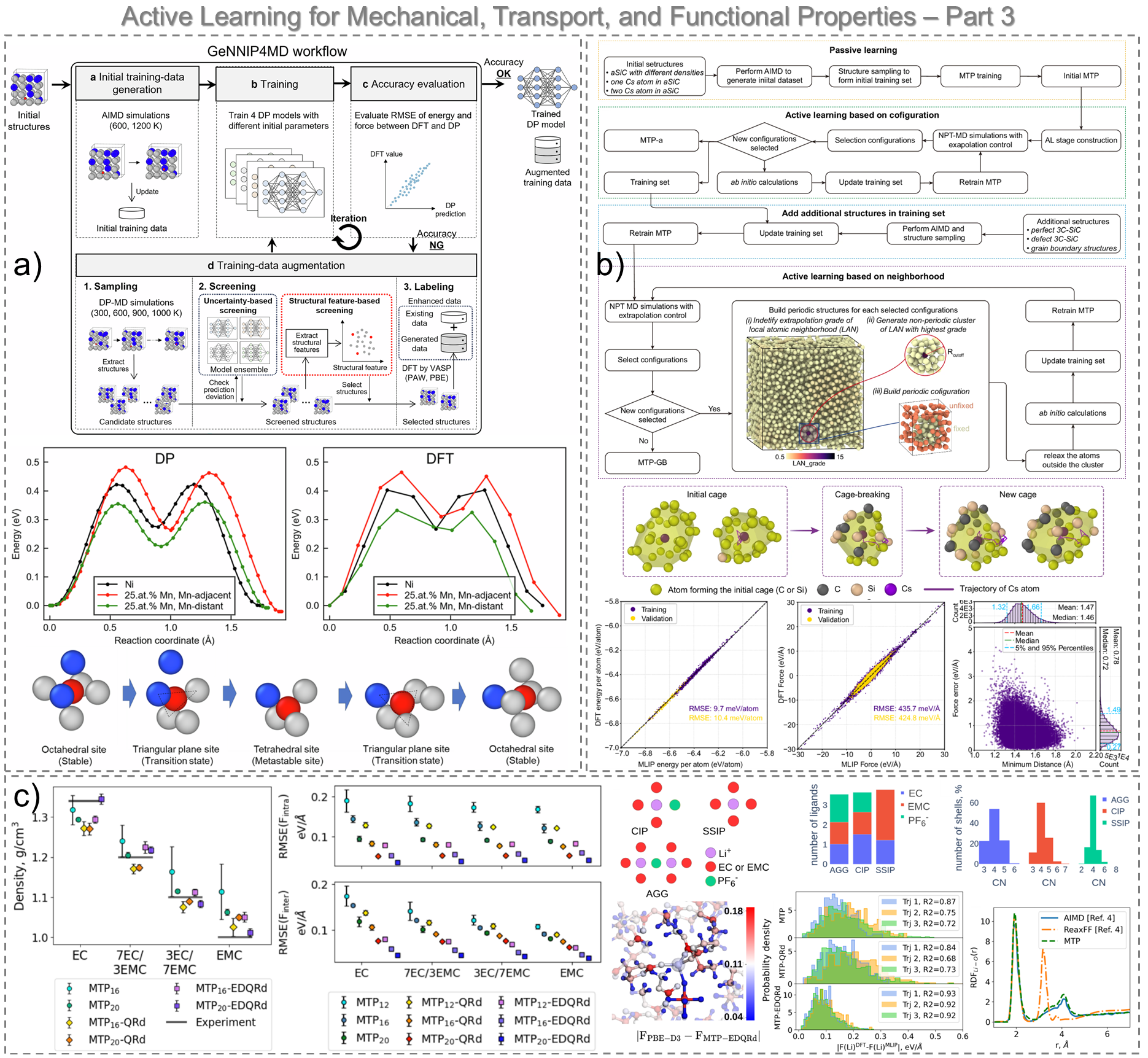}}  \\
\end{minipage}
\caption{\textbf{Active Learning for Ionic and Impurity Transport, cases of Hydrogen in Ni--Mn, Cesium in SiC, and Liquid Electrolytes.}
\textbf{(a)} Hydrogen diffusion in Ni--Mn random alloys (Ref.~\cite{Ito2025}) studied by developing the GeNNIP4MD workflow, in which four deep potentials are trained, then candidate structures from MD are screened by the maximum force deviation among them and then by structural similarity to the training set, and the selected structures are labeled with DFT (top).
Hydrogen-diffusion energy profiles for pure Ni and Ni--25~at.\% Mn from the deep potential and from DFT, with and without Mn adjacent to the barrier site, together with the local atomic configuration along the diffusion path (bottom).
\textbf{(b)} Cesium diffusion in amorphous and polycrystalline SiC (Ref.~\cite{LiJ2025}) studied by the developed MTP workflow of passive learning, configuration-based active learning for amorphous SiC--Cs, addition of further structures, and neighborhood-based active learning for grain boundaries (top).
Schematic illustration of the cage diffusion mechanism of Cs and parity plots of energies and forces for the amorphous-SiC potential and the relation between force error and minimum interatomic distance for C--C pairs (bottom).
\textbf{(c)} Liquid battery electrolytes (Ref.~\cite{Chalykh2026}) investigated by mean densities of EC/EMC mixtures predicted by ensembles of three MLIPs compared with experiment-based literature values, and intra- and intermolecular force errors of the plain MTP and its variants with fixed (MTP-QRd) and environment-dependent (MTP-EDQRd) charges at several levels relative to PBE-D3 and for LiPF$_6$ solutions (left).
The ion-pair types and composition of the training set, the Li$^+$ solvation-shell and coordination statistics, force-error magnitudes, and the Li--O radial distribution function compared with AIMD and ReaxFF (right).
Panels adapted from the cited references.
}
\label{fig:sec7c}
\end{figure*}

Ito \emph{et al.}~\cite{Ito2025} developed a deep-learning interatomic potential for the Ni-Mn-H system using an active learning tool that combines a structural diversity filter (densMAP) with conventional force uncertainty screening, see Fig.~\ref{fig:sec7c}a.
They used the resulting dataset of 17\,777 structures to reproduce the experimentally observed non-monotonic dependence of hydrogen diffusion activation energy on manganese content without the need for empirical fitting, for the first time. 
The study resolves this dependence into two competing mechanisms: Mn-induced lattice expansion that lowers the barrier and Mn-H core repulsion that raises it.
Li \emph{et al.}~\cite{LiJ2025} extended the local-fragment, \enquote{active learning on neighborhood} strategy of Sec.~\ref{sec3}, to cesium diffusion in amorphous and high-energy grain-boundary polycrystalline silicon carbide.
This is relevant to fission product transport in nuclear fuel. 
They resolved a cage-breaking diffusion mechanism with activation energies between 1.32 and 1.85~eV, which fall between literature values for crystalline bulk (5.1--5.5~eV) and disordered boundaries (0.7--2.4~eV).
These values are consistent with experimental ranges.
Chalykh et al.~\cite{Chalykh2026} applied the D-optimality/MaxVol framework to liquid battery electrolytes, see Fig.~\ref{fig:sec7c}c. 
This is an application domain where machine learning-based methods (MLIPs) have historically struggled to remain stable without extensive manual training set augmentation.
Their actively trained models for ethylene carbonate, ethyl methyl carbonate, and their mixtures, as well as LiPF$_6$ solutions, required no manual intervention and produced ionic conductivities with a mean deviation of 11\% from the experiment across the full range of solvent compositions. 
A systematic comparison of the plain model against charge-augmented variants with fixed and environment-dependent point charges showed that the explicit electrostatics matched or exceeded the accuracy of the standard model with fewer parameters. 
The environment-dependent charges also improved simulation stability. 
This evidence suggests that the choice of which physics to include in the potential, rather than just how the training data is acquired, is crucial for the success of machine learning models in this application domain.
This study investigates how far the efficiency gains of active learning can be pushed in strongly ionic systems with long-range interactions.

\subsection{Magnetic degrees of freedom}

\begin{figure*}[!t]
\begin{minipage}[h]{1\linewidth}
\center{ \includegraphics[width=0.97\linewidth]{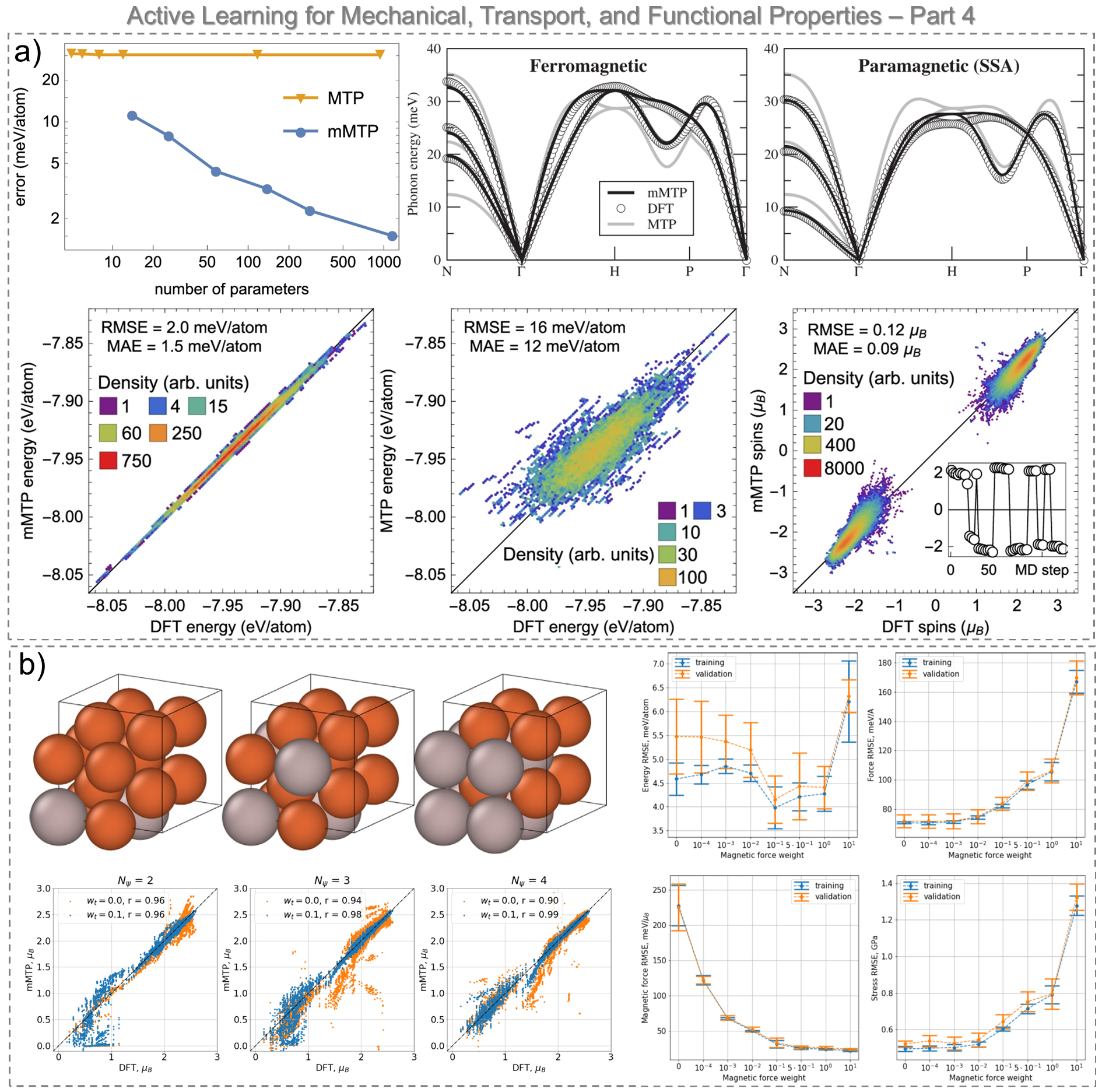}} \\
\end{minipage}
\caption{\textbf{Magnetic Moment Tensor Potentials. Learning Spin and Atomic Degrees of Freedom Together.}
\textbf{(a)} Convergence of magnetic MTPs for bcc Fe (Ref.~\cite{novikov2022}) for the energy error on held-out configurations with the number of parameters for the magnetic (mMTP) and non-magnetic MTP, showing that most of the energy variation in the training set lies in the magnetic degrees of freedom, which only the magnetic potential can capture (left).
Phonon spectra of the ferromagnetic and the paramagnetic (spin-space averaged) state from DFT, mMTP, and the non-magnetic MTP, which cannot distinguish the two states and gives spectra lying between them; and correlations with DFT in disordered-local-moment MD at 800~K for the energies of the mMTP (RMSE 2.0~meV/atom) and of the non-magnetic MTP (RMSE 16~meV/atom) and for the magnetic moments of the mMTP (RMSE 0.12~$\mu_B$) (right).
\textbf{(b)} Fitting mMTPs to magnetic forces for Fe--Al alloys (Ref.~\cite{Kotykhov2024}) showing on the examples of the 16-atom bcc Fe--Al supercells of the training set (Fe$_{14}$Al$_2$, Fe$_{12}$Al$_4$, Fe$_8$Al$_8$) (left).
Training and validation RMSEs of energies, forces, stresses, and magnetic forces from five-fold cross-validation as a function of the magnetic-force weight in the objective function, where increasing the weight lowers the magnetic-force error about tenfold while the other errors stay nearly unchanged up to weights of about 0.1~$\mu_B^2$ (bottom left).
The correlation between the equilibrium Fe magnetic moments from mMTP ensembles and DFT for $N_\psi = 2$, 3, and 4 magnetic basis functions, which is higher for models fitted to magnetic forces ($w_t = 0.1\,\mu_B^2$) than for models fitted without them ($w_t = 0$) (right).
Panels adapted from the cited references.
}
\label{fig:sec7e}
\end{figure*}

Novikov \emph{et al.}~\cite{novikov2022} introduced a class of machine learning interatomic potentials called Magnetic Moment Tensor Potentials (mMTPs), see Fig. \ref{fig:sec7e}a.
These potentials explicitly incorporate both vibrational and magnetic degrees of freedom by coarsely graining the atomic and spin spaces using a two-step minimization process.
The key methodological advancement of mMTPs is that they parameterize the energy in terms of atomic coordinates and magnetic moments, which allows them to distinguish between different magnetic states of a crystal structure.
This capability is missing from conventional MLIPs, which only depend on atomic positions and therefore cannot capture magnetic energy variations.
The authors demonstrated the performance of machine learning-based interatomic potentials (mMTPs) on the prototype bcc iron magnetic system, showing that the magnetic version consistently converges towards density functional theory (DFT) accuracy, whereas the non-magnetic MTP does not.
They achieved an root mean square error (RMSE) of 16~meV/atom in energy, while reproducing phonon spectra for different magnetic states and enabling molecular dynamics simulations with fluctuating magnetic moments.
Kotykhov \emph{et al.}~\cite{Kotykhov2024} improved previously describe approach for fitting magnetic Moment Tensor Potentials (mMTPs) to magnetic forces, which are defined as negative derivatives of energy with respect to magnetic moments, Fig. \ref{fig:sec7e}b.
These forces were obtained from spin-polarized DFT calculations.
The training set consisted of 2632 16-atom bcc Fe-Al configurations with Al concentrations up to 50\%, generated using constrained DFT (cDFT) for non-equilibrium magnetic moments and standard DFT for equilibrium configurations.
They compared two fitting strategies: one using only energies, forces, and stresses (with magnetic force weight $w_i = 0$), and another that additionally fitted to magnetic forces with an optimal weight of $w_i = 0.1\,\mu_B^2$.
The mMTP ensembles were constructed with varying numbers of magnetic basis functions ($N_\psi = 2, 3, 4$).
The results showed that fitting to magnetic forces reduces the RMSE for magnetic forces by a factor of ten (e.g., from $\sim 200$ to $\sim 15$~meV/$\mu_B$) for $N_\psi=4$ without significantly degrading energy, force, or stress accuracy.
More importantly, it ensures 100\% equilibration reliability and 100\% relaxation reliability, whereas models without magnetic force fitting achieved below 100\% reliability.
The mMTPs fitted to magnetic forces also yielded systematically better predictions of formation energies, lattice parameters, and total cell magnetic moments, with RMSEs as low as 1.1 meV/atom, 0.0049~\AA{}, and 0.032~$\mu_B$/Fe for $N_\psi=4$, respectively.
Finite-temperature molecular dynamics simulations at 300~K reproduced the experimentally observed lattice expansion trend of Fe-Al alloys, though with a quantitative underestimation of lattice parameters (RMSE $\sim 0.025$~\AA{} vs experiment) attributed to the PBE functional.
The mMTP was several orders of magnitude faster than spin-polarized DFT ($\sim 10^{-2}$ vs $10^{4}$~s/atom).
This work provided a reliable and efficient framework for modeling magnetic materials with machine-learned potentials, enabling accurate description of coupled magnetic and vibrational degrees of freedom.
By showing that fitting to magnetic forces dramatically improves the reliability of structure relaxation and magnetic moment equilibration while keeping the training set small, the method addresses a critical bottleneck in the development of magnetic MLIPs.
The approach is particularly promising for combination with active learning, potentially reducing the number of costly spin-polarized DFT calculations required to build accurate potentials for magnetic alloys, which are relevant for spintronics, sensors, and other technological applications.

\subsection{Ferroelectric effective Hamiltonians}

\begin{figure*}[!t]
\begin{minipage}[h]{1\linewidth}
\center{ \includegraphics[width=1\linewidth]{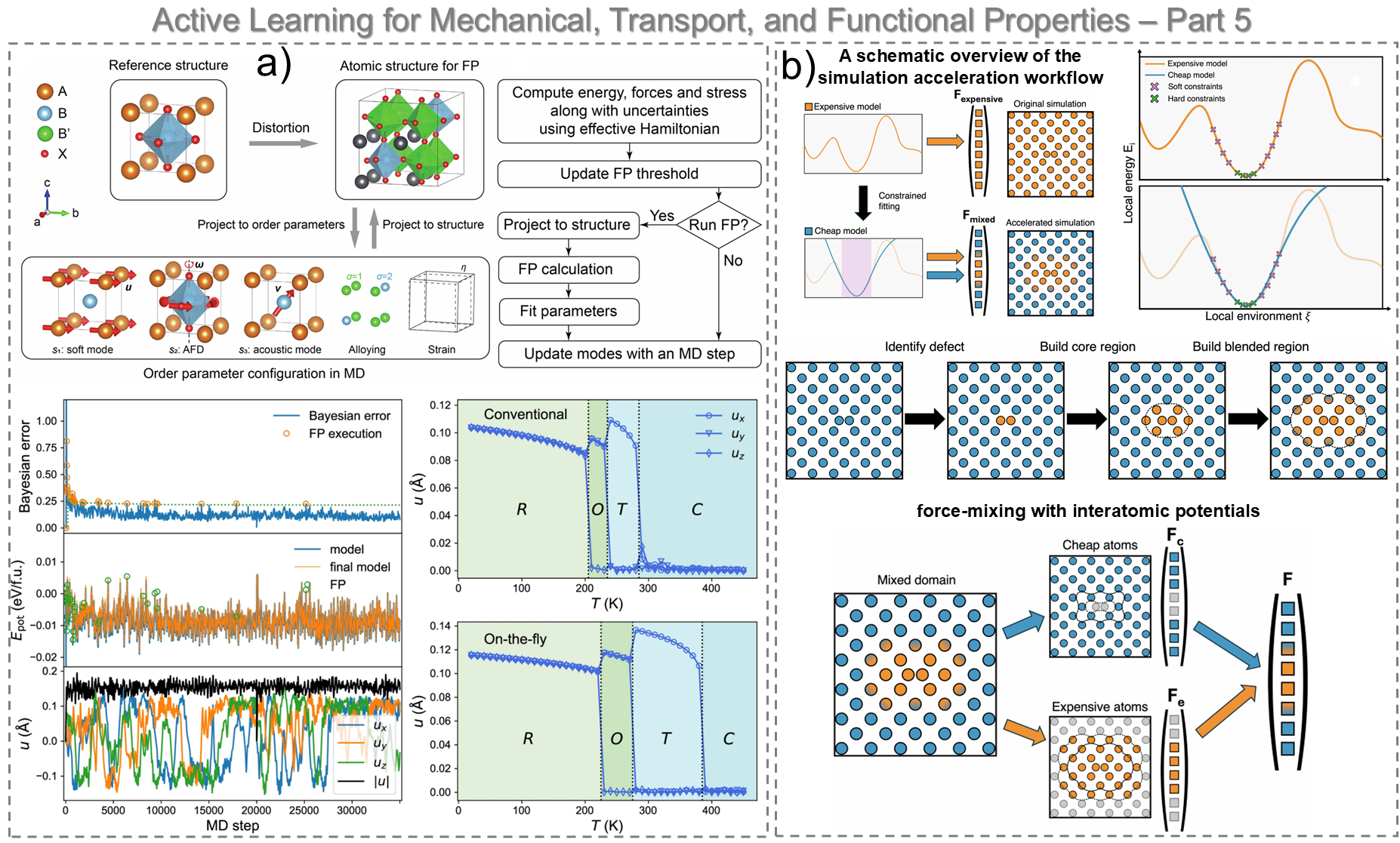}}  \\
\end{minipage}
\caption{\textbf{Scaling Simulations to Very Large Systems with Effective Hamiltonians and Spatial Mixing of Potentials.}
\textbf{(a)} On-the-fly learning of effective Hamiltonians for perovskites (Ref.~\cite{Ma2025}) using the projection between the atomic structures of first-principles calculations and the order-parameter configurations of effective-Hamiltonian MD, and the on-the-fly learning workflow (top).
The temperature-dependent phases of BaTiO$_3$ (rhombohedral, orthorhombic, tetragonal, cubic) obtained with parameters from direct first-principles fitting and from on-the-fly learning (bottom).
\textbf{(b)} The acceleration workflow of ML-MIX spatial mixing of MLIPs (Ref.~\cite{Birks2026}), in which a cheap potential is fitted under constraints to reproduce an accurate, expensive potential locally and then evaluates forces on atoms in bulk-like environments (top).
The construction of these regions around a defect (middle).
The constrained fitting of the cheap model, which approximately matches \enquote{soft} and exactly matches \enquote{hard} constraint data sampled from the expensive potential.
Force mixing, in which the system is split into cheap and expensive subdomains with core, blending, and buffer regions and the partial forces are combined (bottom).
Panels adapted from the cited references.
}
\label{fig:sec7d}
\end{figure*}

Another group of studies applies active learning to a coarser, physically motivated Hamiltonian, instead of a conventional interatomic potential.
This is most often done for ferroelectric phase transitions, where there are a small number of relevant degrees of freedom (soft-mode amplitudes, strain, and site occupation). 
Manual parameterization has traditionally required careful, symmetry-constrained fitting of edges.
Dai \emph{et al.}~\cite{Dai2025} used Gaussian-process Bayesian optimization with an upper-confidence-bound acquisition function to select which distorted BaTiO$_3$ structures to compute next. 
They converged on a nine-parameter effective Hamiltonian within ten iterations, using only 20 total structures.
This is roughly half the 100 structures typically required by conventional fitting.
The results were markedly more accurate than those obtained using the traditional edge-fitting approach, reproducing the three experimental phase-transition temperatures more accurately.
For example, the cubic-tetragonal transition temperature was 385~K, compared to an experimental value of 403~K, while the LDA-based method gave a value of 320~K.
Ma \emph{et al.}~\cite{Ma2025} applied the same on-the-fly Bayesian approach to super-large-scale simulations, training effective Hamiltonians for BaTiO$_3$, multidomain PbTiO$_3$, Pb(Zr$_{0.75}$Ti$_{0.25}$)O$_3$, and a SrTiO$_3$/PbTiO$_3$ bilayer hosting polar skyrmions, see Fig.~\ref{fig:sec7d}a.
Training procedure used only 36 density functional theory (DFT) calculations across approximately 35\,000 Monte Carlo (MD) steps for BaTiO$_3$ alone.
They scaled the resulting models to supercells containing more than ten million atoms using a single CPU core. 
This was roughly three orders of magnitude faster than using deep-potential or on-the-fly machine learning-based interatomic potential (MLIP) molecular dynamics at the same size. 
Additionally, they showed that fitting the same Hamiltonian to pre-generated, non-actively-selected \textit{ab initio} molecular dynamics (AIMD) trajectories gave inaccurate, sampling condition-dependent parameters.

\subsection{Comparison and limitations}

Finally, Birks et al.~\cite{Birks2026}, whose ML-MIX framework accelerates large-scale mechanical-defect simulation by spatially mixing a cheap and an expensive MLIP, rather than by active learning per se, achieved whole-simulation speedups of 4-11$\times$ for point-defect and dislocation-glide problems in Si, Fe, and W.
They explicitly flag the absence of an uncertainty-quantification-driven scheme for automatically sizing the expensive spatial region as an open problem. 
This highlights a practical uncertainty quantification (UQ) challenge also central to active learning: deciding automatically when a more expensive description is needed.

\begin{table*}[htbp]
\centering
\caption{Representative studies of active learning and complementary modeling strategies for mechanical, transport, and functional properties (Sec.~\ref{sec7}).}
\label{tab:sec7}
\small
\setlength{\tabcolsep}{3pt}
\resizebox{1\textwidth}{!}{%
\begin{tabular}{|c|l|c|l|l|c|}
\hline
\multicolumn{1}{|c|}{\textbf{\#}} & \multicolumn{1}{|c|}{\textbf{Authors}} & \multicolumn{1}{|c|}{\textbf{Year}} & \multicolumn{1}{|c|}{\textbf{System / Property}} & \multicolumn{1}{|c|}{\textbf{Key Result}} & \multicolumn{1}{|c|}{\textbf{Fig.}} \\
\hline
1 & Mortazavi et al.~\cite{Mortazavi2023} & 2023 & 2D materials, mechanical failure; minireview & Static DFT/EIPs give qualitatively wrong failure modes & -- \\
2 & Podryabinkin et al.~\cite{Podryabinkin2022} & 2022 & Diamond, AlN, SiC, BC$_2$N, Si; local-fragment AL & Si hardness 11.7--12.5~GPa vs.\ exp.\ 11.1--14~GPa & Fig.~\ref{fig:sec7a}a \\
3 & Jalolov et al.~\cite{Jalolov2024} & 2024 & Polycrystalline diamond; local-fragment MaxVol AL & Bulk modulus vs.\ grain size without full-cell DFT & Fig.~\ref{fig:sec7a}b \\
4 & Dai et al.~\cite{Dai2025} & 2025 & BaTiO$_3$ effective Hamiltonian; BO-driven AL & 20 structures vs.\ $\sim$100 for conventional fitting & -- \\
5 & Ma et al.~\cite{Ma2025} & 2025 & Ferroelectric perovskites; on-the-fly Bayesian AL & 36 DFT calls scale to $>$10M-atom Hamiltonian MD & Fig.~\ref{fig:sec7d}a \\
6 & Angeletti et al.~\cite{Angeletti2025} & 2025 & H diffusion in Mg; on-the-fly AL + foundation FT & Fine-tuned MACE surpasses on-the-fly-only accuracy & Fig.~\ref{fig:sec7b}c \\
7 & Ito et al.~\cite{Ito2025} & 2025 & H diffusion in Ni--Mn; diversity-filtered AL & Reproduces non-monotonic H-diffusion vs.\ Mn content & Fig.~\ref{fig:sec7c}a \\
8 & Li et al.~\cite{LiJ2025} & 2026 & Cs diffusion in SiC; neighborhood-level AL & Resolves cage-breaking mechanism, $E_a$=1.32--1.85~eV & Fig.~\ref{fig:sec7c}b \\
9 & Jalolov et al.~\cite{Jalolov2026} & 2026 & WC--Co composites; ALLCC local-fragment AL & Reproduces brittle-to-ductile transition vs.\ Co content & Fig.~\ref{fig:sec7b}a \\
10 & Chalykh et al.~\cite{Chalykh2026} & 2026 & EC/EMC/LiPF$_6$ electrolytes; D-optimality AL & Ionic conductivity within 11\% of experiment & Fig.~\ref{fig:sec7c}c \\
11 & Volkmer et al.~\cite{Volkmer2026} & 2026 & Al--Mg--Zr solid solutions; on-the-fly Bayesian AL & Elastic moduli within a few GPa of ultrasonic experiment & Fig.~\ref{fig:sec7b}b \\
12 & Birks et al.~\cite{Birks2026} & 2026 & Si, Fe, W defects; spatial MLIP mixing & Flags absent UQ-driven region-sizing as open problem & Fig.~\ref{fig:sec7d}b \\
13 & Novikov et al.~\cite{novikov2022} & 2022 & bcc Fe; magnetic states, phonons, MD & Magnetic degree of freedom included; RMSE 16~meV/atom; basis for magnetic MLIPs & Fig. \ref{fig:sec7e}a \\
14 & Kotykhov et al.~\cite{Kotykhov2024} & 2024 & bcc Fe--Al; magnetic forces, formation energy, lattice & 10$\times$ lower magnetic force RMSE; 100\% reliability; promising for AL & Fig. \ref{fig:sec7e}b \\
\hline
\end{tabular}%
}
\end{table*}

Across the studies summarized in Table~\ref{tab:sec7}, the value of active learning is particularly evident in simulations of mechanical, diffusive, magnetic, and ferroelectric behavior at scales of thousands to millions of atoms, beyond the practical reach of direct DFT calculations.
Local-fragment and effective-Hamiltonian workflows make these simulations possible with models trained on tens to thousands of DFT calculations on smaller structures.

The scale and accuracy of these simulations reflect several distinct contributions.
Active acquisition selects configurations for reference calculations, while local-fragment methods reduce the size of the structures that must be labeled.
In the magnetic studies, the gains come from including magnetic degrees of freedom in the potential~\cite{novikov2022} and magnetic forces in the fitting targets~\cite{Kotykhov2024}.
Including magnetic degrees of freedom enlarges the configuration space that must be sampled and typically requires larger training datasets~\cite{Kotykhov2024}.
This makes active learning particularly important for selecting informative configurations in the joint atomic and magnetic space while limiting the number of costly spin-polarized DFT calculations.
For effective Hamiltonians, active learning selects the training configurations, while coarse-graining reduces the cost of the subsequent simulations~\cite{Ma2025}.
The speedups from ML-MIX come from spatially restricting the use of the expensive potential during simulation~\cite{Birks2026}.
These representation, training, and simulation choices complement active acquisition (Sec.~\ref{sec:complementary}).

This section adds to Sec.~\ref{sec3}--\ref{sec4} by directly demonstrating that the resulting potentials can reproduce genuine experimental observables.
These include orientation-resolved nanohardness, ultrasonic elastic moduli, Arrhenius diffusion curves, ionic conductivities, magnetic moments, and ferroelectric transition temperatures. 
This is a stronger form of validation than the low DFT benchmark errors reported in most other studies surveyed in this review. 
The recurring limitation is the scope of the training data.
Volkmer et al.'s AlZr$_3$ discrepancy \cite{Volkmer2026} and Mortazavi et al.'s broader caution that MLIPs trained on pristine, defect-free configurations are not automatically stable for amorphous or polycrystalline structures \cite{Mortazavi2023} both illustrate that active learning can correct sampling density within a physically anticipated domain.
However, it does not, on its own, anticipate an entirely new phase, precipitate, or microstructural feature. 
This limitation is structurally similar to the out-of-distribution failures discussed in Sec.~\ref{sec4} and~\ref{sec10} for foundation MLIPs.

\section{Active Learning for High-Entropy and Compositionally Complex Materials}
\label{sec8}

High-entropy and compositionally disordered materials, alloys, perovskite oxides, and chalcogenides built from five or more near-equimolar constituents, pose a distinctive challenge for active learning.
The combinatorial space of local chemical environments grows far faster than composition count, and the near-equilibrium, ordered structures that dominate foundation-model pretraining databases are, almost by construction, a poor structural analogue for a randomly or short-range-ordered disordered lattice.

\subsection{Coverage gaps and model representations}

\begin{figure*}[!t]
\begin{minipage}[h]{1\linewidth}
\center{ \includegraphics[width=1\linewidth]{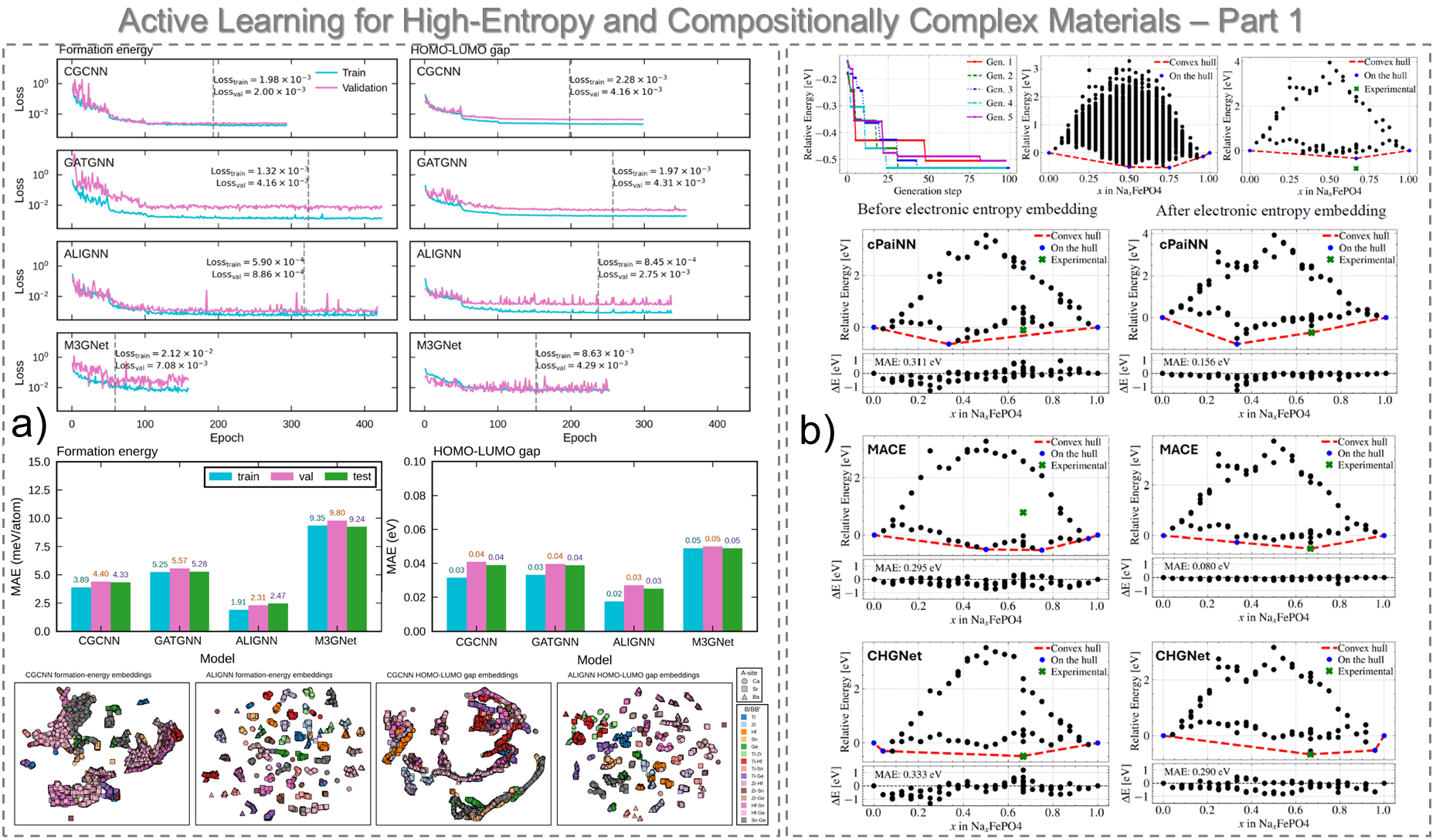}}  \\
\end{minipage}
\caption{\textbf{Coverage Gaps in Compositionally Complex Materials with Ordered-to-Disordered Transfer and Charge Ordering.}
\textbf{(a)} Training and validation loss trajectories of CGCNN, GATGNN, ALIGNN, and M3GNet on ordered single and double perovskites for the formation energy and the HOMO--LUMO gap using ordered-to-disordered transfer learning for high-entropy perovskite oxides (Ref.~\cite{Untarabut2026}), top panel.
Train, validation, and test MAEs of these models on the ordered domain (middle).
UMAP projections of the graph-level embeddings of CGCNN and ALIGNN for both targets, colored by B-site composition (bottom).
\textbf{(b)} Electronic entropy in NaFePO$_4$ (Ref.~\cite{Petersen2026}) obtained by the genetic-algorithm search over Na--vacancy orderings driven by a MACE MLIP, its convex hull, which predicts four stable phases instead of the three observed experimentally, and the DFT-optimized structures (top).
Convex hulls from CHGNet, cPaiNN, and MACE before and after charge-state (Fe$^{2+}$/Fe$^{3+}$) information is embedded in the model representation, with residuals relative to DFT, showing that only after embedding does the experimentally verified phase lie on the hull (bottom).
Panels adapted from the cited references.
}
\label{fig:sec8a}
\end{figure*}

Tahir \emph{et al.}~\cite{Tahir2026}, in their perspective on the MLIP-guided design of high-entropy perovskite cathodes for solid-oxide and protonic ceramic fuel cells, emphasize this point as a central and explicit thesis rather than a mere incidental observation. 
They argue that because public MLIP training datasets and foundation models predominantly consist of stoichiometric and near-equilibrium bulk structures, which severely underrepresent off-stoichiometric, defective, surface and hydrated configurations that are relevant to real fuel cell cathodes, robust deployment depends on active learning and deliberate inclusion of these off-equilibrium configurations. 
They propose a strategy based on a committee of independently trained MLIPs to identify out-of-distribution compositions, as standard neural network potentials lack intrinsic uncertainty estimates.
Untarabut \emph{et al.}~\cite{Untarabut2026} quantified one specific instance of this coverage gap directly by transferring graph-neural network models pretrained on 10\,898 chemically ordered perovskites to disordered high-entropy perovskite oxides, see Fig.~\ref{fig:sec8a}a.
They found that prediction of the formation energy transfers well with little or no additional disordered-domain training. 
However, HOMO-LUMO gap prediction, which is far more sensitive to the local chemical environment, transfers poorly in a zero-shot setting. 
To improve performance, targeted fine-tuning is required on the disordered domain, resulting in an improvement of approximately 87\%.
This property-dependent transferability is similar to the configuration-space coverage arguments presented in Sec.~\ref{sec3}--\ref{sec4}.
Study by Zhou and Komsa~\cite{Zhou2026} of the fine-tuning of the five-component 2D alloy (Mo, Ta, Nb, W, V)S$_2$, discussed in detail in Sec.~\ref{sec4} belongs equally to this section.
Every one of the five benchmarked foundation models gave unsatisfactory zero-shot predictions for the mixing energy of this system. 
The authors explicitly verified that the MPTrj pretraining dataset only contains the end-point structures and a few small supercell alloys for four of the ten binary combinations, and nothing with more elements.
This confirms that the alloy is a genuine, out-of-distribution target for the MACE and CHGNet models. 
MatterSim's training data could not be used to check the results, as it was not available.
Petersen \emph{et al.}~\cite{Petersen2026} identified a related but distinct failure mode in the NaFePO$_4$ battery cathode, see Fig.~\ref{fig:sec8a}b. 
Genetic algorithm search over Na-vacancy orderings driven by a machine learning algorithm (MACE) trained on a sodium-cathode dataset predicted four stable phases instead of the three observed experimentally, and missed the experimental configuration at 66\% Na. 
CHGNet, cPaiNN, and MACE all misordered the relative energies of the low-energy candidates because structural optimization with these machine learning algorithms assigns Fe$^{2+}$ and Fe$^{3+}$ incorrectly and cannot represent the electronic entropy of charge ordering.
The fix, distinguishing between Fe$^{2+}$ and Fe$^{3+}$ environments during training and retraining all three models, placing the experimentally verified phase on the convex hull, is a representation, rather than an acquisition strategy correction.
It illustrates a form of electronic complexity, rather than purely structural compositional complexity, that conventional configurational active learning, well executed, cannot address on its own.

\subsection{Experimental composition search}

A second strand of research uses active learning as a truly experimental, rather than purely computational, tool for discovering high-entropy functional materials with high thermoelectric performance. 
Jang \emph{et al.}~\cite{Jang2026} used Gaussian process Bayesian optimization to optimize a physics-based quality factor that is robust to variable, non-optimal carrier concentrations, in order to guide the real-world synthesis and characterization of high-entropy nonary chalcogenide materials within a space of 16\,206 candidate compositions, see Fig.~\ref{fig:sec8b}a.
Starting from 22 randomly chosen compositions, authors added the ten compositions with the highest expected improvement per batch. 
Six active learning iterations with 80 synthesized samples in total under 0.5\% of the full space were performed. 
This led to the identification of three compositions with a thermoelectric figure of merit $zT>2$.
One of these compositions (AL71) yielded a predicted conversion efficiency of 18.25\%, exceeding a previous record holder's 13.55\% with $zT=2.3$ at 473~K.
Post-hoc SHAP analysis of the trained surrogate revealed known p-type thermoelectric design heuristics, favoring Sn/Ge and disfavoring Bi. 
The model was never explicitly given these design heuristics.
A broader materials-design review by Yuan \emph{et al.}~\cite{Yuan2024} placed high-entropy design within a wider context of machine learning-assisted materials engineering. 
The review highlights a closed-loop active learning campaign that screened millions of candidate compositions for high-entropy Invar alloys. 
This campaign validated 17 new alloys both computationally and experimentally, identifying two with an ultra-low thermal expansion coefficient ($\sim 2 \times 10^{-6}$~K$^{-1}$ at 300~K).

\subsection{Acquisition strategies and observable validation}

Yuan \emph{et al.}~\cite{Yuan2024} also describe a VASP on-the-fly Bayesian force field, which replaced 99\% of first-principles calculations during training.
This approach resulted in a two-order-of-magnitude speedup, consistent with the gains documented in Sec.~\ref{sec3}.

\begin{figure*}[!t]
\begin{minipage}[h]{1\linewidth}
\center{ \includegraphics[width=1\linewidth]{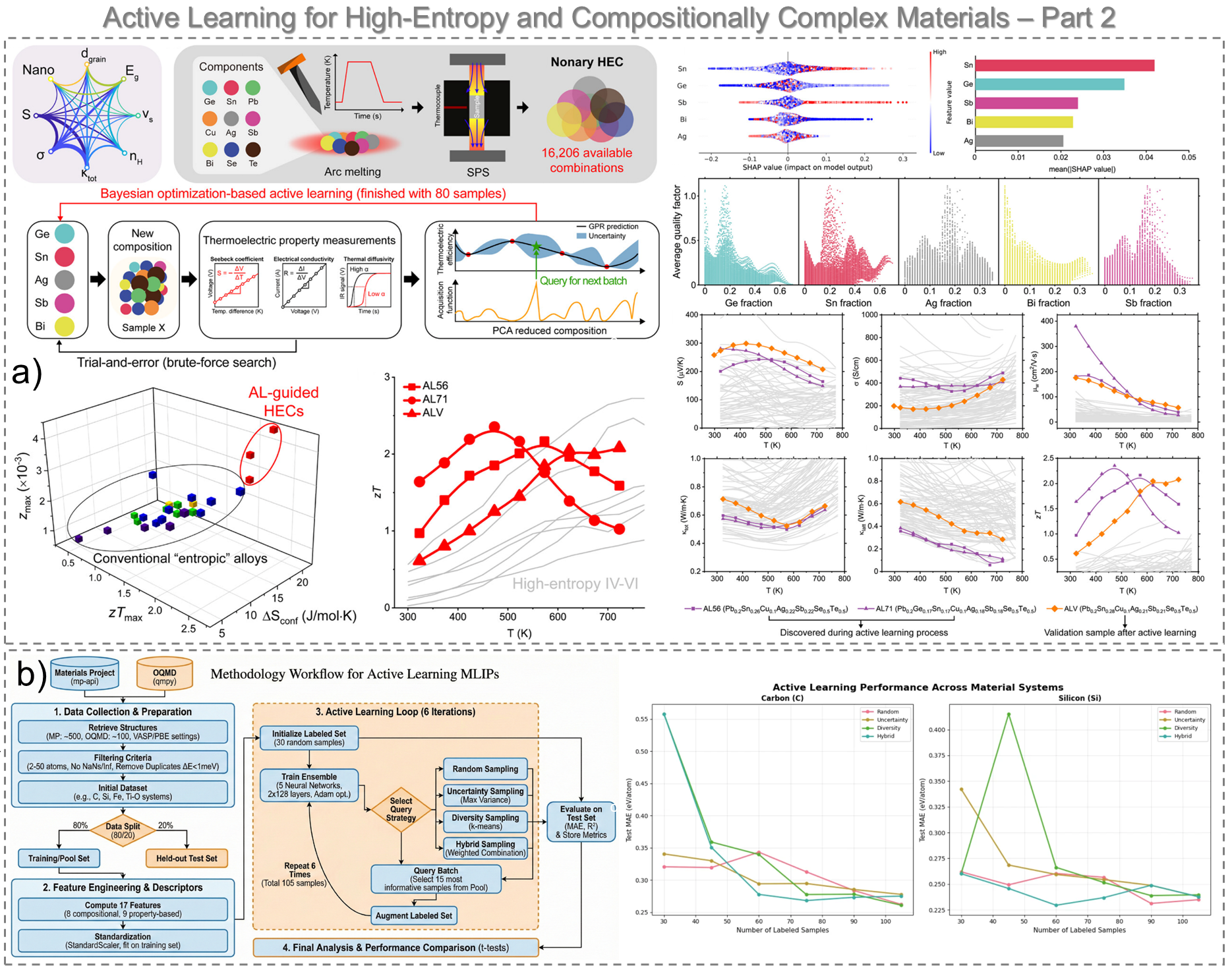}}  \\
\end{minipage}
\caption{\textbf{Active Learning for Experimental Composition Search and Formation-Energy Prediction.}
\textbf{(a)} Active-learning-guided discovery of high-entropy chalcogenide thermoelectrics (Ref.~\cite{Jang2026}) showing the correlations among thermoelectric parameters, the high-throughput synthesis of nine-element chalcogenides (top left).
The closed loop combining conventional trial-and-error with Bayesian-optimization active learning, and the configurational entropy, maximum $zT$, and temperature-dependent $zT$ of the discovered compositions compared with earlier high-entropy thermoelectrics (bottom left).
The design rules extracted from the trained model by SHAP analysis of each element's effect on the average quality factor, together with the measured Seebeck coefficient, electrical and thermal conductivities, weighted mobility, and $zT$ of the discovered samples against those used for training (right).
\textbf{(b)} Pool-based active-learning strategies for formation energies (Ref.~\cite{Khan2026}) representing the loop, which grows the labeled set from 30 to 105 samples in batches of 15 using an ensemble of five neural networks and random, uncertainty, diversity, or hybrid selection (left).
Learning curves of the MAE versus the number of labeled samples for carbon, silicon, iron, and Ti--O, where diversity sampling gives the clearest advantage for Ti--O (right).
Panels adapted from the cited references.
}
\label{fig:sec8b}
\end{figure*}

Khan \emph{et al.}~\cite{Khan2026} report a small-scale comparison of four pool-based query strategies, namely random, uncertainty (committee variance), diversity ($k$-means clustering), and a hybrid approach, for formation-energy prediction in C, Si, Fe, and Ti--O using data from the Materials Project and OQMD, see Fig.~\ref{fig:sec8b}b.
They found that diversity-based sampling was most consistently competitive, and for the Ti--O system, there was a statistically significant 10.9\% reduction in mean absolute error compared to random sampling.

Finally, the broader machine learning-interatomic potential roadmap of Zhang \emph{et al.}~\cite{Zhang2025} dedicated an entire section to chemical complexity in high-entropy materials.
They report active learning-augmented moment tensor potential studies of chemical short-range order (SRO) in systems with up to 573\,672 atoms, cautioning that standard MLIP error metrics do not always correlate with accuracy for chemically specific order parameters such as the short-range Warren--Cowley parameters that determine high-entropy material behavior.
This caution is echoed in the methodological revision made by Zhou and Komsa~\cite{Zhou2026} between the two versions of their high-entropy alloy study, where they replaced an ad hoc cluster vector order metric with a proper multicomponent Warren-Cowley formalism.

\subsection{Comparison and limitations}

\begin{table*}[htbp]
\centering
\caption{Representative studies relevant to active learning for high-entropy and compositionally complex materials (Sec.~\ref{sec8}).}
\label{tab:sec8}
\small
\setlength{\tabcolsep}{3pt}
\resizebox{1\textwidth}{!}{%
\begin{tabular}{|c|l|c|l|l|c|}
\hline
\multicolumn{1}{|c|}{\textbf{\#}} & \multicolumn{1}{|c|}{\textbf{Authors}} & \multicolumn{1}{|c|}{\textbf{Year}} & \multicolumn{1}{|c|}{\textbf{System}} & \multicolumn{1}{|c|}{\textbf{Key Result}} & \multicolumn{1}{|c|}{\textbf{Fig.}} \\
\hline
1 & Zhang et al.~\cite{Zhang2025} & 2025 & High-entropy alloys, 573k-atom SRO; roadmap section & Standard error metrics miss SRO-relevant accuracy & -- \\
2 & Khan et al.~\cite{Khan2026} & 2026 & C, Si, Fe, Ti--O; formation-energy AL (preprint) & Diversity sampling: 10.9\% MAE gain for Ti--O & Fig.~\ref{fig:sec8b}b \\
3 & Jang et al.~\cite{Jang2026} & 2026 & Nonary high-entropy chalcogenides; experimental BO & 80/16{,}206 syntheses find 3 compositions with $zT>2$ & Fig.~\ref{fig:sec8b}a \\
4 & Untarabut et al.~\cite{Untarabut2026} & 2026 & High-entropy perovskite oxides; ordered$\to$disordered TL & Gap prediction needs FT; formation energy does not & Fig.~\ref{fig:sec8a}a \\
5 & Petersen et al.~\cite{Petersen2026} & 2026 & NaFePO$_4$ charge ordering; representation fix & Charge-state embedding puts experimental phase on the hull & Fig.~\ref{fig:sec8a}b \\
6 & Tahir et al.~\cite{Tahir2026} & 2026 & High-entropy perovskite SOFC/PCFC cathodes; perspective & Argues AL essential given off-equilibrium data gap & -- \\
7 & Zhou \& Komsa~\cite{Zhou2026} & 2026 & (Mo,Ta,Nb,W,V)S$_2$ 2D HEA; fine-tuning & MPTrj screening confirms genuine OOD target & -- \\
\hline
\end{tabular}%
}
\end{table*}

For high-entropy and compositionally disordered materials, a central task for active learning is to identify and fill gaps in coverage of the relevant chemical and structural environments (Table~\ref{tab:sec8}).
The strongest evidence is found in cases where the gap and its consequences have been directly measured.
Zhou and Komsa~\cite{Zhou2026} screened the MPTrj pretraining dataset to establish the out-of-distribution status of their target alloy for MACE and CHGNet, while Untarabut \emph{et al.}~\cite{Untarabut2026} demonstrated property-dependent transfer from ordered to disordered structures for formation-energy and HOMO--LUMO gap predictors.
These studies help identify where additional target-specific training data are needed.

Controlled comparisons of acquisition strategies for high-entropy MLIP training remain limited in the studies surveyed here.
The preliminary comparison by Khan \emph{et al.}~\cite{Khan2026}, reported in a preprint, concerns formation-energy prediction in elemental and binary systems.
Its results support diversity sampling for the Ti--O benchmark and motivate analogous comparisons for high-entropy MLIP training.
In experimental composition search, Jang \emph{et al.}~\cite{Jang2026} select compositions for synthesis and use experimental measurements to update a Gaussian process surrogate of thermoelectric quality.
Their success demonstrates that active learning can navigate a large compositional search space with a limited experimental budget.
We will return to this topic in Sec. \ref{sec9} and the Conclusion.

\section{Active Learning in Generative Design and Autonomous Materials Discovery}
\label{sec9}

Beyond training a single interatomic potential, active learning has become a standard closed-loop ingredient in materials-discovery pipelines that couple a generative or search model to a property predictor, an experimental synthesis platform, or a large language model, using newly acquired data to guide subsequent selections.

The four design choices in Sec.~\ref{sec2} also organize these discovery workflows.
The \emph{organization of the feedback loop} determines how candidate generation, property evaluation, and updates are scheduled.
The \emph{acquisition criterion} determines which compositions, structures, molecules, or experimental conditions are selected for evaluation, while the \emph{exploration strategy} determines how these candidates are generated or assembled into a pool.
The \emph{model training and updating regime} includes retraining property predictors on the growing dataset, as in the molecular search of Antoniuk \emph{et al.}~\cite{Antoniuk2025}.
In the frozen language-model workflow of Wang \emph{et al.}~\cite{WangHC2026}, the accumulated observations instead enter the prompt used to select the next candidate, while the model weights remain fixed.

\subsection{Generative structure and molecular search}

\begin{figure*}[!t]
\begin{minipage}[h]{1\linewidth}
\center{ \includegraphics[width=1\linewidth]{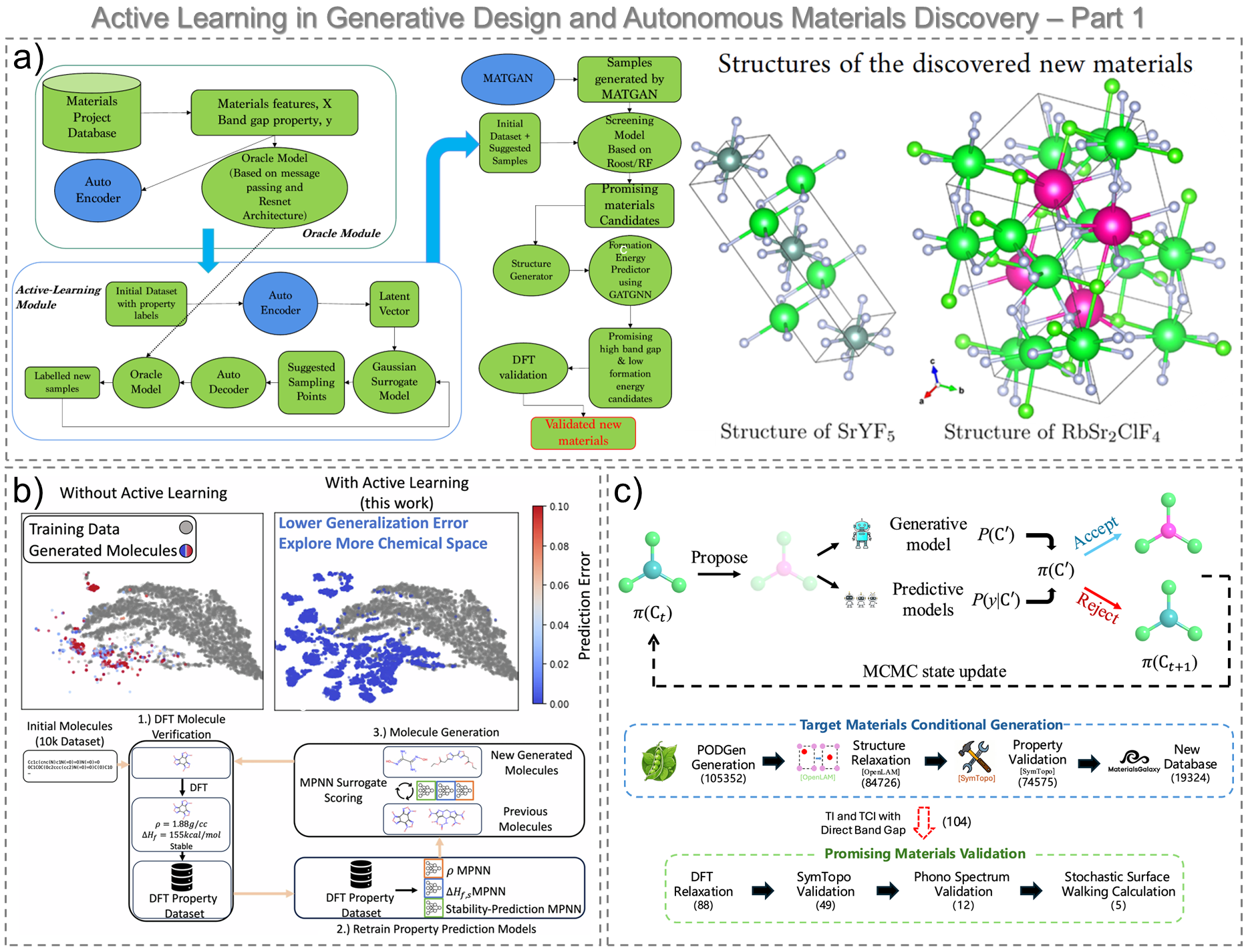}}  \\
\end{minipage}
\caption{\textbf{Active Learning Coupled to Generative Design.}
\textbf{(a)} The framework for active generative design of wide band-gap materials (Ref.~\cite{Xin2021}) in which an active-learning module selects samples for band-gap calculation to train a screening model. 
The MATGAN generative model proposes candidate materials, the screening model filters them, and an oracle model trained on all data is used for evaluation; and structures of newly discovered materials confirmed by DFT.
\textbf{(b)} t-SNE maps of closed-loop molecule generation (Ref.~\cite{Antoniuk2025}) colored by the error of the predicted density before and after active learning, and the pipeline in which the JANUS genetic algorithm generates molecules. Their density, heat of formation, and stability are computed with DFT, and three message-passing networks are retrained on the growing data to guide the next generation.
\textbf{(c)} PODGen conditional generation of crystals (Ref.~\cite{Ye2026}) with one Markov-chain Monte Carlo step, in which the target distribution combines the crystal probability of a generative model with the conditional probability of the target property from a prediction model (top).
The workflow in which generated structures are relaxed, screened, and evaluated, and the most promising candidates are validated by first-principles calculations (bottom).
Panels adapted from the cited references.
}
\label{fig:sec9a}
\end{figure*}

Xin \emph{et al.}~\cite{Xin2021} provided an early framework for this approach, combining Gaussian process Bayesian optimization with an autoencoder and GAN generative model to explore the full, unconstrained chemical space for wide-gap materials, rather than just existing databases (Fig.~\ref{fig:sec9a}a).
Starting with as few as 300 labeled samples, active learning augmentation increased the recovery rate of high-gap candidates relative to a large dataset by up to 55 percentage points, and density functional theory (DFT) validation confirmed several new thermodynamically stable wide-gap compounds, including SrYF$_5$ and phonon-stable SrClF$_3$. 
Antoniuk \emph{et al.}~\cite{Antoniuk2025} closed the loop with a genetic algorithm molecular generator (JANUS) for organic energetic materials, training three message passing property predictors over four rounds of active learning with a fixed DFT evaluation budget of 500, see Fig.~\ref{fig:sec9a}b.
The closed-loop system generated molecules with a density exceeding the maximum density of the training set (2.014 versus 1.963~g/cc), where neither the generator alone nor a static predictor could reduce the RMSE of the prediction on newly generated out-of-distribution molecules after three active learning iterations. 
The held-out RMSE was reduced by 75-83\%. 
Notably, a dedicated stability prediction classifier trained on the actively growing dataset outperformed the standard synthetic accessibility heuristic (SAScore). 
This evidence suggests that the generative model's blind spots, rather than just its target property predictions, are corrected by active learning.
Ye \emph{et al.}~\cite{Ye2026} applied a conceptually related conditional generation pipeline, PODGen, see Fig. \ref{fig:sec9a}c.
It couples an MCMC-guided crystal generator with property classifiers and a universal machine learning-based (ML) surrogate for structural relaxation to the discovery of topological insulators. 
They achieved a roughly fivefold higher hit rate for targeted topological phases under conditionally generated samples compared to unconstrained sampling.
From 105\,352 generated candidates, screening and first-principles validation of a selected subset identified 12 dynamically stable topological materials, five of which lie near global minima of their potential energy surfaces.

\subsection{Language-model-guided acquisition}

A further group of studies couples active learning to large language models (LLMs), either as the surrogate/acquisition model itself or as an orchestrating layer atop conventional predictors.

Wang \emph{et al.}~\cite{WangHC2026} proposed a \enquote{training-free} active learning framework in which a frozen large language model suggests experimental conditions in natural language and a text-similarity model matches each suggestion to a candidate in a fixed pool, without explicit uncertainty quantification, see Fig.~\ref{fig:sec9b}b.
The selected candidate's recorded property value is revealed and added to the observations in the prompt for the next iteration; the language-model weights remain fixed.
Across four diverse material datasets (steel yield strength, polymer-nanocomposite conductivity, perovskite instability, and a novel, pre-training corpus-absent, membrane elastic modulus dataset), this approach identified the best-performing candidate after evaluating less than 30\% of the candidates in the pool in most settings. 
Relative to exhaustive screening of the full pool, this corresponds to avoiding over 70\% of the candidate evaluations.
The membrane dataset benefited particularly from descriptive prompts that included procedural context.
Since memorization of answers from previous training cannot be relied upon in this case, the results support the authors' assertion that the improvements reflect genuine understanding of material trends rather than simple memorization.

\subsection{Experimental discovery loops}

Wang et al.~\cite{WangG2025} in a broader review of large language models in materials science situate this type of LLM-guided acquisition within the emerging paradigm of autonomous experimentation.
They cite the A-Lab platform as an example, which couples density functional theory (DFT)-based phase stability screening, literature-mined synthesis recipes, and an active learning loop that refines failed synthesis attempts using experimental characterization feedback.
This is a demonstration of a closed-loop logic that is applied to physical discovery rather than purely computational, although it is not specific to MLIP.
The A-Lab campaign comprised 355 autonomous synthesis experiments over 17 days, successfully synthesizing 41 out of 58 targeted novel compounds.

Ma, Kumar, Wang, and Amanchukwu~\cite{Amanchukwu2025} provided a direct, experimentally-closed-loop demonstration for discovery of battery electrolytes, see Fig.~\ref{fig:sec9c}a.
Their active learning framework uses sequential Bayesian experimental design with Bayesian model averaging over Gaussian-process surrogates with different covariance kernels.
This approach was chosen to be robust to the noisy and data-scarce labels that are typical of real battery cycling experiments. 
The target property is capacity retention in Cu$|$LiFePO$_4$ anode-free cells.
Starting with only 58 internal cycling data points, the framework explored a virtual search space of one million potential electrolyte formulations. 
After seven active learning campaigns of approximately ten physically synthesized and tested electrolytes each (approximately 70 total real experiments), four distinct new solvent chemistries were identified that rival state-of-the-art electrolytes in cycling performance. 
This is a rare example of an active learning loop that was closed entirely on physical, rather than computational, labels.

\subsection{Complementary predictors and broader evidence}

\begin{figure*}[!t]
\begin{minipage}[h]{1\linewidth}
\center{ \includegraphics[width=1\linewidth]{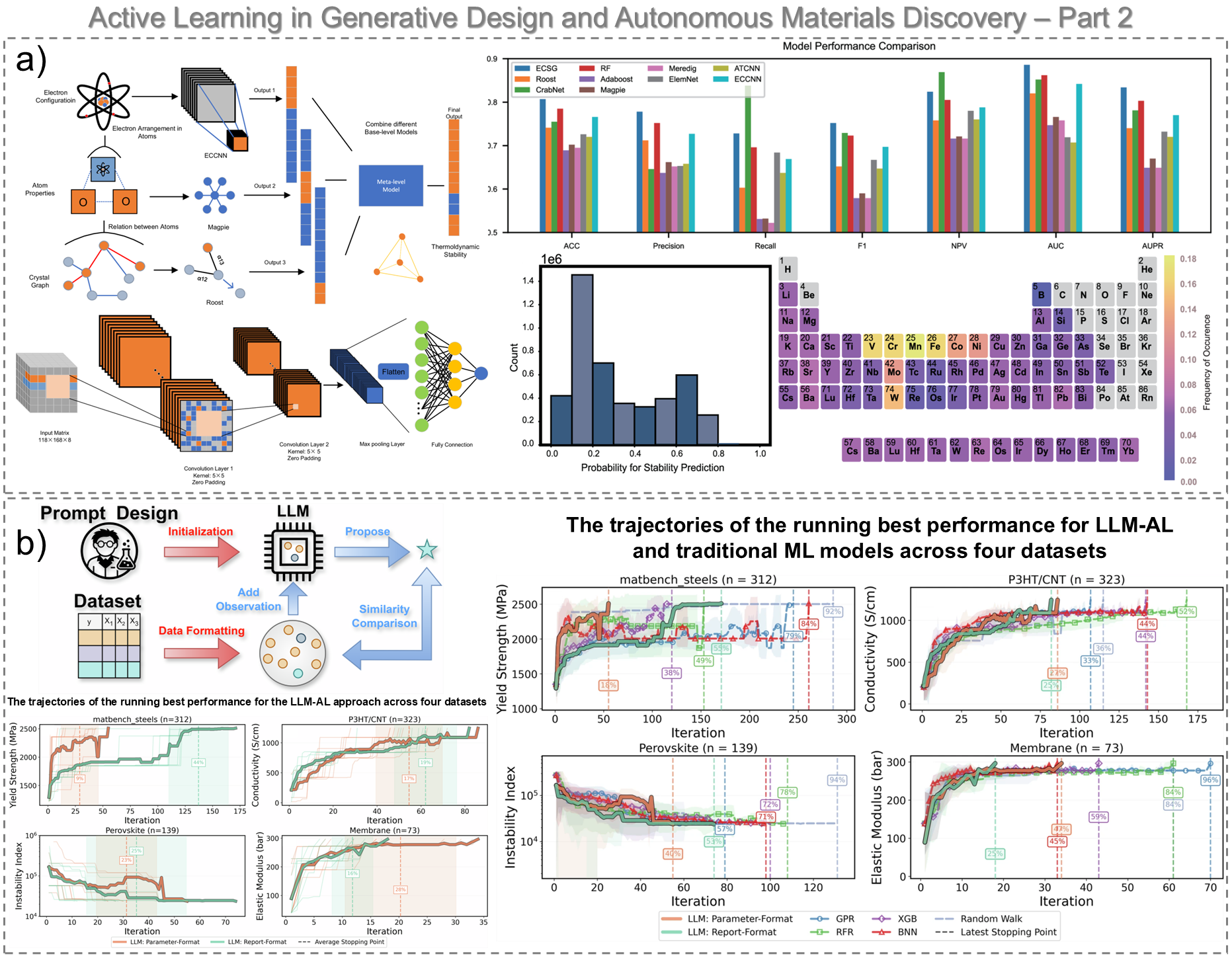}}  \\
\end{minipage}
\caption{\textbf{Data-Efficient Screening Models and Training-Free, LLM-Driven Active Learning.}
\textbf{(a)} Stacked-generalization stability prediction (Ref.~\cite{Zou2024}) with the ECSG framework, in which predictions of three complementary base models, the electron-configuration convolutional network ECCNN, Magpie, and Roost, are combined by a meta-level model, and the ECCNN architecture (left). 
The performance of ECSG and ten comparison models on seven classification metrics (top right).
The application to unexplored perovskite oxides, showing the distribution of predicted stability probabilities and the frequency of elements on the A and B sites of the compounds predicted to be stable (bottom).
\textbf{(b)} Training-free active learning with a large language model (Ref.~\cite{WangHC2026}) in which a dataset-specific prompt and text-formatted data let the model propose the next experiment, which is matched to the closest entry of the data pool and added to the labeled set (left).
Running-best trajectories on four datasets (steel strength, P3HT/CNT conductivity, perovskite stability, and membrane modulus) with parameter-format and report-format prompts (bottom).
The same trajectories compared with Gaussian-process, random-forest, XGBoost, and Bayesian-neural-network acquisition and a random walk, with the fraction of the dataset needed to reach the stopping criterion annotated (right).
Panels adapted from the cited references.
}
\label{fig:sec9b}
\end{figure*}

Zou \emph{et al.}~\cite{Zou2024}, while not explicitly based on an uncertainty-driven acquisition loop, presents a data-efficient approach to classification of thermodynamic stability of materials, Fig. \ref{fig:sec9b}a.
Their stacked-generalization ensemble combines elemental statistics, graph attention, and a novel electron configuration tensor model. The model achieved the same predictive accuracy (AUC of 0.800) with only one-seventh the training data compared to other graph-based models.
It also correctly identified 25 out of 35 double-perovskite oxide candidates as thermodynamically stable, compared to 1-6 out of 35 with a prior comparison model.
This underscores the importance of sample efficient architecture design and active data acquisition as complementary rather than competing approaches to achieving discovery throughput.

\begin{figure*}[!t]
\begin{minipage}[h]{1\linewidth}
\center{ \includegraphics[width=1\linewidth]{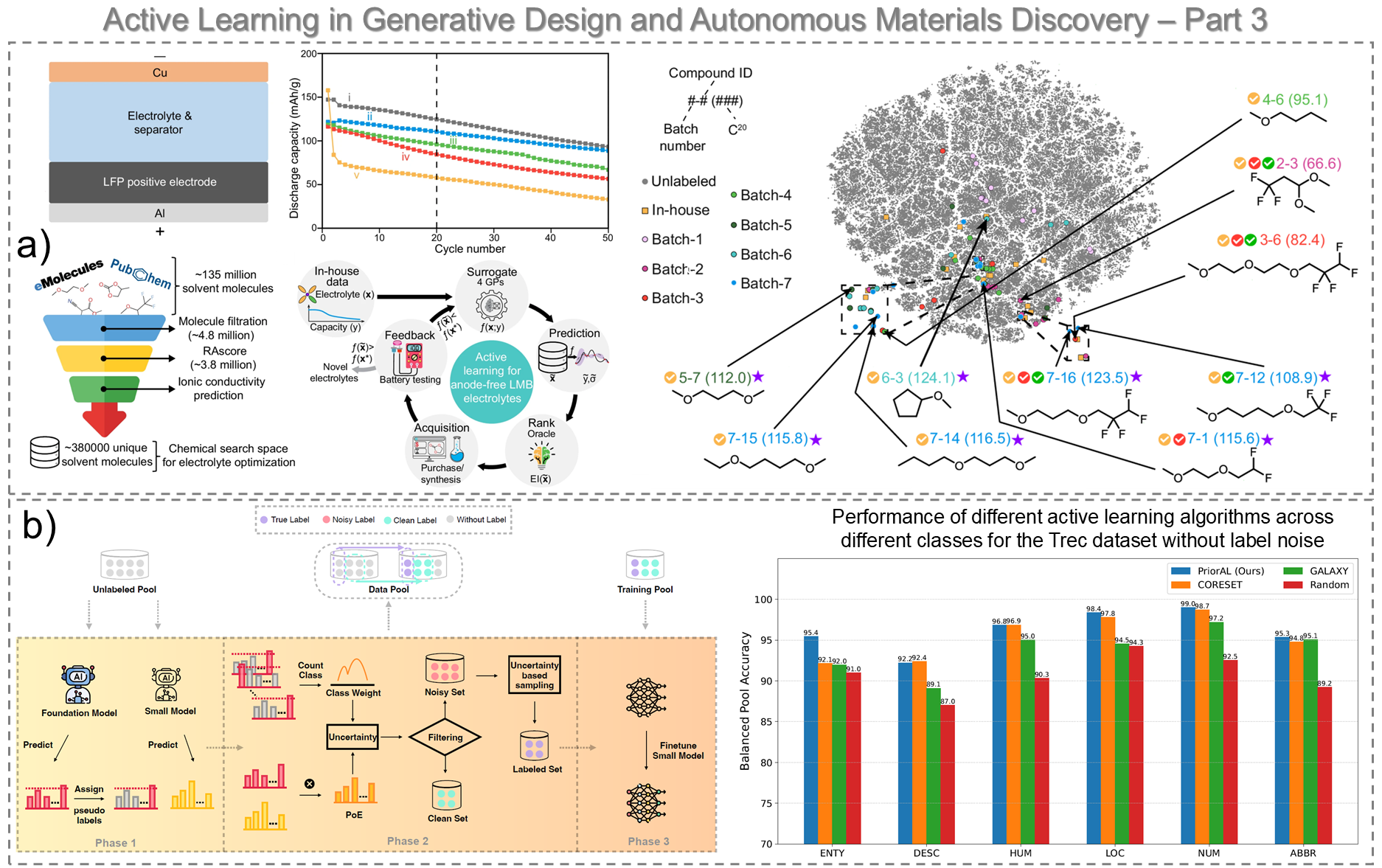}}  \\
\end{minipage}
\caption{\textbf{Closing the Loop with Experiments and Using Foundation-Model Priors for Acquisition.}
\textbf{(a)} Active learning for electrolyte solvents in anode-free lithium-metal batteries (Ref.~\cite{Amanchukwu2025}). 
The Cu$\|$LiFePO$_4$ cell is schematically shown together with illustrative capacity-retention curves and the target property (discharge capacity at the 20th cycle) (top).
The construction of a virtual search space of solvent molecules by filtering, synthesizability scoring, and ionic-conductivity prediction, and the closed loop of surrogate training with Gaussian processes, inference on the search space, ranking by expected improvement, purchase or synthesis, and battery testing (bottom).
A t-SNE map of the complete search space showing where the solvents acquired in each batch lie (right).
\textbf{(b)} Active learning with foundation-model priors under class imbalance (Ref.~\cite{ZhangJ2026}) presenting the framework of prior labeling by a foundation model and a small model, imbalance-aware uncertainty sampling based on the product of experts that separates clean and noisy sets, and training of the small model for the next iteration (left).
Class-wise balanced pool accuracy on the long-tailed TREC dataset at 20\% annotation cost for this method, CORESET, GALAXY, and random selection (right).
Panels adapted from the cited references.
}
\label{fig:sec9c}
\end{figure*}

The pattern of foundation model priors sharpening active learning efficiency is not unique to materials science. 
In a general machine learning setting, Zhang \emph{et al.}~\cite{ZhangJ2026} showed that combining foundation model predictions with a small target model through a class imbalance-aware product of experts scheme reduces annotation cost by more than 50\% relative to the best conventional active learning baseline across 21 imbalanced image and text classification tasks, see Fig.~\ref{fig:sec9c}b.
This evidence suggests that the synergy between foundation models and active learning documented throughout this review is a more general property of how large pretrained models interact with uncertainty-driven data acquisition, rather than an artifact specific to interatomic potentials.

\subsection{Comparison and limitations}

\begin{table*}[htbp]
\centering
\caption{Representative studies of active learning and complementary methods in generative design and autonomous materials discovery (Sec.~\ref{sec9}).}
\label{tab:sec9}
\small
\setlength{\tabcolsep}{3pt}
\resizebox{1\textwidth}{!}{%
\begin{tabular}{|c|l|c|l|l|c|}
\hline
\multicolumn{1}{|c|}{\textbf{\#}} & \multicolumn{1}{|c|}{\textbf{Authors}} & \multicolumn{1}{|c|}{\textbf{Year}} & \multicolumn{1}{|c|}{\textbf{System / Method}} & \multicolumn{1}{|c|}{\textbf{Key Result}} & \multicolumn{1}{|c|}{\textbf{Fig.}} \\
\hline
1 & Xin et al.~\cite{Xin2021} & 2021 & Wide-band-gap materials; BO + autoencoder/GAN & Recovery rate up to +55 pp with 300 initial samples & Fig.~\ref{fig:sec9a}a \\
2 & Zou et al.~\cite{Zou2024} & 2025 & Inorganic compound stability; stacked ensemble & Matches accuracy with 1/7 of training data & Fig.~\ref{fig:sec9b}a \\
3 & Antoniuk et al.~\cite{Antoniuk2025} & 2026 & Organic energetic molecules; JANUS + closed-loop DFT & 75--83\% RMSE reduction over 4 iterations, 500 DFT calls & Fig.~\ref{fig:sec9a}b \\
4 & Wang et al.~\cite{WangG2025} & 2025 & Autonomous synthesis (A-Lab); LLM review & 41/58 novel compounds over 355 experiments, 17 days & -- \\
5 & Ye et al.~\cite{Ye2026} & 2026 & Topological insulators; conditional generation + MLIP & $\sim$5$\times$ higher hit rate vs.\ unconstrained generation & Fig.~\ref{fig:sec9a}c \\
6 & Wang et al.~\cite{WangHC2026} & 2026 & 4 materials datasets; training-free LLM-AL & $>70\%$ fewer evaluations vs.\ exhaustive screening (most settings) & Fig.~\ref{fig:sec9b}b \\
7 & Ma et al.~\cite{Amanchukwu2025} & 2025 & Battery electrolytes; GP Bayesian model averaging & 4 new solvents from 58 initial + $\sim$70 real experiments & Fig.~\ref{fig:sec9c}a \\
\hline
\end{tabular}%
}
\end{table*}

The studies surveyed in this section show how active learning extends beyond interatomic potential fitting to generative composition/structure search, molecular design, and experiment selection, alongside complementary improvements in candidate generation and property prediction (Table~\ref{tab:sec9}).
What sets these discovery workflows apart is that their acquisition goals are downstream outcomes, such as finding stable, high-performing candidates.
Discovery yield and the number of evaluations needed to reach a target therefore provide task-specific measures of success, consistent with the validation concerns raised in Sec.~\ref{sec2}.
As Nair and Foppa~\cite{Nair2026} discuss, discovery rates can improve even when the surrogate model's global prediction error does not decrease.
These studies demonstrate improvements in candidate discovery, predictor accuracy, and data efficiency. 
The magnitude of each improvement depends on the specific objective and comparison baseline. 
For example, Wang et al.~\cite{WangHC2026} reported reductions in the number of iterations required to find the best-performing candidate by over 70\%, compared to traditional machine learning-based active learning methods such as RFR, XGBoost, Gaussian process regression (GPR), and Bayesian neural networks (BNN) with a upper confidence bound (UCB) acquisition function, consistent with the literature's 40-60\% benchmark for data usage for the P3HT/CNT and perovskite datasets.
These data are comparable to those documented for machine learning-based interatomic potential (MLIP) training in Sec.~\ref{sec3}--\ref{sec4}.

These gains reflect improvements to different components of the discovery workflow.
Zou \emph{et al.}~\cite{Zou2024} improve the predictor's data efficiency through a stacked ensemble, while Ye \emph{et al.}~\cite{Ye2026} increase the hit rate approximately fivefold through conditional candidate generation, an exploration strategy.
Antoniuk \emph{et al.}~\cite{Antoniuk2025} show that adding active learning to JANUS enables generation of molecules with better combinations of density and heat of formation than those in the training data, surpassing the JANUS and REINVENT generators without active learning in this comparison.

Among the approaches surveyed here, LLM-guided acquisition remains at an early stage of validation.
Wang \emph{et al.}~\cite{WangHC2026} evaluate candidate selection on four fixed datasets, including repeated runs and different initial observations.
Their workflow reached the optimum in fewer iterations than conventional surrogate models using upper-confidence-bound acquisition in most tested settings, with performance depending on the prompt format and dataset.
Given the limited transparency of model pretraining, evaluation on additional previously unpublished datasets and in prospective experimental campaigns is important for establishing when LLM-guided acquisition can serve as a general-purpose alternative to conventional acquisition strategies.

\section{Benchmarking Foundation Models. Why Active Learning Still Matters?}
\label{sec10}

The sections above discuss the efficiency gains of active learning in terms of each individual study.
However, a separate body of work asks a more difficult question: with the rapid proliferation of large, pre-trained foundation models, is targeted active learning or fine-tuning really necessary, and if so, where and to what extent?

\subsection{Performance across properties and applications}

Liu and Mo~\cite{Liu2024} provided an early, sobering answer at the level of model selection rather than training.
They sampled 2300 already-trained MLIP models from six different architectures and benchmarked each one against 60 different physical property metrics for elemental silicon.
It was found that only 2.8--5\% of the models reached a 0.1~eV benchmark for defect formation energies and only 1--13\% reached the DFT-derived benchmarks for elastic constants.
For 46\% of the five-property combinations, none of the sampled models met all five benchmarks (against 2\% for pairs of properties).
Different architectures led to different results, with neural-network potentials providing the best results for defect formation energies.
Moment tensor and Gaussian-approximation potentials showed the best overall performance for forces on rare-event atoms.
This is a \enquote{curse of dimensionality} in MLIP performance: simultaneously meeting several property benchmarks becomes increasingly difficult for the sampled models.
These trade-offs suggest that active learning targeted at a single property may leave other properties poorly described.
The authors propose iterative training and analysis, adding or reweighting data for challenging properties and then reassessing performance across the property set.

\begin{figure*}[!t]
\begin{minipage}[h]{1\linewidth}
\center{ \includegraphics[width=1\linewidth]{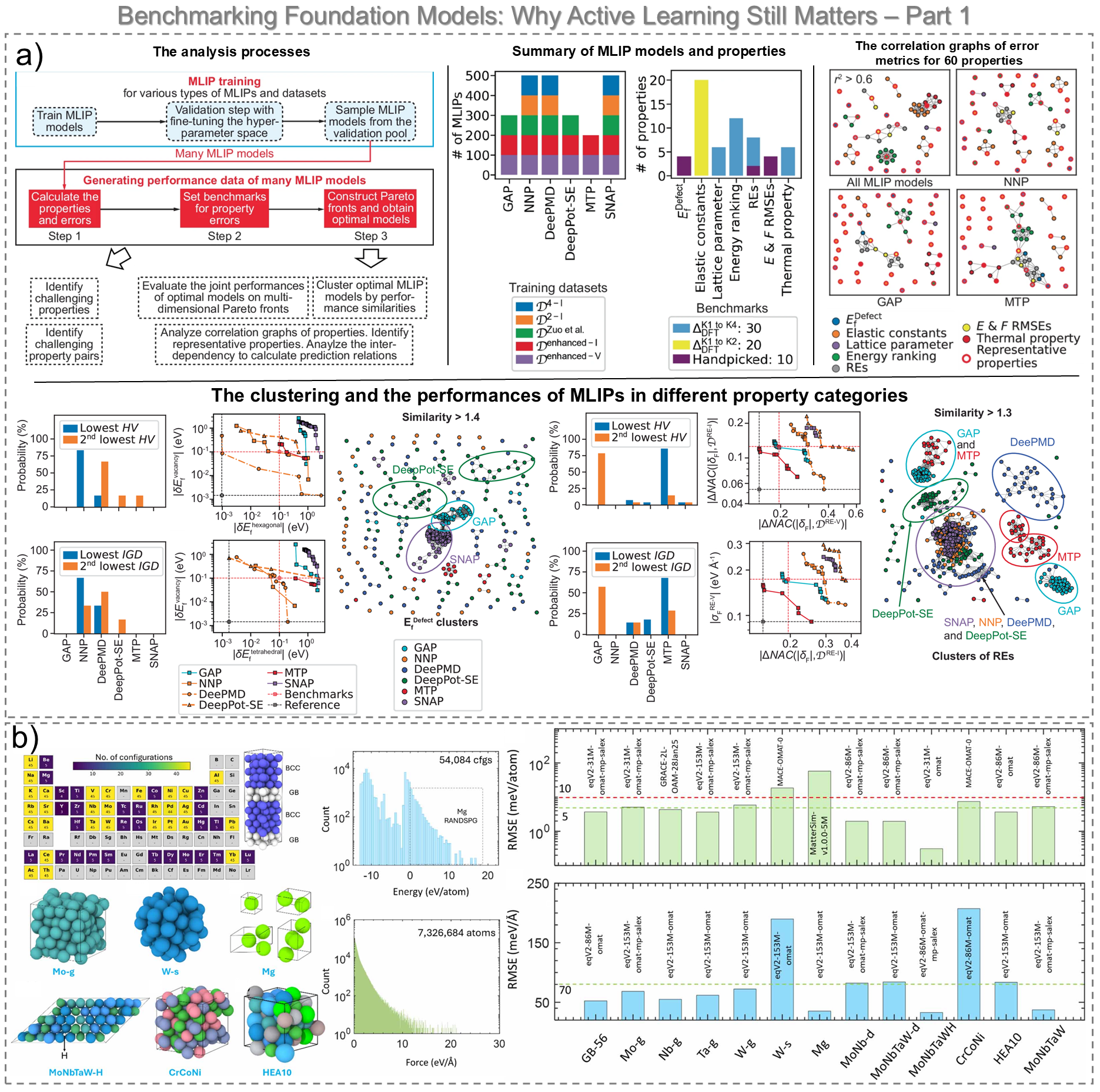}}  \\
\end{minipage}
\caption{\textbf{How Good Are Trained Universal Potentials Across Many Properties and Defects?}
\textbf{(a)} High-dimensional analysis of MLIP performance for silicon (Ref.~\cite{Liu2024}) from MLIP training to sampling a pool of models and evaluating them on many properties to identify challenging properties and correlations (left). 
The 2300 sampled models of six types trained on five datasets and the 60 properties in seven categories with their benchmarks (middle).
Pareto fronts, the fractions of optimal models of each type, and clusters of optimal models for the defect-formation-energy and rare-event categories (right).
Correlation graphs of the error metrics ($r^2>0.6$) for all models and for GAP, NNP, and MTP separately (bottom).
\textbf{(b)} Universal MLIPs for defects in metals and alloys (Ref.~\cite{Shuang2025}) with presented used DFT datasets, including grain boundaries in 56 elements, general-purpose Mo and spherical W-cluster datasets, Mg structures, H near a screw dislocation in MoNbTaW, CrCoNi, and a ten-element alloy, with their energy and force distributions (left).
The energy and force RMSE of the best-performing universal model for each dataset.
Panels adapted from the cited references (right).
}
\label{fig:sec10a}
\end{figure*}

Shuang \emph{et al.}~\cite{Shuang2025} compared 26 pre-trained universal MLIPs to 54\,084 reference configurations from density functional theory (DFT) spanning grain boundaries, point defects, and solute-defect interactions in metals and random alloys, see Fig. ~\ref{fig:sec10a}b.
They found that the best zero-shot foundation model matched or outperformed specialized, actively trained moment tensor and ACE potentials on the general-purpose tungsten dataset.
This surprising result suggests that cross-element generalization can sometimes replace system-specific active learning. 
However, atomic clusters with free surfaces were poorly described by all tested models, with the best RMSE of $\sim190$ meV/\AA{}.
Even the most accurate model had force errors above 100 meV/\AA{} for grain boundaries in magnetic metals such as Cr and Ni.
This indicates that targeted correction, either through active learning or fine-tuning, may still be necessary in these cases.
Kempen, Cheula, and Andersen~\cite{Kempen2025}, as shown in Fig.~\ref{fig:sec10b}a, extended this benchmarking approach to heterogeneous catalysis.
They systematically evaluated the zero-shot performance of 80 different fundamental MLIPs on adsorption, reaction, and vibrational property benchmarks for alloyed metal, oxide, and metal oxide interface surfaces.
They found that current-generation models, such as eSEN, Orb, and UMA, can already predict perovskite-oxide vacancy formation energies and nanocluster zero-point energies with high accuracy.
However, they also found that a significant number of these models fail catastrophically when applied to magnetic surfaces.
Here, the approach proposed in Refs.\cite{Kotykhov2024,novikov2022} might be successfully applied, see Sec.~\ref{sec7}.
Additionally, they discovered that relaxing the structure using the MLIP, instead of simply evaluating a single point on a DFT-relaxed structure, systematically increases the prediction error for even the most accurate models.
They also found that low-cost, task-specific models trained directly on target data sets are competitive with or even superior to more general-purpose, foundational MLIP models when used in a zero-shot setting.
This finding confirms the overall conclusion that no single model or training strategy (from scratch, fine-tuning, or zero-shot deployment) dominates all applications.

Du \emph{et al.}~\cite{Du2025} benchmarked twelve universal machine learning potentials (MLIPs) specifically for solid-state electrolyte research against a large active learning-generated dataset of 59\,868 structures spanning 18 sulfide, oxide, and halide systems.
They found that MatterSim model, which was trained partly through active learning with a dedicated data exploration tool, provided the best overall performance in terms of formation energy, elastic moduli, and phonon properties.
Other models showed complementary strengths and weaknesses, namely MACE excelled at thermodynamics but had larger errors in elastic moduli, ORB v2 was fast but systematically biased.

\subsection{Distribution shift and limits of fine-tuning}

\begin{figure*}[!t]
\begin{minipage}[h]{1\linewidth}
\center{ \includegraphics[width=1\linewidth]{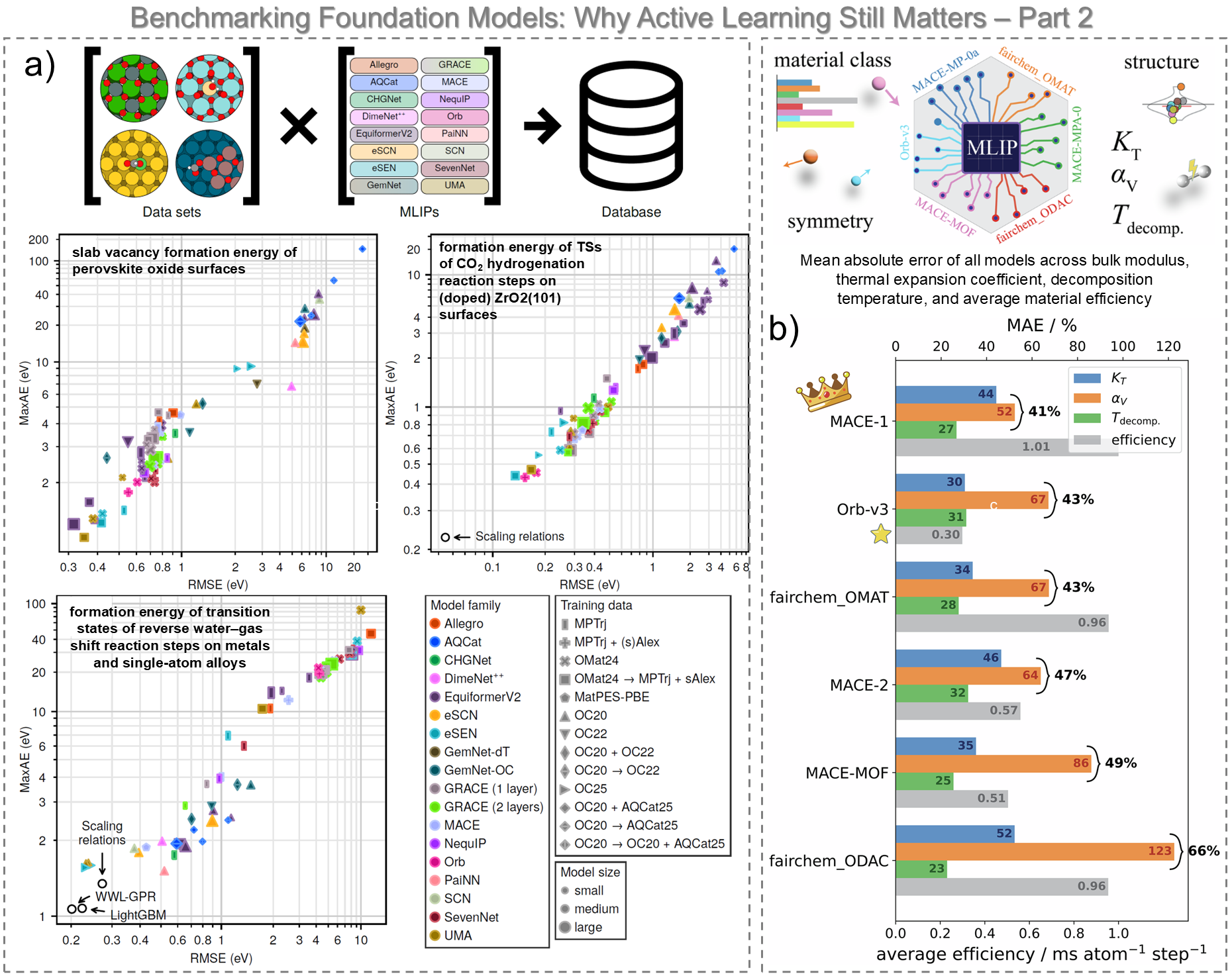}}  \\
\end{minipage}
\caption{\textbf{Where Zero-Shot Foundation Potentials Fall Short. Heterogeneous Catalysis and Mechanical Properties.}
\textbf{(a)} Foundational MLIPs for heterogeneous catalysis (Ref.~\cite{Kempen2025}) where the six catalysis datasets and 80 pretrained foundational MLIPs assessed (top).
The prediction performance of these models, grouped by family and training data, for slab vacancy formation energies of perovskite-oxide surfaces and for transition-state formation energies of CO$_2$ hydrogenation on (doped) ZrO$_2$(101) and of the reverse water--gas shift on metals and single-atom alloys (bottom).
\textbf{(b)} A graphical overview of the benchmark of six universal MLIPs (Ref.~\cite{Stracke2026}) on the bulk modulus, thermal expansion, and decomposition temperature of nine metal--organic frameworks and four inorganic compounds (top). 
The mean absolute errors of all models for these three properties and on average, with the best-performing and the fastest models marked (bottom).
Panels adapted from the cited references.
}
\label{fig:sec10b}
\end{figure*}

Stracke \emph{et al.}~\cite{Stracke2026} presented the most direct warning regarding fine-tuning, see Fig. \ref{fig:sec10b}b.
They benchmarked six foundation ML models across three architecture types for predicting mechanical properties (bulk modulus, thermal expansion, and decomposition temperature) of nine metal-organic frameworks and four inorganic compounds. 
They found that all models systematically underestimated the bulk modulus and overestimated thermal expansion (average absolute errors of 40\% and 76\%, respectively), which is consistent with the potential energy surface softening reported by Deng \emph{et al.}~\cite{Deng2024} for other properties. 
However, a domain-specific fine-tuned model (MACE-MOF, fine-tuned on 127 MOFs) showed no measurable improvement over MACE-MP-0a on the tested MOF subset and had reduced transferability to inorganic compounds.
In the comparison of MACE-MPA-0 with MACE-MP-0a, enlarging the pre-training dataset by millions of structures also reduced accuracy for structures already well represented in the original training set.
These results provide a counterpoint to the fine-tuning gains discussed in Sec.~\ref{sec4}.
Fine-tuning can help when the target chemistry or configurations are underrepresented and the additional data are informative, but its success also depends on how the model is updated.

Kreiman and Krishnapriyan~\cite{Kreiman2025}, introduced in Sec.~\ref{sec:complementary}, formally decompose the reasons why even a foundation model trained on 951\,000 structures exhibits 2-10$\times$ larger force error under distribution shift.
This evidence shows that dataset scale alone, without targeted, distribution-aware correction, is not sufficient to close the generalization gap that active learning was designed to address.

\subsection{Observable-specific validation}

Tahmasbi \emph{et al.}~\cite{Tahmasbi2026} benchmarked six machine learning-based interatomic potentials (MLIPs) on 15 elementary systems using both the equation-of-state accuracy and the minima-hopping global structure search method.
They found that potential-energy surface smoothness (search efficiency) and structural/energetic fidelity are not correlated.
A model can navigate the potential energy landscape efficiently, while still incorrectly ranking competing low-energy structures.
All models performed poorly on pairwise energy ordering accuracy, a stringent test of fine energy resolution.
This motivates active-learning-guided fine-tuning focused on competing low-energy structures.

Creed \emph{et al.}~\cite{Creed2026} reported in a wide-ranging perspective six open questions about the foundation of MLIP models. 
They suggest that compact datasets of 400\,000 structures, carefully selected via a two-stage process similar to DIRECT, can match or even exceed models trained on much larger, less curated datasets. 
Additionally, they note that multi-fidelity active learning, mixing 90\% cheap and 10\% expensive functional data, can be as effective as training on eight times as much pure high-fidelity data.
Most importantly for this review, they highlight benchmark results in which migration-barrier errors of $0.74\pm0.06$~eV coexist with energy errors that are two orders of magnitude smaller.
This illustrates why average energy errors can be poor proxies for the accuracy of physical observables.
As both Nair and Foppa~\cite{Nair2026} argue, we need to shift towards observable-targeted validation rather than aggregate-error-targeted validation.

\subsection{Implications for active learning}

\begin{table*}[htbp]
\centering
\caption{Representative studies of foundation MLIP performance and validation, with related benchmarks of system-specific potentials (Sec.~\ref{sec10}).}
\label{tab:sec10}
\small
\setlength{\tabcolsep}{3pt}
\resizebox{1\textwidth}{!}{%
\begin{tabular}{|c|l|c|l|l|c|}
\hline
\multicolumn{1}{|c|}{\textbf{\#}} & \multicolumn{1}{|c|}{\textbf{Authors}} & \multicolumn{1}{|c|}{\textbf{Year}} & \multicolumn{1}{|c|}{\textbf{Benchmark Scope}} & \multicolumn{1}{|c|}{\textbf{Key Finding}} & \multicolumn{1}{|c|}{\textbf{Fig.}} \\
\hline
1 & Liu \& Mo~\cite{Liu2024} & 2024 & 2300 MLIP models, Si, 60 property metrics & No model meets all 5 benchmarks for 46\% of property sets & Fig.~\ref{fig:sec10a}a \\
2 & Deng et al.~\cite{Deng2024} & 2024 & M3GNet, CHGNet, MACE-MP-0; surfaces/defects/barriers & Systematic PES softening away from equilibrium & -- \\
3 & Kreiman \& Krishnapriyan~\cite{Kreiman2025} & 2026 & 7 foundation models; distribution-shift diagnosis & 2--10$\times$ larger force error under shift & -- \\
4 & Shuang et al.~\cite{Shuang2025} & 2025 & 26 uMLIPs, 54k configs; metal/alloy defects & Best uMLIP matches specialized MTP/ACE on W defects & Fig.~\ref{fig:sec10a}b \\
5 & Beiersdorfer et al.~\cite{Beiersdorfer2026} & 2026 & GAP vs.\ MACE; liquid electrolyte & Architecture affects speed of AL failure-mode fixes & -- \\
6 & Creed et al.~\cite{Creed2026} & 2026 & Foundation-model open questions & 0.008~eV energy error; 0.74~eV barrier error & -- \\
7 & Du et al.~\cite{Du2025} & 2025 & 12 uMLIPs, 18 solid electrolytes & MatterSim (partly AL-trained) best overall & -- \\
8 & Li \& Walsh~\cite{LiZ2026} & 2026 & 7 foundation models; latent-space geometry & Shared manifold despite architectural diversity & -- \\
9 & Tahmasbi et al.~\cite{Tahmasbi2026} & 2026 & 6 uMLIPs, 15 elements; EOS + minima hopping & PES smoothness decoupled from energy-ranking accuracy & -- \\
10 & Stracke et al.~\cite{Stracke2026} & 2026 & 6 uMLIPs, 13 materials; mechanical properties & MACE-MOF: no MOF gain over MACE-MP-0a; reduced transferability & Fig.~\ref{fig:sec10b}b \\
11 & Kempen et al.~\cite{Kempen2025} & 2025 & 80 foundational MLIPs; heterogeneous catalysis & Catastrophic failure on magnetic surfaces; no single best model & Fig.~\ref{fig:sec10b}a \\
\hline
\end{tabular}%
}
\end{table*}

These benchmarks underscore the continuing importance of active learning for foundation MLIPs.
Even broadly pretrained models retain weaknesses in high-energy environments~\cite{Deng2024}, unfamiliar configurations~\cite{Kreiman2025}, and the prediction of specific physical observables~\cite{Tahmasbi2026,Creed2026}.
Active learning provides a systematic way to target additional reference calculations toward poorly described configurations and refine the model iteratively.
This makes it particularly valuable when improving performance for a specific application under a limited reference-calculation budget.

The strong zero-shot performance reported by Shuang \emph{et al.}~\cite{Shuang2025} and the adverse fine-tuning results of Stracke \emph{et al.}~\cite{Stracke2026} also show why the need for additional training should be assessed against the intended observables, using the pre-trained model as a baseline.
As discussed in Sec.~\ref{sec4}, fine-tuning determines how a model is updated, whereas active acquisition determines which reference data are added.
To assess the benefit of active selection, comparisons should hold the starting model, training procedure, and reference-label budget fixed and measure accuracy in the target observables.

\section{Conclusions and Outlook}

\textit{What has active learning already achieved?} 
Across the more than seventy studies reviewed in this paper, active learning has repeatedly demonstrated substantial savings in reference calculations, including reductions of one to several orders of magnitude relative to naive \emph{ab initio} molecular-dynamics sampling for system-specific potentials (Sec.~\ref{sec3}).
These benefits extend to active fine-tuning of foundation models (Sec.~\ref{sec4}), reactive pathways and spectroscopic observables beyond the practical reach of direct \emph{ab initio} dynamics (Sec.~\ref{sec6}), and materials-discovery searches for rare, high-performance candidates (Sec.~\ref{sec8}--\ref{sec9}).
In discovery workflows, success is measured by the candidates found and the evaluations needed to find them, alongside predictor accuracy.
These successes span Gaussian processes, moment tensor potentials, linear SNAP models, kernel methods, and a wide range of equivariant graph-neural-network potentials, using acquisition criteria such as query-by-committee, Bayesian posterior variance, D-optimality/MaxVol, and kernel or representation distances computed on frozen foundation-model embeddings.
Uncertainty-biased dynamics contributes through exploration, extending the set of configurations available for acquisition (Sec.~\ref{sec:exploration}).
Beyond DFT-benchmark accuracy, actively learned models have reproduced experimental Arrhenius diffusion curves (Sec.~\ref{sec7}), ferroelectric transition temperatures (Sec.~\ref{sec7}), superionic phase transitions (Sec.~\ref{sec3}), and reaction quantum yields (Sec.~\ref{sec6}).
In several cases, it has resolved mechanisms (competing hydrogen-diffusion pathways in Ni--Mn alloys~\cite{Ito2025}, a long-timescale photoisomerization re-equilibration~\cite{Martyka2025}) that had not been captured by previous methods. 

\textit{What are the current pitfalls and limitations?} 
As Table~\ref{tab:predict} shows, active learning has a stronger foundation as a tool to reduce the cost of achieving a pre-defined, single-task accuracy goal than as a general solution for ensuring dataset completeness, representativeness, or transferability. 
Actively selected training data are generally not independent and identically distributed samples from the target distribution, which can introduce active-learning bias.
Sampling strategies that focus on informativeness alone, as documented in Nair \emph{et al.}~\cite{Nair2026}, have been shown to result in models that perform worse than random sampling in at least one case.
No single uncertainty quantification method dominates across architectures and scenarios.
Query-by-committee, while simple and widely used, has shown weak and round-to-round unstable correlations with true error~\cite{VargaUmbrichA2026}, and reliable calibration in the unlabeled pool regime where acquisition decisions are made remains largely unsolved \cite{Park2026, Nair2026}.
Validated committees nevertheless remain a practical option for suitable tasks~\cite{Shuang2025}, as discussed in Sec.~\ref{sec5}.
Fine-tuning foundation MLIPs can greatly reduce data requirements (Sec.~\ref{sec4}), but its benefits depend on both the starting model and the update procedure.
For example, MACE-MOF showed no improvement over MACE-MP-0a on the tested metal-organic frameworks and reduced transferability to inorganic compounds~\cite{Stracke2026}.
Fine-tuned models may also inherit rather than correct systematic artifacts from the model or dataset they were trained on~\cite{HanserothA2026}, and distilled models may fall short of the teacher's performance~\cite{ZhangW2026}.
Finally, average error metrics (energy MAE, force RMSE), which are reported in almost every study surveyed here, have been shown to be poor proxies for observable-specific errors.
These include migration barriers, quantum yield, and short-range order parameters, which actually determine the scientific or engineering value downstream \cite{Creed2026,Zhang2025}. 
Validation should therefore assess the physical observables of interest directly.

\textit{Do we really need to minimize aggregate MLIP errors?}
Most of the studies discussed here are driven by the goal of minimizing prediction error relative to first-principles calculations.
This pursuit is essentially a mathematical optimization problem, where metrics like MAE or RMSE are reduced as much as possible, often without explicit regard for whether the resulting model captures the underlying physics. 
However, for genuine physical understanding, this purely numerical framing is insufficient and can even be misleading: driving aggregate test errors toward zero does not guarantee that the model reproduces the correct mechanisms, scaling relations, or thermodynamic behavior of the system.

A physically correct machine learning model is far more useful than one that achieves the lowest possible numerical error but fails to describe the actual physical process.
For instance, a model with an error orders of magnitude larger than the best-performing MLIP may still be preferable if it correctly captures bond breaking, reaction pathways, or adsorption trends, whereas a low-error model that violates physical constraints may yield incorrect mechanistic conclusions.
Therefore, the field should shift its emphasis from pure error minimization toward physical fidelity, ensuring that MLIPs are not only accurate but also physically consistent and interpretable.
This requires the development of validation protocols (similar to those discussed in Sec.~\ref{sec10}) that explicitly probe the physical phenomena of interest, rather than relying solely on aggregate error statistics.

\begin{table*}[htbp]
\centering
\caption{Demonstrated benefits and limitations of active learning and complementary training strategies for interatomic-potential and materials-discovery workflows.}
\label{tab:predict}
\small
\setlength{\tabcolsep}{4pt}
\resizebox{1\textwidth}{!}{%
\begin{tabular}{|l|l|l|}
\hline
  \multicolumn{1}{|c|}{\textbf{Claim}}
  & \multicolumn{1}{|c|}{\textbf{Evidence in this review}}
  & \multicolumn{1}{|c|}{\textbf{What is \emph{not} thereby guaranteed}} \\
\hline
  Fewer DFT calls for a fixed accuracy target
  & One to several orders of magnitude vs.\ naive AIMD, Sec.~\ref{sec3}
  & Coverage outside the sampled trajectory/task \\
  Improved materials-discovery efficiency
  & Higher hit rates; fewer evaluations to reach targets, Sec.~\ref{sec8}--\ref{sec9}
  & Uniform savings across targets and comparison baselines \\
  Convergence of average energy/force error
  & Routinely reported and reproduced, Sec.~\ref{sec3}--\ref{sec8}
  & Accuracy of specific downstream observables (barriers, SRO) \cite{Creed2026,Zhang2025} \\
  Reproduction of a targeted experimental observable
  & Diffusion, phase transitions, spectra, Sec.~\ref{sec7},~\ref{sec6}
  & Generalization to untrained phases, defects, or chemistries \\
  Fine-tuning can match or exceed from-scratch accuracy
  & 2--10\% of from-scratch data, Sec.~\ref{sec4}
  & Gains from active selection; retained transferability \\
  Uncertainty can flag unreliable predictions
  & Successful Bayesian, kernel, and committee examples, Sec.~\ref{sec3},~\ref{sec5}
  & Calibration; acquisition gains on a new task \\
  A well-converged model is representative of its task
  & Not directly tested by most surveyed studies
  & Freedom from active-learning bias / i.i.d.\ violation \cite{Nair2026} \\
\hline
\end{tabular}%
}
\end{table*}

\textit{What is the realistic near-term opportunity?} 
Several of the gaps identified above can be addressed using methods that have already been demonstrated in the literature, rather than requiring the development of new theory.
For example, pretrained gradient and activation kernels are a compelling alternative to query-by-committee: the MACE-based studies in Sec.~\ref{sec5} demonstrate acquisition gains in reactive chemistry and improved accuracy at lower per-round computational cost on a catalyst benchmark.
Extending the chunked, linearly scaling feature-space implementation of neural tangent kernel (NTK) acquisition~\cite{VargaUmbrichB2026} to other architectures and application domains presents a concrete opportunity.
Coupling active learning explicitly to the physical observable of interest is another priority.
For example, exploration can target migration pathways using the nudged elastic band method or configurations relevant to short-range order in high-entropy materials, with acquisition selecting the configurations that need reference calculations.
Validating the resulting barriers or order parameters directly would address the metric mismatch highlighted by Creed \emph{et al.} and the roadmap of Zhang \emph{et al.}~\cite{Creed2026,Zhang2025}.
Systematic comparisons at matched labeling budgets should become standard practice, building on the controlled comparisons reviewed here~\cite{Lee2026,VargaUmbrichA2026,Achar2025}.
Informativeness-only and diversity-aware acquisition should be compared within each training regime, from scratch and fine-tuning, while holding the exploration strategy and model-update schedule fixed.
For high-entropy and compositionally disordered materials (Sec.~\ref{sec8}), it is important to check which parts of the target compositional and configurational space are represented in a foundation model's pretraining data.
Screening the MPTrj dataset, as done by Zhou and Komsa~\cite{Zhou2026}, provides an example of identifying such coverage gaps.
These checks can guide targeted exploration and additional reference calculations, helping prevent unrecognized extrapolation.

\textit{What would constitute a true active-learning-driven materials-discovery framework?}
Active learning's repeatedly demonstrated value lies in directing reference calculations toward the needs of a target application, with validation establishing whether the resulting model and dataset are adequate for that task.
An effective workflow links the following stages: candidate generation (MD, metadynamics, generative models, or experimental design) $\rightarrow$ selection using an acquisition criterion $\rightarrow$ targeted, budget-constrained reference labeling $\rightarrow$ model or prompt update $\rightarrow$ validation against a task-relevant metric $\rightarrow$ deployment.
New observations feed back into subsequent rounds, while task-specific validation guides the decision to continue learning or deploy.
Such a workflow makes the four design choices in Sec.~\ref{sec2} explicit: the organization of the feedback loop, acquisition criterion, exploration strategy, and model training and updating regime.
A well-executed active learning workflow keeps these choices auditable, reporting the number of DFT calculations or experiments saved and the comparison baseline, how the acquisition criterion was evaluated, the representativeness of the dataset relative to the deployment task, and the specific observable validated.
For uncertainty-based acquisition, this evaluation should also assess calibration of the reported uncertainty.
Read this way, more than seventy studies surveyed here contain most of the elements of such a framework: on-the-fly feedback loops, pool-based selection for foundation-model fine-tuning, acquisition based on kernels and representations, reaction-aware exploration, and a mature, self-critical culture of benchmarking willing to report when active learning does not help.
What remains is to consistently assemble these elements and hold every new application of active learning to the standard of task-specific, critically validated evidence set by the most rigorous studies surveyed.
This standard extends beyond interatomic potentials, given the shared data-generation challenges of atomistic simulations in chemistry, materials science, and biology.

\section*{Acknowledgements}
The study was supported by the Russian Science Foundation grant 25-23-01058  (https://rscf.ru/en/project/25-23-01058/).

\bibliography{TEXT_Active_Learning}
\bibliographystyle{apsrev4-2}

\end{document}